\documentclass[a4paper,12pt]{article}
\pdfoutput = 1

\usepackage{jheppub}

\title{\boldmath Deriving the long-string CFT in AdS$_3$}

\abstract{We revisit the path integral formulation of bosonic string theory in locally-AdS$_3$ spacetimes. Through a careful analysis of the worldsheet sigma model, we write down an effective theory of long strings living near the boundary of AdS$_3$. By directly computing the partition function of the long-string sector, we find that the worldsheet path integral naturally organizes itself into the Coulomb-gas expansion of a 2D conformal field theory on the asymptotic boundary, with the number of coulomb integrals being dual to the dimension of the moduli space of worldsheet instantons in the bulk. As such, we provide a derivation of the CFT dual of long strings in AdS$_3$ for completely generic spacetime and worldsheet topologies.}

\author{Bob Knighton}
\affiliation{Department of Applied Mathematics \& Theoretical Physics, University of Cambridge,\\
Wilberforce Road, Cambridge CB3 0WA, United Kingdom}
\emailAdd{rik23@cam.ac.uk}

\usepackage{bm} % Bold Greek letters
\usepackage[dvipsnames]{xcolor}
\definecolor{green_maf}{RGB}{28, 166, 46}
\definecolor{blue_mrg}{RGB}{12, 143, 145}
\definecolor{detail}{RGB}{110,110,110}
\usepackage{amsmath} % Mathematics
\usepackage{nccmath} % Mathematics
\usepackage{amssymb} % \mathbb{•}
\usepackage{amsthm} % For proof environment
\usepackage{mathtools} % To be able to define bra, ket and braket
\usepackage[utf8]{inputenc} % For special characters in an Inspire reference
\usepackage{braket}
\usepackage{enumerate}
\usepackage[many]{tcolorbox}
\usepackage[inline]{enumitem}
\usepackage{comment}
\usepackage{soul} % To strikeout text
\usepackage{tikz}
\usepackage{csquotes}
\usepackage{mathrsfs}
\usepackage{tensor}
\usepackage{dsfont}

\newtcolorbox{empheqboxed}{colback=gray!30, 
 colframe=white,
 width=\textwidth,
 sharpish corners,
 top=-2mm, % default value 2mm
 bottom=0pt
}

\hypersetup{
		pdfencoding=unicode,
		colorlinks=true,
		urlcolor=Maroon,
		linkcolor=RoyalBlue,
		citecolor=Maroon,
		pdfstartview=FitH,
		linktocpage=true
}

\usetikzlibrary{calc,decorations.markings}
\usetikzlibrary{decorations.pathmorphing}
\usetikzlibrary{shapes,backgrounds}
\usetikzlibrary{fadings}
\usetikzlibrary{cd}
\usetikzlibrary{decorations.pathreplacing,calligraphy}
\usetikzlibrary{hobby}

\tikzset{
	partial ellipse/.style args={#1:#2:#3}{
		insert path={+ (#1:#3) arc (#1:#2:#3)}
	}
}

\tikzset{
  every overlay node/.style={
    draw=black,fill=white,rounded corners,anchor=north west,
  },
}

\tikzfading
[
  name=fade out,
  inner color=transparent!0,
  outer color=transparent!100
]
\newif\ifdetails
\detailstrue

\def\be{\begin{equation}}
\def\ee{\end{equation}}

\begin{document}

\maketitle

\newpage

\section{Introduction}

In the context of the AdS/CFT correspondence \cite{Maldacena:1997re,Witten:1998qj}, locally-$\text{AdS}_3$ spacetimes supported by pure NS-NS flux provide a particularly interesting corner in the space of holographic backgrounds in string theory. On the one hand, the RNS formalism of the worldsheet theory admits a tractable description in terms of a WZW model on $\text{SL}(2,\mathbb{R})$ (or the coset $\text{SL}(2,\mathbb{C})/\text{SU}(2)$ in Euclidean signature) \cite{Maldacena:2000hw}. On the other hand, the worldsheet spectrum of these backgrounds contains a continuum of states, arising from so-called `long strings' -- worldsheet configurations which can wrap the conformal boundary of $\text{AdS}_3$ with finite cost of energy. This feature of the worldsheet theory implies that the tentative dual CFT must also admit a continuous spectrum, and thus cannot be a traditional, compact CFT. Both of these features -- the tractability of the worldsheet theory and the non-compactness of the dual CFT -- are unique to $\text{AdS}_3$ compactifications, and are both spoiled by the addition of any amount of RR-flux.

Since the continuum of states in the dual CFT arises from the existence of long strings in the bulk, one can obtain information about scattering of non-normalizable states in the spacetime CFT by scattering long strings. One can in fact consider a further simplification, in which all intermediate strings states also wrap the conformal boundary, i.e. that one does not allow short strings in the sum over histories of the worldsheet path integral. This essentially boils down to studying a subsector -- the `long-string sector' -- of the bulk worldsheet theory in which all worldsheets live close to the asymptotic boundary.\footnote{In \cite{Knighton:2023mhq,Knighton:2024qxd,Sriprachyakul:2024gyl}, the term `near-boundary limit' was used for the same concept.} This subsector should in turn be dual to a subsector of the dual CFT, which is referred to in the literature as the `long-string CFT' \cite{Seiberg:1999xz}.

Since closed bosonic string theory only contains in its massless spectrum the graviton, dilaton, and Kalb-Ramond 2-form, bosonic backgrounds of the form $\text{AdS}_3\times C$ provide a toy model for pure NS-NS superstring backgrounds. For such backgrounds, it has long been argued \cite{Hosomichi:1999uj,Hikida:2000ry,Argurio:2000tb} that the long-string CFT is described by the symmetric product orbifold of the CFT of a single long-string, which in turn is given by a sigma model in the compact space $C$, as well as a linear dilaton $\phi$, whose momentum states furnish the continuum of long strings in the bulk \cite{Seiberg:1999xz}. In \cite{Eberhardt:2021vsx},\footnote{See also \cite{Balthazar:2021xeh} for a similar proposal for superstring backgrounds of the form $\text{AdS}_3\times\text{S}^3_b$ with $L_{\text{AdS}}<\ell_s$.} this proposal was refined, and it was proposed that the dual CFT of bosonic string theory on $\text{AdS}_3\times C$ (not just the long-string sector) is given by the deformed symmetric orbifold\footnote{Strictly speaking, the dual CFT is a grand canonical ensemble of theories with varying values of $N$, see Section \ref{sec:dual-cft}.}
\begin{equation}\label{eq:dual-cft-introduction}
\text{Sym}^N(\mathbb{R}_{\mathcal{Q}}\times C)+\mu\int\sigma_{2,\alpha}\,,
\end{equation}
where $\mathbb{R}_{\mathcal{Q}}$ is a linear dilaton whose background charge $\mathcal{Q}$ is determined by the AdS$_3$ radius in string units. The deformation operator $\sigma_{2,\alpha}$ is a twist field in the 2-cycle twisted sector of the symmetric orbifold, which carries a certain momentum $\alpha$ in the $\phi$-direction. Evidence for the above proposal is a full matching of the continuous spectra on both sides \cite{Eberhardt:2019qcl}, the agreement residues of poles of sphere 2-, 3-, and 4-point functions \cite{Dei:2021xgh,Dei:2021yom,Eberhardt:2021vsx,Bufalini:2022toj,Dei:2022pkr}, and later sphere $n$-point functions \cite{Knighton:2023mhq,Knighton:2024qxd} and even $n$-point function at higher-genus \cite{Hikida:2023jyc} (see also \cite{Sriprachyakul:2024gyl,Yu:2024kxr} for progress in the supersymmetric case).

In the present work, we are less interested in the equality of correlation functions of non-normalizable operators in the boundary/bulk, and more interested in the equality between partition functions
\begin{equation}
\mathfrak{Z}_{\text{CFT}}=\mathfrak{Z}_{\text{string}}\,.
\end{equation}
If the symmetric orbifold \eqref{eq:dual-cft-introduction} indeed describes strings in $\text{AdS}_3\times C$, then it should in principle be possible to verify the equality of the partition functions. On the CFT side, this would consistitute formulating the theory on a Riemann surface $X$ and computing the path integral. In the bulk, one would instead compute the worldsheet path integral on a spacetime of the form $M\times C$, where $M$ is a hyperbolic 3-manifold whose conformal boundary is the surface $X$.\footnote{In addition, one would in principle be instructed to sum over topological classes of 3-manifolds $M$ with the same conformal boundary.}

The full string partition function on locally-$\text{AdS}_3$ backgrounds has indeed been computed in special cases, for example global $\text{AdS}_3$ and thermal $\text{AdS}_3$ \cite{Gawedzki:1991yu,Maldacena:2000hw,Maldacena:2000kv}. However, for more complicated geometries, such as higher-genus handlebodies, the computation of the full string partition function is much more complicated. Instead, it is conceptually easier to compute the path integral of the long-string subsector -- that is, the path integral of worldsheets which are constrained to wrap the asymptotic boundary of $M$. One reason for this simplification is that, since the long-strings live near the boundary, the target space they see is not the full 3-manifold $M$, but rather the foliation $X\times\mathbb{R}_{\geq 0}$. The long-string partition function is given perturbatively by the expansion
\begin{equation}
\mathfrak{Z}^{(0)}_{\text{string}}=\exp\left(\sum_{g=0}^{\infty}g_s^{2g-2}\mathcal{F}_g^{(0)}\right)\,,
\end{equation}
where $\mathcal{F}_g^{(0)}$ is the genus-$g$ free energy of the long-string subsector.

On the CFT side, on the other hand, one can consider computing $\mathfrak{Z}_{\text{CFT}}$ perturbatively in $\mu$. Strictly speaking, since the marginal operator $\sigma_{2,\alpha}$ is not normalizable, a perturbative expansion in $\mu$ cannot yield the correct result for the partition function. This is analogous to the case of Liouville theory, for which the coulomb-gas expansion provides only a part of the full correlation functions \cite{Dorn:1994xn,Zamolodchikov:1995aa}. That said, the perturbative series in $\mu$ still defines an observable in the CFT \eqref{eq:dual-cft-introduction}, and so we can formally write down the perturbative expansion
\begin{equation}\label{eq:perturbative-series-introduction}
\mathfrak{Z}_{\text{CFT}}^{\text{pert}}=\sum_{m=0}^{\infty}\frac{(-\mu)^m}{m!}\Braket{\left(\int_X\sigma_{2,\alpha}\right)^m}\,.
\end{equation}
At every order in perturbation theory, the correlators are in principle calculable by employing the covering space method of Lunin and Mathur \cite{Lunin:2000yv,Lunin:2001pw}, however the $m$ integrals over the operator insertions provide a large amount of computational complexity.

In the present work, we calculate the perturbative long-string partition function $\mathfrak{Z}_{\text{string}}^{(0)}$. Inspired by the analysis of \cite{Seiberg:1999xz}, we first provide a semiclassical computation of the long-string partition function using the Nambu-Goto action as the basis of the worldsheet theory. After that, we carefully compute the Polyakov path integral, which provides important quantum corrections to the Nambu-Goto analysis. In the end, we arrive at an answer for the partition function $\mathfrak{Z}_{\text{string}}^{(0)}$, which organizes itself into a sum over `worldsheet-instanton' sectors. The novel feature of our analysis is that it is valid for any Riemann surface $X$, and is exact at all orders in string perturbation theory. By comparing to the perturbative CFT partition function, we find a precise match
\begin{equation}\label{eq:pert-ads/cft}
\mathfrak{Z}_{\text{string}}^{(0)}=\mathfrak{Z}_{\text{CFT}}^{\text{pert}}\,,
\end{equation}
provided that one relates the deformation parameter $\mu$ to the string coupling $g_s$. 

Crucial to the equality of the two sides is the surprising feature that the worldsheet path integral over the moduli space $\mathcal{M}_g$ completely localizes to the points where there exists a holomorphic map $\gamma:\Sigma\to X$. This feature of the worldsheet theory was first noticed in \cite{Maldacena:2001km}, was made precise for correlators in global $\text{AdS}_3$ in the seminal work of Eberhardt-Gaberdiel-Gopakumar \cite{Eberhardt:2019ywk} and its extension \cite{Eberhardt:2020akk} to higher-genus worldsheets, and has been essential in establishing the duality between `tensionless' string theory on $\text{AdS}_3\times\text{S}^3\times\mathbb{T}^4$ and the symmetric orbifold of $\mathbb{T}^4$ \cite{Gaberdiel:2018rqv,Eberhardt:2018ouy,Giribet:2018ada,Dei:2020zui,Gaberdiel:2020ycd,Knighton:2020kuh,Eberhardt:2020bgq,Eberhardt:2021jvj,Gaberdiel:2021kkp,Gaberdiel:2022oeu,McStay:2023thk,Dei:2023ivl,Gaberdiel:2023lco,Knighton:2024ybs,Knighton:2024qxd}.

The equality of the two partition functions \eqref{eq:pert-ads/cft} is the main result of this paper, and we find a simple dictionary between the two sides:
\begin{itemize}

    \item The worldsheet moduli space integral localizes to the covering spaces used to compute the symmetric orbifold correlator in \eqref{eq:perturbative-series-introduction}, as was predicted in \cite{Pakman:2009zz}.

    \item The gauge-fixed worldsheet path integral can be reduced to an integral over a moduli space of `worldsheet instantons', i.e. topological classes of maps $x:\Sigma\to X$ for which the worldsheet wraps around the boundary of $M$ a given number of times. The integral over the moduli space of such maps is found to precisely reproduce the integral of the $m$ integrals in the $\mu^m$ term in the Coulomb gas expansion of \eqref{eq:perturbative-series-introduction}.

    \item The linear dilaton $\phi$ appearing in the boundary CFT is dual to the radial coordinate near the conformal bounday of $M$ \cite{Seiberg:1999xz}.

    \item The dual CFT is not a symmetric orbifold CFT with a fixed value of $N$, but rather a grand canonical ensemble with chemical potential set by the string coupling \cite{Kim:2015gak,Eberhardt:2021vsx}.

\end{itemize}

\vspace{1cm}

This paper is organized as follows. In Section \ref{sec:effective-action}, we provide a generalized version of the derivation of \cite{Seiberg:1999xz} of the long-string CFT on a general hyperbolic 3-manifold with boundary. The results of Section \ref{sec:effective-action} are valid semiclassically, i.e. for strings whose tension is large compared to the curvature of the locally-$\text{AdS}_3$ spacetime. In Section \ref{sec:sigma-model}, we repeat the analysis starting with the Polyakov path integral, which, while technically more involved, yields a quantum-exact description of the long-string theory. In Section \ref{sec:dual-cft}, we explain how the long-string partition function derived in Sections \ref{sec:effective-action} and \ref{sec:sigma-model} reproduces the coulomb-gas expansion of the dual CFT proposed by Eberahrdt \cite{Eberhardt:2021vsx}. We emphasize that the dual `CFT' is based on the grand canonical ensemble of the symmetric orbifold $\text{Sym}^N(\mathcal{S})$ for the non-compact seed theory $\mathcal{S}=\mathbb{R}_{\mathcal{Q}}\times C$, and identify the appropriate twist-2 deformation directly from the worldsheet result. In Section \ref{sec:Weyl-anomaly}, we demonstrate how the worldsheet theory behaves under a Weyl transformation of the metric on the asymptotic boundary, and give further evidence that the dual CFT is a grand canonical ensemble of CFTs with varying central charge. In Section \ref{sec:backgrounds}, we comment on the worldsheet partition functions on some specific 3-dimensional backgrounds, such as Euclidean thermal $\text{AdS}_3$, the Poincar\'e ball (Euclidean global $\text{AdS}_3$), and Euclidean wormholes. Finally, we close with a discussion and a list of potential future directions in Section \ref{sec:discussion}.

\section{The effective action of a long string}\label{sec:effective-action}

In this section, we derive the semiclassical action of long strings which wrap the boundary of a general locally (Euclidean) $\text{AdS}_3$ spacetime a given number of times. As a semiclassical analysis, the results are only valid in the limit that the string length $\sqrt{\alpha'}$ is small compared to the characteristic length scale set by the cosmological constant. For the case of a long-string which winds a single time, this analysis was already presented in \cite{Seiberg:1999xz}. We review their results, which are conceptually simpler, before moving on to the case of a multiply-wound long string, which is more subtle.

\subsection[Asymptotics in AdS\texorpdfstring{$_3$}{3}]{\boldmath Asymptotics in AdS\texorpdfstring{$_3$}{3}}

Let $M$ be a three-manifold which admits a metric $G$ of constant negative curvature
\begin{equation}
R_G=-\frac{6}{L^2}\,.
\end{equation}
The asymptotic boundary $X=\partial M$ inherits a conformal class of metrics from the metric on $M$. If $g_{ij}\mathrm{d}x^i\mathrm{d}x^j$ is a metric on $X$, then near the boundary we can construct $M$ as a foliation of copies of $X$, labeled by a coordinate $r\geq 0$. There is then, near the boundary, a unique choice of metric on $M$ which has constant negative curvature (see Lemma 5.2 of \cite{Graham:1991jqw}):
\begin{equation}\label{eq:fefferman-graham}
\mathrm{d}s^2=\frac{L^2}{r^2}\left(\mathrm{d}r^2+g_{ij}(x)\mathrm{d}x^i\mathrm{d}x^j-r^2P_{ij}(x)\mathrm{d}x^i\mathrm{d}x^j+\cdots\right)\,.
\end{equation}
Here, $P_{ij}$ is a symmetric tensor on the boundary $X$. For $D>3$, the Einstein equations can be used to determine $P_{ij}$ uniquely in terms of the boundary metric $g_{ij}$. For $D=2$, Einstein's equations only determine the trace
\begin{equation}\label{eq:p-trace}
g^{ij}P_{ij}=\frac{1}{2}R\,,
\end{equation}
where $R$ is the scalar curvature of $g_{ij}$. That the metric on $X$ is ambiguous up to a Weyl factor is seen from the fact that the metric constructed on $M$ is invariant under the transformation
\begin{equation}
g_{ij}\to e^{2\omega}g_{ij}\,,\quad r\to e^{\omega}r+\cdots\,,
\end{equation}
where, again, the ellipses denote terms contributing at higher order in $r$.

The Kalb-Ramond field which supports the curvature on $M$ is determined by the $\beta$-function equations on the worldsheet:
\begin{equation}
\begin{split}
R_{\mu\nu}-\frac{1}{4}H_{\mu\rho\sigma}H\indices{_\nu^\rho^\sigma}&=0\,,\\
\nabla^{\omega}H_{\omega\mu\nu}&=0\,.
\end{split}
\end{equation}
where $H=\mathrm{d}B$ is the field strength of the Kalb-Ramond field.\footnote{For general backgrounds, these equations are only the lowest-order terms in an $\alpha'$ expansion. For locally-$\text{AdS}_3$ spacetimes, the worldsheet sigma model is a WZW model based on the group $\text{SL}(2,\mathbb{R})$ \cite{Maldacena:2000hw}, and as such the lowest-order $\beta$-function equations imply full conformal invariance of the worldsheet CFT.} In order solve the first equation for $H$, we first note that the $\text{AdS}_3$ Ricci tensor satisfies
\begin{equation}
R_{\mu\nu}=-\frac{2}{L^2}G_{\mu\nu}\,.
\end{equation}
Furthermore, we can make the ansatz $H_{\mu\nu\rho}=i\sqrt{G}f\varepsilon_{\mu\nu\rho}$ for some unknown function $f$, since this is the most general form of a 3-form in three-dimensions. The constraint $\nabla^{\omega}H_{\omega\mu\nu}=0$ implies that $f$ is constant, and plugging the ansatz into the Einstein equation gives the unique solution\footnote{There is a sign ambiguity since the manifold $(M,G)$ does not have a specified orientation.}
\begin{equation}
\begin{split}
H_{\mu\nu\rho}&=\frac{2i}{L}\sqrt{G}\varepsilon_{\mu\nu\rho}\\
&\approx\frac{2iL^2}{r^3}
\sqrt{g}\left(1-\frac{r^2}{2}R+\cdots\right)\varepsilon_{\mu\nu\rho}\,,
\end{split}
\end{equation}
where in the second line we have used the expansion \eqref{eq:fefferman-graham}, the identity \eqref{eq:p-trace}, and the expansion $\text{det}(A+\varepsilon B)=\det(A)(1+\varepsilon\text{tr}[A^{-1}B]+\cdots)$. Finally, we can integrate to find the $B$-field
\begin{equation}\label{eq:fefferman-graham-b-field}
B=iL^2\sqrt{g}\left(\frac{1}{r^2}+\frac{1}{2}\log\frac{r}{r_0}R\right)\varepsilon_{ij}\mathrm{d}x^i\wedge\mathrm{d}x^j+\cdots
\end{equation}
for some arbitrary length scale $r_0$.\footnote{The $B$ field is ambiguous up to an element of the de Rham cohomology $\text{H}^2(M,\mathbb{R})$. Near the asymptotic boundary, the topology of $M$ looks like $X\times\mathbb{R}_{\geq 0}$, and so we have
\begin{equation}
\text{H}^2(M,\mathbb{R})\approx\text{H}^2(X\times\mathbb{R}_{\geq 0},\mathbb{R})\cong\mathbb{R}\,,
\end{equation}
and so near the boundary of $M$, there is a one-parameter family of solutions for $B$, which is taken into account by the length scale $r_0$.}

\subsection{The theory of a single long string}

Before taking on the case of a generic long string, let us review the work of Seiberg and Witten \cite{Seiberg:1999xz}, which showed that the action of a long string wrapping the boundary of $\text{AdS}_3$ a single time is described by a linear dilaton CFT $\mathbb{R}_{\mathcal{Q}}$ with background charge $\mathcal{Q}$ to be determined below.

Let us consider $M$ to be global Euclidean $\text{AdS}_3$, i.e. the Poincar\'e ball. Despite the existence of a simple globally-defined metric on $M$, we keep the metric of the boundary sphere generic. A long string which wraps the boundary of $\text{AdS}_3$ can be parametrized by the fields
\begin{equation}
Y^{\mu}(\sigma):=(x^i(\sigma),\phi(\sigma))\,,
\end{equation}
where $\sigma^{a}$ denotes the coordinates on the worldsheet. As the target space is supported by a Kalb-Ramond field, the Nambu-Goto action is
\begin{equation}
S_{\text{NG}}=\frac{1}{2\pi\alpha'}\int\mathrm{d}^2\sigma\sqrt{\text{det}(Y^*G)}+\frac{i}{2\pi\alpha'}\int Y^*B\,.
\end{equation}
Our goal now is to write this action in a near-boundary approximation.

If we consider a string which winds the boundary of $\text{AdS}_3$ once, the string must be a sphere, since any worldsheet of higher topology must necessarily wind the boundary multiple times. Now, the field $x^i(\sigma)$ represents a map from the worldsheet to the boundary, and thus a map $x:\text{S}^2\to\text{S}^2$. Up to diffeomorphisms on the worldsheet, there is one equivalence class of such maps, and so due to the diffeomorphism symmetry of the worldsheet theory, we can fix a gauge by choosing a particular representative. The simplest possible gauge is to fix
\begin{equation}\label{eq:single-long-string-gauge}
\sigma^i=x^i\,,
\end{equation}
where $i=1,2$. That is, we identify the coordinates on the worldsheet with their image on the boundary. As such, we will temporarily use $i,j$ for both target-space and worldsheet indices. Since the map $x$ is invertible, this is a globally well-defined gauge choice. Thus, in the case of a singly-wound long string, the worldsheet configuration depends only on the radial profile $r(x)$ of the worldsheet embedded in the target space (as well as the motion of the string in the compact directions, more on this later).

In the gauge \eqref{eq:single-long-string-gauge}, the pullback of the target space metric \eqref{eq:fefferman-graham} to the worldsheet is
\begin{equation}
(Y^*G)_{ij}=\frac{L^2}{r^2}(\partial_{i}r\partial_{j}r+g_{ij}(x)-r^2P_{ij}(x)+\cdots)\,.
\end{equation}
Expanding the determinant of this matrix in $r$, we can thus compute the worldsheet area to be
\begin{equation}\label{eq:area-pullback-semiclassical}
\int\mathrm{d}^2x\sqrt{\text{det}(Y^*G)}=L^2\int\mathrm{d}^2x\sqrt{g}\left(\frac{1}{r^2}+\frac{1}{2r^2}g^{ij}\partial_i r\partial_jr-\frac{1}{4}R+\cdots\right)\,.
\end{equation}
Here, the Ricci scalar $R$ is obtained from the trace of $P_{ij}$ in the expansion of the determinant of $Y^*G$. Similarly, the contribution of the Kalb-Ramond field is
\begin{equation}\label{eq:kb-pullback-semiclassical}
i\int Y^*B=-L^2\int\mathrm{d}^2x\sqrt{g}\left(\frac{1}{r^2}+\frac{1}{2}\log\frac{r}{r_0}R+\cdots\right)\,.
\end{equation}
Here, we must be careful to demand that the worldsheet winds the boundary with positive orientation, since the sign of the pullback $Y^*B$ to the worldsheet changes under a chance of orientation of the map $x$.

Putting together \eqref{eq:area-pullback-semiclassical} and \eqref{eq:kb-pullback-semiclassical}, the Nambu-Goto action of a singly-wound long string reads
\begin{equation}
S_{\text{NG}}=\frac{k}{2\pi}\int\mathrm{d}^2x\left(\frac{1}{2r^2}g^{ij}\partial_i r\partial_jr-\frac{1}{2}\log\frac{r}{r_0}R-\frac{1}{4}R+\cdots\right)\,,
\end{equation}
where we have defined $k=L^2/\alpha'$. Note that the divergent $1/r^2$ term, which appeared in both the action and the $B$-field, canceled between the two contributions. This cancellation between the string tension and the force felt under the Kalb-Ramond background is the reason for the existence of long strings, and occurs only when the string winds the boundary with positive orientation.\footnote{This analysis can be generalized to long BPS D$p$-branes in $\text{AdS}_{p+2}$ spacetimes supported by pure RR-flux \cite{Seiberg:1999xz}.}

This action defines the dynamics of a classical long string as a function of its radial profile $r$. In order to obtain a more standard quadratic action, it is convenient to define a Liouville field $\phi$ via
\begin{equation}\label{eq:phi-semiclassical}
r=r_0e^{-\phi/\sqrt{2k}}\,.
\end{equation}
Written in terms of the field $\phi$, the worldsheet action becomes
\begin{equation}\label{eq:single-string-action}
S_{\text{NG}}=\frac{1}{4\pi}\int\mathrm{d}^2x\sqrt{g}\left(\frac{1}{2}g^{ij}\partial_i\phi\partial_j\phi+\sqrt{\frac{k}{2}}R\phi-\frac{k}{2}R+\cdots\right)\,.
\end{equation}
This is the action of a linear dilaton CFT living on the boundary of $\text{AdS}_3$. The linear dilaton has background charge
\begin{equation}
\mathcal{Q}=-\sqrt{2k}
\end{equation}
and thus, in the limit of large $k$, where this analysis is valid, the central charge of the theory is approximately
\begin{equation}
c\approx 3\mathcal{Q}^2=6k\,,
\end{equation}
which is consistent with the analysis of Brown and Henneaux \cite{Brown:1986nw}. As we will see in Section \ref{sec:sigma-model}, the exact CFT, including quantum corrections, has background charge \cite{Seiberg:1999xz,Balthazar:2021xeh,Eberhardt:2021vsx}
\begin{equation}\label{eq:background-charge-correction}
\mathcal{Q}=-\sqrt{\frac{2(k-3)^2}{k-2}}\,.
\end{equation}

\subsection{Multiply-wound long strings}

We now turn our attention to strings which wind around the conformal boundary a fixed number $N$ of times. We will also relax the condition that the conformal boundary and the worldsheet are both spheres, and work in full generality, denoting by $G$ the genus of the conformal boundary of $M$, and by $g$ the genus of the worldsheet $\Sigma$.

The saving grace which made the analysis for the singly-wound long string simple is that one can use diffeomorphism invariance to uniquely determine the map $x:\text{S}^2\to\text{S}^2$ by identifying the worldsheet coordinates with the boundary coordinates. Indeed, this is always possible so long as $N=1$ and the genera of the boundary and the worldsheet agree, i.e. $G=g$. However, for generic values of $N,G,g$, things are not so simple. Let us denote by $\text{Map}_N(\Sigma,X)$ the space of all smooth, orienation-preserving maps $x:\Sigma\to X$ which have degree $N$, i.e. such that the generic point in $X$ has $N$ pre-images.

In the neighborhood of a generic point, we can use diffeomorphism symmetry on the worldsheet to fix the map $x$. Indeed, a diffeomorphism $\widetilde{\sigma}(\sigma)$ on the worldsheet changes the map $x$ to
\begin{equation}
\tilde{x}^{i}(\sigma)=x^i(\tilde{\sigma}(\sigma))\,.
\end{equation}
If we fix a point $\sigma_0$ on the worldsheet, we can consider the Taylor expansion of $x$ around $\sigma_0$:
\begin{equation}
x^i(\sigma)=x^i_0+\frac{\partial x^i}{\partial\sigma^a}\bigg|_{\sigma_0}(\sigma-\sigma_0)^{\alpha}+\frac{1}{2}\frac{\partial^2x^i}{\partial\sigma^a\partial\sigma^b}\bigg|_{\sigma_0}(\sigma-\sigma_0)^{a}(\sigma-\sigma_0)^{b}+\cdots\,.
\end{equation}
Choosing the diffeomorphism $\tilde{\sigma}$ to satisfy
\begin{equation}
\frac{\partial\tilde{\sigma}^a}{\partial\sigma^b}=\frac{\partial\sigma^a}{\partial x^b}\bigg|_{x^i_0}
\end{equation}
leads to $\tilde{x}$ having the Taylor expansion
\begin{equation}
\tilde{x}^i(\sigma)=\tilde{x}_0^i+(\sigma-\sigma_0)^i+\cdots\,,
\end{equation}
where $\tilde{x}^i_0=x^i(\tilde{\sigma}(\sigma_0))$. Thus, we can use diffeomorphism invariance to locally identify the worldsheet coordinates with the boundary coordinates. However, this only works in the neighborhoods of \textit{generic} points on the worldsheet. Specifically, we assumed that the matrix $\partial x^i/\partial\sigma^a|_{\sigma_0}$ was invertible. While this is almost always true, for general values of $G,g,N$, there will be a set of isolated points $\xi_\ell$ (`branch points') on the boundary $X$ for which the Jacobian $\partial x^i/\partial\sigma^{a}$ is non-invertible. The number $m$ of such points, counted with multiplicity, is determined by the Riemann-Hurwitz formula:
\begin{equation}\label{eq:Riemann-Hurwitz}
m=N(2-2G)-(2-2g)\,.
\end{equation}

The branch points $\xi_\ell$ represent obstructions to using diffeomorphism symmetry to fully gauge-fix the map $x:\Sigma\to X$. Indeed, the moduli space of possible winding configurations (\textit{worldsheet instantons} \cite{Maldacena:2001km}) of a long string around the conformal boundary of $M$
\begin{equation}\label{eq:nambu-goto-moduli-space}
\text{Map}_N(\Sigma,X)/\text{Diff}(\Sigma)
\end{equation}
has real dimension $2m$ (we compute this dimension in Section \ref{sec:sigma-model}). Given a set $\xi_\ell$ of branch points, there is a discrete set of covering maps $x:\Sigma\to X$ branched over $\xi_\ell$. Physically, this is to say that the angular profile $x^i(\sigma)$ of the worldsheet near the boundary of $X$ is completely determined by the data of the branch points, up to a discrete choice of branched covering map $x$. Since in the path integral we want to integrate over all such embeddings of the worldsheet, we must integrate over the locations of all of the branch points for a given value of $N$. In principle one must also carefully treat the points in the moduli space \eqref{eq:nambu-goto-moduli-space} where two branch points collide. As we discuss in Section \ref{sec:dual-cft}, this is reflected in the choice of prescription in the computation of Coulomb integrals in the long-string CFT.

With the preliminary discussion out of the way, we can now write down the action of a long string winding $N$ times around the boundary of $M$. For a fixed set of branch points $\xi_\ell$, we can use diffeomorphism invariance to gauge-fix a map $x:\Sigma\to X$, which we will leave arbitrary. The induced metric on the worldsheet is
\begin{equation}
(Y^*G)_{ab}=\frac{L^2}{r^2}\left(\partial_{a}r\partial_{b}r+(x^*g)_{ab}-r^2(x^*P)_{ab}+\cdots\right)\,,
\end{equation}
where $x^*g$ and $x^*P$ are the pullbacks under the map $x$, i.e.
\begin{equation}
(x^*g)_{ab}=\partial_{a}x^i\partial_{b}x^jg_{ij}\,,\quad (x^*P)_{ab}=\partial_{a}x^i\partial_{b}x^jP_{ij}\,.
\end{equation}
For ease of notation, we define the auxiliary metric
\begin{equation}
h=x^*g\,.
\end{equation}
Following precisely the same logic as in the singly-wound case, we can now write the Nambu-Goto action in the form
\begin{equation}
S_{\text{NG}}=\frac{k}{2\pi}\int_{\Sigma}\mathrm{d}^2\sigma\sqrt{h}\left(\frac{1}{2r^2}h^{ab}\partial_{a}r\partial_{b}r-\frac{1}{2}\log\frac{r}{r_0}R(x)-\frac{1}{4}R(x)\right)\,,
\end{equation}
where we emphasize that $R(x)$ is computed with respect to the boundary metric $g$, while $h$ is the pullback of $g$ to the worldsheet. Introducing the Liouville field $\phi$ as in \eqref{eq:phi-semiclassical}, we arrive at the action
\begin{equation}
S_{\text{NG}}=\frac{1}{4\pi}\int_{\Sigma}\mathrm{d}^2\sigma\sqrt{h}\left(\frac{1}{2}h^{ab}\partial_{a}\phi\partial_{b}\phi+\sqrt{\frac{k}{2}}R(x)\phi-\frac{k}{2}R(x)\right)\,.
\end{equation}
This action looks deceptively similar to the action \eqref{eq:single-string-action} of a singly-wound long string. There is, however, a fundamental difference, namely that the curvature that the scalar $\phi$ couples to is \textit{not} the curvature computed from the worldsheet metric $h$, but rather the curvature computed from the boundary metric $g$, pulled back to $\Sigma$. The former curvature (let us call it $R_h$) is related to the latter by the relation
\begin{equation}
\sqrt{h}R(x)=\sqrt{h}R_h+4\pi\sum_{\ell=1}^{m}\delta^{(2)}(\sigma,\zeta_\ell)\,,
\end{equation}
where $\zeta_\ell$ are the pre-images of the branch points $\xi_{\ell}$ under $x$.

Written purely in terms of the worldsheet metric $h$, then, the action of the long-string CFT takes the form
\begin{equation}
S_{\text{NG}}=\frac{1}{4\pi}\int_{\Sigma}\mathrm{d}^2\sigma\sqrt{h}\left(\frac{1}{2}h^{ab}\partial_{a}\phi\partial_{b}\phi+\sqrt{\frac{k}{2}}R_h\phi-\frac{k}{2}R_h\right)+\frac{km}{2}-\sqrt{\frac{k}{2}}\sum_{\ell=1}^{m}\phi(\zeta_\ell)\,,
\end{equation}
or
\begin{equation}
e^{-S_{\text{NG}}}=e^{kN(1-G)}e^{-S[\phi]}\prod_{\ell=1}^{m}e^{-\sqrt{\frac{k}{2}}\phi}(\zeta_\ell)\,,
\end{equation}
where 
\begin{equation}
S[\phi]=\frac{1}{4\pi}\int_{\Sigma}\mathrm{d}^2\sigma\sqrt{h}\left(\frac{1}{2}h^{ab}\partial_a\phi\partial_b\phi-\frac{\mathcal{Q}}{2}R_h\right)
\end{equation}
is the standard action for a linear dilaton of background charge $\mathcal{Q}=-\sqrt{2k}$. The prefactor of $e^{kN(1-G)}$ comes from integrating over the scalar curvature term in the Nambu-Goto action, through the Gauss-Bonnet theorem.

The semiclassical free energy of the long-string is now found by integrating over the space of radial profiles $\phi$, as well as the locations $\xi_\ell$ of the branch points on the boundary $X$. The result for a fixed worldsheet genus $g$ then reads\footnote{The notation $\mathcal{F}_g^{(0)}$ for the long-string free energy is chosen for consistency with Section \ref{sec:path-integral}.}
\begin{equation}
\begin{split}
\mathcal{F}_{g}^{(0)}\approx e^{k(1-g)}\sum_{m=0}^{\infty}&\frac{e^{km/2}}{m!}
\left(\int\prod_{\ell=1}^{m}\mathrm{d}^2\xi_{\ell}\sqrt{g}\right)\\
&\times\sum_{\substack{\text{covering maps}\\\gamma:\Sigma:\to X}}\int\mathcal{D}\phi\,e^{-S[\phi]}\prod_{\ell=1}^{m}\,e^{-\sqrt{\frac{k}{2}}\phi}(\zeta_\ell)\,,
\end{split}
\end{equation}
where we emphasize that the action of the Liouville field $\phi$ is defined with respect to the pullback metric $x^*g$, which in turn depends on the locations of the branch points $\xi_1,\ldots,\xi_m$, up to diffeomorphism on $\Sigma$. The prefactor of $e^{k(1-g)+km/2}$ is simply the prefactor $e^{kN(1-G)}$ after using the Riemann-Hurwitz formula. We note also that the factor of $1/m!$ comes from the fact that any permutation of the points $\xi_1,\ldots,\xi_m$ represent the same point in the moduli space \eqref{eq:nambu-goto-moduli-space}. Of course, as mentioned above, one should in principle be careful in treating the points of integration upon which two of the branch points collide, but this subtlety will not concern us here.

The path integral over $\phi$ is immediately recognized as the correlation function of $m$ states of momentum $\sqrt{k/2}$ in the linear dilaton CFT, and so we can write
\begin{equation}
\begin{split}
\mathcal{F}_{g}^{(0)}\approx e^{k(1-g)}\sum_{m=0}^{\infty}\frac{e^{km/2}}{m!}\int&\left(\prod_{\ell=1}^{m}\mathrm{d}^2\xi_{\ell}\sqrt{g}\right)\\
&\times\sum_{\substack{\text{covering maps}\\\gamma:\Sigma:\to X}}\Braket{\prod_{\ell=1}^{m}\,e^{-\sqrt{\frac{k}{2}}\phi}(\zeta_\ell)}_{(\Sigma,x^*g)}\,.
\end{split}
\end{equation}
Here, the subscript $(X,x^*g)$ reminds us that the correlation function is computed on the worldsheet $\Sigma$ with the pullback metric $x^*g$. As we will see below, this semiclassical expression is correct for large values of $k$ (small values of $\alpha'$), but receives nontrivial corrections for small values of $k$. Specifically, as mentioned above, the exact linear dilaton background charge is given by equation \eqref{eq:background-charge-correction}, and the form of the local $\phi$ insertions is corrected to
\begin{equation}
e^{-\sqrt{\frac{k}{2}}\phi}\to e^{-\sqrt{\frac{k-2}{2}}\phi}\,.
\end{equation}

\subsection{Including compact directions}

So far, we have only considered semiclassical strings moving on a hyperbolic 3-manifold $M$ with asymptotic boundary $X$. However, in order to have a consistent worldsheet theory, we must work with a 26-dimensional compactified spacetime $M\times C$, where $C$ is a 23 dimensional background. In order to satisfy the low-energy supergravity equations to lowest order in $\alpha'$, the total curvature of $M\times C$ must vanish.

Let $\ell_{AB}$ and $(B_C)_{AB}$ be the metric and Kalb-Ramond field on $C$ with some local coordinates $y^A$. The metric on the target space $M\times C$ is
\begin{equation}
G_{\text{tot}}=
\begin{pmatrix}
G & 0\\
0 & \ell
\end{pmatrix}\,,
\end{equation}
while the total Kalb-Ramond field is simply $B+B_C$, viewed as a 2-form on the total space $M\times C$. On the worldsheet, the fundamental field is the coordinate field
\begin{equation}
Y^{\mu}(\sigma)=(x^i(\sigma),\phi(\sigma),y^A(\sigma))\,.
\end{equation}
The pullback metric $Y^*G_{\text{tot}}$ is simply
\begin{equation}
(Y^*G_{\text{tot}})_{ab}=\frac{L^2}{r^2}(\partial_{a}r\partial_{b}r+h_{ab}-r^2(x^*P)_{ab}+\cdots)+\ell_{AB}\partial_{a}y^A\partial_{b}y^B\,.
\end{equation}
Assuming the manifold $C$ is compact and smooth, its metric is bounded and so we can consider a small-$r$ expansion of the worldsheet area. Specifically, we have
\begin{equation}
\sqrt{\text{det}(Y^*G_{\text{tot}})}=\sqrt{h}\left(\frac{L^2}{r^2}+\frac{L^2}{2r^2}(\partial r)^2-\frac{L^2}{4}R+h^{ab}\ell_{AB}\partial_{a}y^A\partial_{b}y^B+\cdots\right)\,.
\end{equation}
Defining again $k=L^2/\alpha'$ and $r=r_0e^{-\phi/\sqrt{2k}}$, we can write the Nambu-Goto action for a long string on $M\times C$ as
\begin{equation}
\begin{split}
S_{\text{NG}}=&\frac{1}{4\pi}\int\mathrm{d}^2\sigma\sqrt{h}\left(\frac{1}{2}h^{ab}\partial_{a}\phi\partial_{b}\phi+\sqrt{\frac{k}{2}}R\phi-\frac{k}{2}R+\cdots\right)\\
&\hspace{1cm}+\frac{1}{2\pi\alpha'}\int\mathrm{d}^2\sigma\sqrt{h}\,h^{ab}\ell_{AB}\partial_{a}y^A\partial_{b}y^B+\frac{i}{2\pi\alpha'}\int y^*B_C\,.
\end{split}
\end{equation}
The first line is the usual linear dilaton CFT associated with a long string on $M$, while the second term is the action of a nonlinear sigma model on the compact CFT $C$, evaluated with the worldsheet metric $h=x^*g$. Writing $R_g$ in terms of the worldsheet curvature $R_h$ and integrating over the branch points gives the semiclassical path integral
\begin{equation}
\begin{split}
\mathcal{F}_{g,m}^{(0)}\approx \sum_{m=0}^{\infty}e^{kN(1-G)}g_s^{2g-2}&\left(\frac{1}{m!}\int\prod_{\ell=1}^{m}\mathrm{d}^2\xi_{\ell}\sqrt{g}\right)\\
&\times\sum_{\substack{\text{covering maps}\\\gamma:\Sigma:\to X}}\Braket{\prod_{\ell=1}^{m}\,e^{-\sqrt{\frac{k}{2}}\phi}(\zeta_\ell)}_{(\Sigma,x^*g)}Z_{C}(\Sigma,x^*g)\,,
\end{split}
\end{equation}
where $Z_C$ is the partition function of the compact CFT $C$.

\section{The Polyakov path integral}\label{sec:sigma-model}

The analysis of the previous section demonstrated in a simple way how the long-string partition function in $\text{AdS}_3$ naturally arises as a set of correlation functions of a linear dilaton theory on covering spaces of the asymptotic boundary. However, the path integral was only performed at the semiclassical ($k\gg 1$) level, and in particular various anomalies in the path integral measure were ignored. In this section we repeat the derivation of the long-string theory using the formalism of the Polyakov path integral. This analysis is significantly more technical than that of the previous section, and we will find that the end result of the computation is the same in the semiclassical limit $k\gg 1$, but differs for small values of $k$.

\subsection{The worldsheet sigma model}

As in the case of the semiclassical analysis above, the starting point for the worldsheet theory is the small-$r$ expansion of the metric and Kalb-Ramond field, which we repeat for convenience:
\begin{equation}
\begin{gathered}
\mathrm{d}s^2=L^2\left(\mathrm{d}\Phi^2+e^{2\Phi}g_{ij}\mathrm{d}x^i\mathrm{d}x^j-P_{ij}+\cdots\right)\,,\\
B=iL^2\sqrt{g}\left(e^{2\Phi}-\frac{1}{2}\Phi\,R\right)\varepsilon_{ij}\mathrm{d}x^i\wedge\mathrm{d}x^j+\cdots\,.
\end{gathered}
\end{equation}
Here, we have defined $r=r_0e^{-\Phi}$, so that the conformal boundary of $M$ is located at $\Phi\to \infty$. We remind the reader that $g_{ij}$ is the induced metric on the boundary $X$ of $M$, and $P_{ij}$ is a symmetric 2-tensor on $X$ which satisfies $g^{ij}P_{ij}=\frac{1}{2}R$, and $R$ is the scalar curvature on $X$ derived from $g_{ij}$.

The Polyakov action in this background is given by
\begin{equation}
\begin{split}
S&=\frac{1}{4\pi\alpha'}\int_{\Sigma}\mathrm{d}^2\sigma\left(\sqrt{h}h^{ab}G_{\mu\nu}(Y)\partial_aY^{\mu}\partial_bY^{\nu}+i\varepsilon^{ab}B_{\mu\nu}(X)\partial_aY^{\mu}\partial_bY^{\nu}\right)\\
&=\frac{k}{4\pi}\int_{\Sigma}\mathrm{d}^2\sigma\,\Big(\sqrt{h}h^{ab}\partial_a\Phi\partial_b\Phi+x^*(\sqrt{g}R)\Phi\\
&\hspace{3cm}+r_0^{-2}e^{2\Phi}\left(\sqrt{h}h^{ab}g_{ij}-\sqrt{g}\varepsilon^{ab}\varepsilon_{ij}\right)\partial_ax^i\partial_bx^j\\
&\hspace{7cm}-\sqrt{h}h^{ab}P_{ij}\partial_ax^i\partial_bx^j+\mathcal{O}(e^{-2\Phi})\Big)\,,
\end{split}
\end{equation}
where
\begin{equation}
\begin{split}
x^*(\sqrt{g}R)&=\frac{1}{2}\varepsilon^{ab}\varepsilon_{ij}\partial_ax^i\partial_bx^j\sqrt{g(x)}R(x)\\
&=\text{det}(\partial_ax^i)\sqrt{g(x)}R(x)
\end{split}
\end{equation}
is the pullback of the scalar density $\sqrt{g}R$ from the boundary of $\text{AdS}_3$ to the worldsheet. The approximation is valid so long as we consider the limit of large $\Phi$. We have furthermore defined
\begin{equation}
k=\frac{L^2}{\alpha'}=\frac{L^2}{\ell_s^2}\,,
\end{equation}
which quantifies the string tension in the units $L=1$.\footnote{The number $k$ corresponds to the usual notion of the level of the underlying $\text{SL}(2,\mathbb{R})$ WZW model.}

\subsection{The path integral measure}

The path integral measure which defines the quantum theory of the string in this background is defined implicitly by the inner products
\begin{equation}
\begin{split}
||\delta\Phi||^2&=\int_{\Sigma}\mathrm{d}^2\sigma\sqrt{h}\,\delta\Phi\,\delta\Phi\\
||\delta x||^2&=\frac{L^2}{r_0^2}\int_{\Sigma}\mathrm{d}^2\sigma\sqrt{h}\,e^{2\Phi}\left(g_{ij}-e^{-2\Phi}P_{ij}+\cdots\right)\delta x^i\delta x^j\,.
\end{split}
\end{equation}
Since the inner product $||\delta x||^2$ depends on $\Phi$, the path integral measure does not factorize between the radial and boundary coordinates of $\text{AdS}_3$. In the limit of large $\Phi$, we can instead work with the measure
\begin{equation}
||\delta x||_{\text{new}}^2=\int_{\Sigma}\mathrm{d}^2\sigma\sqrt{h}\,g_{ij}\delta x^i\delta x^j
\end{equation}
at the expense of picking up a Jacobian
\begin{equation}\label{eq:x-measure-transformation}
\mathcal{D}x=\mathcal{D}x_{\text{new}}\mathscr{J}\,.
\end{equation}
The Jacobian $\mathscr{J}$ depends on the scalar $\Phi$ and the boundary metric $g$, and takes the form (see Appendix \ref{app:measure})
\begin{equation}\label{eq:x-measure-jacobian}
\mathscr{J}=\exp\left(\frac{1}{2\pi}\int_{\Sigma}\mathrm{d}^2\sigma\left(\sqrt{h}h^{ab}\partial_a\Phi\,\partial_b\Phi+\frac{1}{2}\sqrt{h}R_h\Phi+f\Phi+\mathcal{O}(e^{-2\Phi})\right)\right)\,,
\end{equation}
where $R_h$ is the worldsheet scalar curvature constructed from $h$. Here, $f$ is a scalar density depending on the coordinates $x$, whose form we will not need now, but which we will determine later.

We thus conclude that the effect of the redefinition of the path integral measure is a modification of the action to:
\begin{equation}
\begin{split}
S&=\frac{1}{4\pi}\int_{\Sigma}\mathrm{d}^2\sigma\,\Big((k-2)\sqrt{h}h^{ab}\partial_a\Phi\partial_b\Phi-\sqrt{h}R_h\Phi+kx^*(\sqrt{g}R)\Phi-2f\Phi\\
&\hspace{5cm}+kr_0^{-2}e^{2\Phi}\left(\sqrt{h}h^{ab}g_{ij}-\sqrt{g}\varepsilon^{ab}\varepsilon_{ij}\right)\partial_ax^i\partial_bx^j\\
&\hspace{8cm}-\sqrt{h}h^{ab}P_{ij}\partial_ax^i\partial_bx^j+\mathcal{O}(e^{-2\Phi})\Big)\,.
\end{split}
\end{equation}
Finally, we can rescale the radial scalar by defining
\begin{equation}
\phi=\sqrt{2(k-2)}\,\Phi\,,
\end{equation}
so that the kinetic term for $\phi$ is canonically normalized, and we land on the action
\begin{equation}
\begin{split}
S&=\frac{1}{4\pi}\int_{\Sigma}\mathrm{d}^2\sigma\bigg(\frac{1}{2}\sqrt{h}h^{ab}\partial_a\phi\partial_b\phi-\frac{Q}{2}\sqrt{h}R_h\phi+\frac{Qk}{2}x^*(\sqrt{g}R)\phi-Qf\phi\\
&\hspace{4.5cm}+ke^{Q\phi}\left(\sqrt{h}h^{ab}g_{ij}-\sqrt{g}\,\varepsilon^{ab}\varepsilon_{ij}\right)\partial_ax^i\partial_bx^j\\
&\hspace{7cm}-\sqrt{h}h^{ab}P_{ij}\partial_ax^i\partial_bx^j+\mathcal{O}(e^{-Q\phi})\Big)\,,
\end{split}
\end{equation}
where $Q=\sqrt{2/(k-2)}$.

\subsection{Introducing Lagrange multipliers}

The above action is a functional of the fundamental fields $\Phi,x^i$ of the worldsheet theory, as well as of two 2D metrics: the worldsheet metric $h$ and the boundary metric $g$. Recall that any 2D metric naturally defines a complex structure. Specifically, in terms of $h$ and $g$, we can define the complex structures
\begin{equation}\label{eq:complex-structures}
J\indices{^a_b}=\sqrt{h}h^{ac}\varepsilon_{bc}\,,\quad\mathcal{J}\indices{^i_j}=\sqrt{g}g^{ik}\varepsilon_{jk}\,.
\end{equation}
These act on the tangent spaces of the worldsheet $\Sigma$ and the conformal boundary $X$ of $\text{AdS}_3$, respectively. In terms of these complex structures, we can rewrite the worldsheet action as
\begin{equation}\label{eq:second-order-action-factorized}
\begin{split}
S&=\frac{1}{4\pi}\int_{\Sigma}\mathrm{d}^2\sigma\bigg(\frac{1}{2}\sqrt{h}h^{ab}\partial_a\phi\partial_b\phi-\frac{Q}{2}\sqrt{h}R_h\phi+\frac{Qk}{2}x^*(\sqrt{g}R)\phi-Qf\phi\\
&\hspace{4.25cm}+kr_0^{-2}e^{Q\phi}\sqrt{h}h^{ab}g_{ij}\partial_ax^i\left(\partial_b x^j+\mathcal{J}\indices{^j_k}\partial_cx^kJ\indices{^c_b}\right)\\
&\hspace{7cm}-\sqrt{h}h^{ab}P_{ij}\partial_ax^i\partial_bx^j+\mathcal{O}(e^{-Q\phi})\Big)\,.
\end{split}
\end{equation}
Let us take stock of this expression and write it in a more geometric fashion. The field $x$ is a map from the worldsheet $\Sigma$ to the boundary $X$, and $\partial_ax^i$ is the local coordinate expression for the differential $\mathrm{d}x$ of this map. The differential evaluated at a point $p\in\Sigma$ can be thought of as a linear map
\begin{equation}
\mathrm{d}x(p):\text{T}_p\Sigma\to \text{T}_{x(p)}X\,.
\end{equation}
Equivalently, $\mathrm{d}x$ is a section of the vector bundle $\text{T}^*\Sigma\otimes x^*(\text{T}X)$ over the worldsheet.

Next, the complex structures $J$ and $\mathcal{J}$ are pointwise-linear maps on the tangent spaces of $\Sigma$ and $X$, respectively
\begin{equation}
J(p):\text{T}_p\Sigma\to \text{T}_p\Sigma\,,\quad \mathcal{J}(x):\text{T}_xX\to \text{T}_xX\,.
\end{equation}
The combination $\mathcal{J}\indices{^j_k}\partial_cx^kJ\indices{^c_b}$ appearing in the action \eqref{eq:second-order-action-factorized} can be interpreted as the linear map
\begin{equation}
(\mathcal{J}\mathrm{d}xJ)(p):\text{T}_p\Sigma\xrightarrow{J(p)}\text{T}_p\Sigma\xrightarrow{\mathrm{d}x(p)}\text{T}_{x(p)}X\xrightarrow{\mathcal{J}(x(p))}\text{T}_{x(p)}X\,,
\end{equation}
and thus $\mathcal{J}\mathrm{d}xJ$ also defines a section of the bundle $\text{T}^*\Sigma\otimes x^*(\text{T}X)$.

Sections of $\text{T}^*\Sigma\otimes x^*(\text{T}X)$ are intuitively fields which transform as one-forms on $\Sigma$ and as vector fields on $X$. Given two sections $v_a^i$ and $w_a^i$, there is a natural inner product between them induced by the metric tensors on $\Sigma$ and $X$, namely
\begin{equation}
\braket{v,w}=h^{ab}g_{ij}v_a^iw_b^j\,.
\end{equation}
Thus, we can write the worldsheet action in a more covariant form as
\begin{equation}\label{eq:second-order-action-covariant}
\begin{split}
S&=\frac{1}{4\pi}\int_{\Sigma}\mathrm{d}^2\sigma\bigg(\frac{1}{2}\sqrt{h}h^{ab}\partial_a\phi\partial_b\phi-\frac{Q}{2}\sqrt{h}R_h\phi+\frac{Qk}{2}x^*(\sqrt{g}R)\phi-Qf\phi\\
&\hspace{2.5cm}+kr_0^{-2}e^{Q\phi}\sqrt{h}\braket{\mathrm{d}x,\mathrm{d}x+\mathcal{J}\mathrm{d}xJ}-k\sqrt{h}\braket{\mathrm{d}x,\mathrm{d}x}^{(2)}+\mathcal{O}(e^{-Q\phi})\bigg)\,.
\end{split}
\end{equation}
Note that in the second line we have defined the shorthand
\begin{equation}
\braket{v,w}^{(2)}=h^{ab}P_{ij}v_a^iw_b^j\,,
\end{equation}
We should keep in mind that the bilinear form $\braket{\cdot,\cdot}^{(2)}$ is not necessarily an inner-product, since $P_{ij}$ is not required to be non-degenerate.

For the sake of ease of notation and for later convenience, let us define the linear operator $s$ as
\begin{equation}
s(w)=\frac{1}{2}(w+\mathcal{J}wJ)\,,
\end{equation}
where $w^i_a$ is any section of $\text{T}^*\Sigma\otimes x^*(\text{T}X)$. A quick calculation shows that $s$ is a projection operator, i.e. $s^2=s$. Furthermore, one can show that, with respect to the inner product $\braket{\cdot,\cdot}$ defined above, $s$ is self-adjoint. Thus, we have
\begin{equation}
\braket{v,w+\mathcal{J}wJ}=2\braket{v,s(w)}=2\braket{s(v),s(w)}\,,
\end{equation}
which allows us to write the action \eqref{eq:second-order-action-covariant} in the symmetric form
\begin{equation}\label{eq:second-order-action-symmetric}
\begin{split}
S&=\frac{1}{4\pi}\int_{\Sigma}\mathrm{d}^2\sigma\bigg(\frac{1}{2}\sqrt{h}h^{ab}\partial_a\phi\partial_b\phi-\frac{Q}{2}\sqrt{h}R_h\phi+\frac{Qk}{2}x^*(\sqrt{g}R)\phi-Qf\phi\\
&\hspace{2.5cm}+4kr_0^{-2}e^{Q\phi}\sqrt{h}\braket{Dx,Dx}-k\sqrt{h}\braket{\mathrm{d}x,\mathrm{d}x}^{(2)}+\mathcal{O}(e^{-Q\phi})\bigg)\,,
\end{split}
\end{equation}
where we have defined the differential operator
\begin{equation}
Dx:=s(\mathrm{d}x)=\frac{1}{2}\left(\mathrm{d}x+\mathcal{J}\mathrm{d}xJ\right)\,.
\end{equation}
The first line of this action has the interpretation of a linear dilaton CFT which couples to the curvature of both the worldsheet $\Sigma$ and the boundary $X$. The second line almost looks like the action of a sigma model on $X$ whose fundamental field is the coordinate $x$. However, the term proportional to $e^{Q\phi}$ is problematic, as it diverges as we approach the asymptotic boundary, where our approximation is valid. In order to study the worldsheet theory in this limit, a standard trick is to introduce a Lagrange multiplier $\beta^i_a$ which lives in $\text{T}^*\Sigma\otimes x^*(\text{T}X)$ and write the action in a first-order form.\footnote{Strictly speaking, we should consider a Lagrange multiplier which satisfies the `self-duality' constraint $s(\beta)=\beta$, see for example \cite{Witten:1988xj}.}

Consider specifically the action
\begin{equation}
S_{\beta}=\frac{1}{4\pi}\int_{\Sigma}\mathrm{d}^2\sigma\sqrt{h}\left(\braket{\beta,Dx}-\frac{r_0^2}{16k}e^{-Q\phi}\braket{\beta,\beta}\right)\,.
\end{equation}
Classically, one can integrate out $\beta$ and substitute the solution $\beta=8kr_0^{-2}e^{Q\phi}Dx$ and find
\begin{equation}
S_{\beta}\xrightarrow{\text{on-shell}}\frac{1}{4\pi}\int_{\Sigma}\mathrm{d}^2\sigma\,4kr_0^{-2}e^{Q\phi}\sqrt{h}\braket{Dx,Dx}\,.
\end{equation}
This is exactly the divergent kinetic term appearing in the worldsheet sigma model. Thus, we can replace the action \eqref{eq:second-order-action-covariant} with the `first-order' action
\begin{equation}\label{eq:first-order-action}
\begin{split}
S&=\frac{1}{4\pi}\int_{\Sigma}\mathrm{d}^2\sigma\bigg(\frac{1}{2}\sqrt{h}h^{ab}\partial_a\phi\partial_b\phi-\frac{Q}{2}\sqrt{h}R_h\phi+\frac{Qk}{2}x^*(\sqrt{g}R)\phi-Qf\phi\\
&\hspace{2.5cm}+\sqrt{h}\braket{\beta,Dx}-k\sqrt{h}\braket{\mathrm{d}x,\mathrm{d}x}^{(2)}-\nu e^{-Q\phi}\sqrt{h}\braket{\beta,\beta}+\mathcal{O}(e^{-Q\phi})\bigg)\,,
\end{split}
\end{equation}
where we have defined the constant
\begin{equation}
\nu=\frac{r_0^2}{16k}\,.
\end{equation}
We also emphasize that the $\mathcal{O}(e^{-Q\phi})$ terms are independent of $\beta$. In the quantum theory, we include the field $\beta$ at the level of the path integral. Since $\beta$ is a section of $\text{T}^*\Sigma\otimes x^*(\text{T}X)$, the path integral measure is defined with respect to the norm
\begin{equation}
||\delta\beta||^2=\int_{\Sigma}\mathrm{d}^2\sigma\sqrt{h}h^{ab}g_{ij}\beta^i_a\beta^j_b\,.
\end{equation}
For the rest of the paper, we will take the action \eqref{eq:first-order-action} as defining the near-boundary limit of the worldsheet theory on an asymptotically-$\text{AdS}_3$ spacetime.

\subsection{Relation to the Wakimoto representation}

Before moving on to computing the worldsheet path integral, we should take a moment to relate the action \eqref{eq:first-order-action} to the standard first-order action which is familiar to string theory on $\text{AdS}_3$.

Global Euclidean $\text{AdS}_3$ can be written in Poincar\'e coordinates, in which the metric tensor and B-field are
\begin{equation}
\mathrm{d}s^2=\frac{L^2}{r^2}(\mathrm{d}r^2+\delta_{ij}\mathrm{d}x^i\mathrm{d}x^j)\,,\quad B=-\frac{L^2}{r^2}\varepsilon_{ij}\mathrm{d}x^i\wedge\mathrm{d}x^j\,.
\end{equation}
That is, one chooses a gauge in the Fefferman-Graham expansion for which $g_{ij}=\delta_{ij}$ and all higher-order corrections vanish. In this gauge, both the spacetime curvature $\sqrt{g}R$ and the scalar density $f$ vanish, and the worldsheet action \eqref{eq:first-order-action} becomes
\begin{equation}
\begin{split}
S&=\frac{1}{4\pi}\int\mathrm{d}^2\sigma\sqrt{h}\left(\frac{1}{2}h^{ab}\partial_a\phi\partial_b\phi-\frac{Q}{2}R_h\phi+h^{ab}\delta_{ij}\beta^i_aDx^j_b-\nu e^{-Q\phi}h^{ab}\delta_{ij}\beta^i_a\beta^j_b\right)\,.
\end{split}
\end{equation}
Now, we can define complex coordinates
\begin{equation}
\gamma=x^1+ix^2\,,\quad\bar{\gamma}=x^1-ix^2\,,
\end{equation}
as well as complex coordinates on the worldsheet such that $h_{z\bar{z}}=e^{\omega}$ for some (local) Weyl factor $\omega$. In this coordinate system, we have
\begin{equation}
\begin{gathered}
Dx^\gamma_z=Dx^{\bar\gamma}_{\bar{z}}=0\\
Dx^{\bar\gamma}_z=\partial\bar\gamma\,,\quad Dx^{\gamma}_{\bar{z}}=\overline{\partial}\gamma\,,
\end{gathered}
\end{equation}
where $\partial=\partial_z$, $\overline{\partial}=\partial_{\bar{z}}$. Therefore, in local complex coordinates,
\begin{equation}
Dx=
\begin{pmatrix}
0 & \overline{\partial}\gamma\\
\partial\bar\gamma & 0
\end{pmatrix}\,.
\end{equation}
In this coordinate system, the worldsheet action takes the form
\begin{equation}
S=\frac{1}{2\pi}\int_{\Sigma}\mathrm{d}^2z\left(\frac{1}{2}\partial\phi\,\overline{\partial}\phi-\frac{Q}{4}\sqrt{h}R_h\phi+\beta^{\bar\gamma}_z\overline{\partial}\gamma+\beta^\gamma_{\bar{z}}\partial\bar{\gamma}-\nu e^{-Q\phi}(\beta^{\bar\gamma}_z\beta^{\gamma}_{\bar{z}}+\beta^{\gamma}_z\beta^{\bar\gamma}_{\bar z})\right)\,.
\end{equation}
Notice that the components $\beta^{\gamma}_z$ and $\beta^{\bar\gamma}_{\bar{z}}$ decouple from the dynamics completely, and so integrating them out has no effect on the theory. Writing $\beta:=\beta_z^{\bar\gamma}$ and $\bar\beta:=\beta_{\bar{z}}^{\gamma}$, we arrive at the action
\begin{equation}\label{eq:wakimoto-main-text}
S=\frac{1}{2\pi}\int_{\Sigma}\mathrm{d}^2z\left(\frac{1}{2}\partial\phi\,\overline{\partial}\phi-\frac{Q}{4}\sqrt{h}R_h\phi+\beta\overline{\partial}\gamma+\bar\beta\partial\bar{\gamma}-\nu e^{-Q\phi}\beta\bar\beta\right)\,,
\end{equation}
which is precisely the first-order action of string theory in global Euclidean $\text{AdS}_3$ \cite{Giveon:1998ns,Kutasov:1999xu}, also known as the Wakimoto representation of $\mathfrak{sl}(2,\mathbb{R})_k$ \cite{Wakimoto:1986gf}.

While the action \eqref{eq:wakimoto-main-text} is usually the starting point for studying $\text{AdS}_3$ string theory in a pure NS-NS background, it should be pointed out that it cannot be the whole story. The primary reason for this is that, when the conformal boundary of $\text{AdS}_3$ has non-vanishing Euler number, the boundary curvature (called $R_X$ above) must be non-vanishing somewhere by the Gauss-Bonnet theorem. This is reflected in the fact that Poincar\'e coordinates, while globally well-defined from the point of view of the bulk of $\text{AdS}_3$, are singular at the point at infinity on the boundary sphere. In the case of global Euclidean $\text{AdS}_3$, this difficulty can be remedied by the introduction of a screening operator \cite{Dei:2023ivl,Hikida:2023jyc,Knighton:2023mhq} which effectively `compactifies' the boundary, see also the discussion in Section \ref{sec:global-ads3}. For more complicated boundary topologies, the situation is much worse, and one must take into consideration the global structure of the boundary manifold. This is what is achieved by working up front with the covariant action \eqref{eq:first-order-action}.

\subsection{Computing the path integral}\label{sec:path-integral}

Now that we have derived a covariant action for a long string propagating in the spacetime $M\times C$, we can evaluate the Polyakov path integral and compute its free energy. In doing this, we must be careful about the path integral measure. In our model, the measures for the fields $\phi,x,\beta$ are implicitly defined by the norms
\begin{equation}
\begin{split}
||\delta\phi||^2&=\int_{\Sigma}\mathrm{d}^2z\sqrt{h}\,\delta\phi\,\delta\phi\,,\\
||\delta x||^2&=\int_{\Sigma}\mathrm{d}^2z\sqrt{h}\,g_{ij}\delta x^i\delta x^j\,,\\
||\delta\beta||^2&=\int_{\Sigma}\mathrm{d}^2z\sqrt{h}h^{ab}g_{ij}\beta_a^i\beta_b^j\,.
\end{split}
\end{equation}
The norm for $\phi$ is the standard norm for a linear dilaton CFT, while the norms for $x$ and $\beta$ are the standard norms for fields living in the spaces
\begin{equation}
x\in\text{Map}(\Sigma\to X)\,,\quad \beta\in\Gamma(\text{T}^*\Sigma\otimes x^*(\text{T}X))\,.
\end{equation}
As it stands, the path integral of the theory on $M$ is not well-defined, as the path integral measures are not invariant under Weyl transformations. Indeed, the above action defines a CFT of central charge \cite{Maldacena:2000hw}
\begin{equation}
c(M)=\frac{3k}{k-2}\,.
\end{equation}
In order to cancel the Weyl anomaly, we need to consider a worldsheet theory on $M\times C$, where $C$ is some CFT of central charge
\begin{equation}
c(C)=26-\frac{3k}{k-2}\,,
\end{equation}
so as to cancel the $c=-26$ Weyl anomaly in the integral over worldsheet metrics.

With this in place, the genus $g$ free energy of a string propagating in the background $M\times C$ is given by the path integral
\begin{equation}
\mathcal{F}_g=\int\frac{\mathcal{D}(h,\phi,x,\beta)}{|\text{Diff}\times\text{Weyl}|}e^{-S[\phi,x,\beta]}Z_C[\Sigma,h]\,,
\end{equation}
where $Z_C[\Sigma,h]$ is the partition function of the compact CFT $C$ on the worldsheet $(\Sigma,h)$. Fully computing this path integral is beyond the scope of the present work. Instead, we will simplify the computation by computing only the subsector of the path integral which is dominated by large values of $\phi$. In this case, we can compute the path integral in an expansion
\begin{equation}\label{eq:string-partition-expansion}
\mathcal{F}_g=\int\frac{\mathcal{D}(h,\phi,x,\beta)}{|\text{Diff}\times\text{Weyl}|}e^{-S_0[\phi,x,\beta]}Z_C[\Sigma,h]\left(1+\mathcal{O}(e^{-Q\phi})\right)\,,
\end{equation}
where $S_0$ is the `free' action describing long strings:
\begin{tcolorbox}
\begin{equation}\label{eq:sigma-model-action-long-string}
\begin{split}
S_0&=\frac{1}{4\pi}\int_{\Sigma}\mathrm{d}^2\sigma\,\Big(\frac{1}{2}\sqrt{h}h^{ab}\partial_a\phi\partial_b\phi-\frac{Q}{2}\sqrt{h}R_h\phi+\frac{Qk}{2}x^*(\sqrt{g}R)\phi\\
&\hspace{4.5cm}-Qf\phi+\sqrt{h}\braket{\beta,Dx}-k\braket{\mathrm{d}x,\mathrm{d}x}^{(2)}\Big)\,.
\end{split}
\end{equation}
\end{tcolorbox}

In principle, the corrections in \eqref{eq:string-partition-expansion} can be calculated order-by-order in $e^{-Q\phi}$. However, in this paper, we will only be interested in computing the leading-order contribution:
\begin{equation}\label{eq:string-partition-leading}
\mathcal{F}_g^{(0)}=\int\frac{\mathcal{D}(h,\phi,x,\beta)}{|\text{Diff}\times\text{Weyl}|}e^{-S_0[\phi,x,\beta]}Z_C[\Sigma,h]\,.
\end{equation}
This is the definition of the `long string path integral' whose evaluation we will explore in the remainder of this section. Precisely because the action $S_0$ governing the free energy $\mathcal{F}_g^{(0)}$ is quadratic in the worldsheet fields, we will find that the path integral can indeed be exactly computed. The computation of $\mathcal{F}_g^{(0)}$ proceeds in three steps:
\begin{enumerate}

    \item[1)] Integrating out the Lagrange multiplier $\beta$.

    \item[2)] Gauge-fixing the $\text{Diff}\times\text{Weyl}$ symmetry of the worldsheet theory.

    \item[3)] Summing over instanton sectors and integrating over a finite number of worldsheet moduli.

\end{enumerate}

\subsubsection*{Step 1: Integrating out the Lagrange multiplier}

The simplifying feature of the near-boundary approximation $\Phi\to\infty$ is that the action for the Lagrange multiplier $\beta$ becomes linear. In this limit, we can integrate out $\beta$ in the path integral making use of the formal identity
\begin{equation}\label{eq:delta-function-integral}
\int\mathcal{D}\beta\exp\left(-\frac{1}{4\pi}\int_{\Sigma}\mathrm{d}^2\sigma\,\sqrt{h}h^{ab}g_{ij}\beta_a^i\omega_b^j\right)=\delta(\omega)\,,
\end{equation}
where $\omega$ is any section of $\text{T}^*\Sigma\otimes x^*(\text{T}X)$. The functional delta function is defined in such a way that
\begin{equation}
\int\mathcal{D}\omega\,\delta(\omega)F[\omega]=F[0]\,,
\end{equation}
where $f$ is some functional on the space of sections of $\text{T}^*\Sigma\otimes x^*(\text{T}X)$.\footnote{Again, the integrals over $\beta$ and $\omega$ is over the space of \textit{self-dual} sections of $\text{T}^*\Sigma\otimes x^*(\text{T}\Sigma)$, i.e. those sections satisfying $s(\beta)=\beta$ and $s(\omega)=\omega$.} Thus, after integrating out $\beta$, the near-boundary string partition function takes the form
\begin{equation}\label{eq:long-string-free-energy}
\mathcal{F}^{(0)}_{g}=\int\frac{\mathcal{D}(h,\phi,x)}{|\text{Diff}\times\text{Weyl}|}\,\delta(Dx)\,e^{-S^{(2)}[x]}e^{-S_{\text{LD}}[\phi]}Z_C[\Sigma,h]\,,
\end{equation}
where $S_{\text{LD}}$ is an action for $\phi$ which we will write below. As we show in Appendix \ref{app:measure}, on the locus of support $Dx=0$, the scalar density $f$ appearing in the sigma model action can be written as the pullback of the spacetime curvature, i.e.
\begin{equation}\label{eq:f-substitution}
f=x^*(\sqrt{g}R)\,.
\end{equation}
From this, we can read off the action of the radial scalar $\phi$ on the locus $Dx=0$ to be
\begin{equation}\label{eq:linear-dilaton-action-string}
S_{\text{LD}}[\phi]=\frac{1}{4\pi}\int_{\Sigma}\mathrm{d}^2\sigma\left(\frac{1}{2}\sqrt{h}h^{ab}\partial_a\phi\partial_b\phi-\frac{Q}{2}\sqrt{h}R_h\phi+\frac{1}{Q}x^*(\sqrt{g}R)\phi\right)\,.
\end{equation}

The action $S^{(2)}$ in equation \eqref{eq:long-string-free-energy} is the remaining part of the kinetic action for $x$, namely
\begin{equation}
S^{(2)}[x]=-\frac{k}{4\pi}\int_{\Sigma}\mathrm{d}^2\sigma\sqrt{h}h^{ab}P_{ij}\partial_ax^i\partial_bx^j\,.
\end{equation}
The delta function in the partition function enforces the condition $Dx=0$ on the field $x$. This condition is equivalent to the statement that the map $x:\Sigma\to X$ satisfies
\begin{equation}\label{eq:covariant-holomorphic}
\mathrm{d}xJ=\mathcal{J}\mathrm{d}x\,,
\end{equation}
which comes from the definition of $Dx$ as well as the fact that complex structures satisfy $J^2=-1$. In local complex coordinates, this is nothing more than the condition
\begin{equation}
\overline{\partial}\gamma=0\,.
\end{equation}
That is, the near-boundary worldsheet partition function localizes precisely onto those maps $x:\Sigma\to X$ which are holomorphic with respect to the worldsheet and boundary complex structures. Equivalently, the near-boundary path integral localizes to \textit{orientation-preserving conformal maps} from the worldsheet to the boundary.

On the support of the delta function, it is possible to simplify the kinetic action $S^{(2)}[x]$ using equation \eqref{eq:covariant-holomorphic}. Indeed, we have
\begin{equation}
\begin{split}
\sqrt{h}h^{ab}P_{ij}\partial_ax^i\partial_bx^j&=J\indices{^a_c}\varepsilon^{cb}P_{ij}\partial_ax^i\partial_bx^j\\
&=\varepsilon^{cb}P_{ij}\mathcal{J}\indices{^i_k}\partial_cx^k\partial_bx^j\\
&=\sqrt{g}\varepsilon^{cb}g^{i\ell}P_{ij}\varepsilon_{k\ell}\partial_cx^k\partial_bx^j\\
&=\sqrt{g}\,\text{tr}[g^{-1}P]\text{det}(\mathrm{d}x)=\frac{1}{2}x^*(\sqrt{g}R)\,,
\end{split}
\end{equation}
where we have used the definitions of $J,\mathcal{J}$ in terms of $h,g$ \eqref{eq:complex-structures}, in going to the last line we used various properties of 2D matrices, and in the last equality we used the property $\text{tr}[g^{-1}P]=\frac{1}{2}R$ of the Fefferman-Graham metric. The end result is that the contribution $S^{(2)}$ to the full long-string action is simply
\begin{equation}
\begin{split}
S^{(2)}&=-\frac{k}{8\pi}\int_{\Sigma}\mathrm{d}^2\sigma\,x^*(\sqrt{g}R)\\
&=-\frac{kN}{8\pi}\int_X\mathrm{d}^2x\sqrt{g}R=-kN(1-G)\,,
\end{split}
\end{equation}
where $N$ is the degree of the map $x$. The upshot is that we can ignore the kinetic term $S^{(2)}$ by replacing it with its value on the support $Dx=0$ of the delta function, and so the near-boundary partition function is
\begin{equation}\label{eq:partition-function-delta}
\mathcal{F}^{(0)}_{g}=\int\frac{\mathcal{D}(h,\phi,x)}{|\text{Diff}\times\text{Weyl}|}e^{kN(1-G)}\,\delta(Dx)\,e^{-S_{\text{LD}}[\phi]}Z_C[\Sigma,h]\,.
\end{equation}
In what follows, we will usually drop the $e^{kN(1-G)}$ term, keeping in mind that we should re-introduce it in the final answer.

\subsubsection*{Step 2: Gauge-fixing}

The goal now is to compute the integral over metrics $h$ and maps $x:\Sigma\to X$ appearing in the partition function \eqref{eq:partition-function-delta}. The usual trick in bosonic string theory is to use the $\text{Diff}\times\text{Weyl}$ symmetry to pick a gauge for the worldsheet metric, and reduce the integration over metrics to an integral over the moduli space $\mathcal{M}_g$ of complex structures on $\Sigma$. This would yield the result
\begin{equation}\label{eq:naive-bc-partition}
\mathcal{F}^{(0)}_{g}=\int_{\mathcal{M}_g}\left\langle\prod_{\alpha=1}^{3g-3}\left|\int_{\Sigma}b\mu_{\alpha}\right|^2\right\rangle\int\mathcal{D}x\,\mathcal{D}\phi\,\delta(Dx)e^{-S_{\text{LD}}[\phi]}Z_C[\Sigma,h]\,,
\end{equation}
where the measures $\mathcal{D}x,\mathcal{D}\phi$, the action $S_{\text{LD}}$, and the partition function $Z_C$ are computed with respect to a reference metric $h$ for each point in the moduli space, which is usually put into the conformal gauge. The correlator defining the measure of the moduli space is a correlation function in the usual $b,c$ conformal ghost system, which has central charge $c=-26$.

While the above expression works well for something like flat space string theory, it turns out to be rather cumbersome for computing the long-string free energy $\mathcal{F}^{(0)}_{g}$. The issue is the delta function $\delta(Dx)$, which restricts the integral to holomorphic maps $x:\Sigma\to X$. For a given complex structure $J\in\mathcal{M}_g$ on the worldsheet, one is not guaranteed that a holomorphic map $x$ even exists. Indeed, for a fixed complex structure on the worldsheet, the expected (virtual) dimension of the space $\mathcal{H}_N(\sigma\to X)$ of holomorphic maps $x:\Sigma\to X$ of degree $N$ can be calculated from the Riemann-Roch theorem\footnote{The tangent space $\text{T}_x^{(1,0)}\mathcal{H}_N(X\to\Sigma)$ is naturally isomorphic to space $\Gamma(x^*(\text{T}^{(1,0)}X),\Sigma)$ of holomorphic sections of $x^*(\text{T}^{(1,0)}X)$, and so by the Riemann-Roch formula one has
\begin{equation}
\text{vdim}\,\Gamma(x^*(\text{T}^{(1,0)}x),\Sigma)=1-g+\text{deg}(x^*(\text{T}^{(1,0)}x))=1-g+N(2-2G)\,,
\end{equation}
where the virtual dimension is defined to be $\text{dim}\,\Gamma(\mathcal{L},\Sigma)-\text{dim}\,\Gamma(K_{\Sigma}\otimes\mathcal{L}^{-1},\Sigma)$ for a line bundle $\mathcal{L}$.}:
\begin{equation}\label{eq:holomorphic-virtual-dimension}
\text{vdim}\,\mathcal{H}_N(\Sigma\to X)=1-g+N(2-2G)\,,
\end{equation}
where $g$ is the genus of the worldsheet $\Sigma$ and $G$ is the genus of the boundary $X$. Since $N$ is always non-negative, assuming $g,G\geq 2$, the expected dimension of $\mathcal{H}_N(\Sigma\to X)$ is negative, and thus there are generically \textit{no} holomorphic maps from the worldsheet to the boundary, and so the delta function $\delta(Dx)$ has no support.

This is not to say that there are \textit{never} holomorphic maps. The negative virtual dimension \eqref{eq:holomorphic-virtual-dimension} simply means that, in order for the delta function $\delta(Dx)$ to have support, one needs to fine-tune $g-1+N(2G-2)$ of the worldsheet moduli. That is to say that the integral over the moduli space $\mathcal{M}_g$ will localize to a subspace of dimension
\begin{equation}\label{eq:full-moduli-space-dimension}
3g-3+\text{vdim}\,\mathcal{H}_N(\Sigma\to X)=2g-2+N(2-2G)\,,
\end{equation}
which is positive so long as $g$ is large enough.

The point of the above discussion is to emphasize that the partition function \eqref{eq:naive-bc-partition}, while in principle the correct quantity to compute, is more subtle than one might have expected, as the integral over $\mathcal{M}_g$ localizes to a subspace where the worldsheet can holomorphically map to the boundary. While we expect that it is nevertheless possible to carry out the integration, we will opt for a different approach in evaluating \eqref{eq:partition-function-delta}, although we will comment on the fate of the expression \eqref{eq:naive-bc-partition} for $G=0,1$ in Section \ref{sec:backgrounds}.

The trick is to consider the integral over metrics first, and then integrate over the remaining space of maps. The delta function $\delta(Dx)$ can either be thought of as a constraint on the space of maps $x:\Sigma\to X$ given a complex structure $J$, or it can be seen as a constraint on the complex structure $J$ given the map $x$. Indeed, the condition $Dx=0$ is equivalent to demanding that $J$ is the pullback of the complex structure $\mathcal{J}$ under the map $x$.

Demanding that $J$ is the pullback of $\mathcal{J}$ fixes the worldsheet metric $h$ to be in the same conformal class as the pullback metric $x^*g$:
\begin{equation}
h=e^{2\omega}x^*g\,.
\end{equation}
Thus, using the delta function in $\mathcal{F}^{(0)}_g$, we can reduce the integral over metrics to an integral over a single Weyl factor $\omega$. Since the path integral is Weyl-invariant (i.e. since the worldsheet theory is critical), we can formally cancel the integral over $\omega$ with the volume of the group of Weyl transformations, and simply gauge-fix $\omega=0$ in the rest of the path integral. The upshot of this discussion is that we can write
\begin{equation}
\mathcal{F}^{(0)}_g=\int\frac{\mathcal{D}x}{|\text{Diff}(\Sigma)|}\int\mathcal{D}\phi\,e^{-S_{\text{LD}}[\phi]}Z_C[\Sigma,x^*g]\,,
\end{equation}
where the path integral measures for $x,\phi$, as well as the partition function for the compact CFT $C$, are evaluated with respect to the pullback metric $h=x^*g$.

\subsubsection*{Step 3: The moduli space integral}

Apart from the path integral of the linear dilaton $\phi$, the remaining integral in $\mathcal{F}_g^{(0)}$ is over the space
\begin{equation}\label{eq:gauge-fixed-moduli-space}
\mathcal{M}_g(X):=\text{Map}(\Sigma,X)/\text{Diff}(\Sigma)
\end{equation}
of orientation-preserving maps from the worldsheet to the conformal boundary of $M$. As mentioned in Section \ref{sec:effective-action}, this space is disconnected, and consists of a connected component for every integer $N\geq 1$, which counts the number of times the worldsheet wraps the boundary. The component $\mathcal{M}_g(X,N)$ of the moduli space \eqref{eq:gauge-fixed-moduli-space} consisting of maps of degree $N$ has complex dimension
\begin{equation}
\text{dim}_{\mathbb{C}}(\mathcal{M}_g(X,N))=2g-2+N(2-2G)\,,
\end{equation}
in agreement with equation \eqref{eq:full-moduli-space-dimension}. By the Riemann-Hurwitz formula, this is equal to the number $m$ of branch points of $x$, counted with multiplicity. In local complex coordinates, these are points $\xi_\ell$ on the conformal boundary $X$ around which
\begin{equation}
\gamma(z)\sim \xi_\ell+\mathcal{O}((z-\zeta_\ell)^2)
\end{equation}
for a corresponding set of pre-images $\zeta_\ell$.\footnote{A given map $x$ may have branch points of higher-order, but these lie at the boundary strata of the moduli space. We will however return to them later when we discuss the dual CFT.} These branch points $\xi_\ell$ provide a set of local coordinates on $\mathcal{M}_g(X,N)$. For a given set of branch points $\xi_{\ell}$, there may be a discrete number of allowed covering maps. Thus, we can write the integral over $\mathcal{M}_g(X,N)$ as an integral over the coordinates of the branch points, as well as a sum over the possible covering maps for each set of branch points. That is, we can make the replacement\footnote{While we do not derive this measure explicitly, we note that it is the most natural result one could write down which is constistent with diffeomorphism and Weyl transformations on the boundary.}
\begin{equation}
\int\frac{\mathcal{D}x}{|\text{Diff}(\Sigma)|}=\sum_{m=0}^{\infty}\frac{1}{m!}\int\prod_{\ell=1}^{m}\mathrm{d}^2\xi_\ell\sqrt{g(\xi_\ell)}\sum_{\substack{\text{covering maps}\\x:\Sigma\to X}}\frac{1}{|\text{Aut}(x)|}\,.
\end{equation}
The automorphism factor arises from the fact that the moduli space $\mathcal{M}_g(X,N)$ has orbifold singularities at the points where the map $x$ has a nontrivial automorphism group. We will ignore this factor below, but keep in our mind that it is in principle there.

In this set of coordiantes for the moduli space of maps, we can write the genus $g$ free energy as
\begin{equation}
\mathcal{F}_g^{(0)}=e^{k(1-g)}\sum_{m=0}^{\infty}\frac{e^{km/2}}{m!}\int\prod_{\ell=1}^{m}\mathrm{d}^2\xi_\ell\sqrt{g(\xi_\ell)}\sum_{\substack{\text{covering maps}\\x:\Sigma\to X}}\int\mathcal{D}\phi\,e^{-S_{\text{LD}}[\phi]}Z_C[\Sigma,x^*g]\,.
\end{equation}
Here, the prefactor of $e^{k(1-g)+km/2}$ comes from the factor of $e^{Nk(1-G)}$ which we dropped previously, where we have used the Riemann-Hurwitz formula \eqref{eq:Riemann-Hurwitz} to wirte it in terms of $g$ and $m$.

Now, the linear dilaton action $S_{\text{LD}}[\phi]$ is a function of both the worldsheet curvature $R_h$ and the boundary curvature $R$. However, since we have gauged-fixed $h=x^*g$, we can use the identity
\begin{equation}
x^*(\sqrt{g}R)=\sqrt{h}R_h+4\pi\sum_{\ell=1}^{m}\delta^{(2)}(\sigma,\zeta_\ell)
\end{equation}
which holds for $h=x^*g$. Plugging this into the expression into \eqref{eq:linear-dilaton-action-string}, we have
\begin{equation}
S_{\text{LD}}[\phi]=\frac{1}{4\pi}\int_{\Sigma}\mathrm{d}^2\sigma\sqrt{h}\left(\frac{1}{2}h^{ab}\partial_a\phi\partial_b\phi-\frac{\mathcal{Q}}{2}R_h\phi\right)+\frac{1}{Q}\sum_{\ell=1}^{m}\phi(\zeta_\ell)\,,
\end{equation}
where we have defined
\begin{equation}
\mathcal{Q}=Q-\frac{2}{Q}=-\sqrt{\frac{2(k-3)^2}{k-2}}\,.
\end{equation}
Thus, for a given map $x$, the path integral over the radial field $\phi$ is given by
\begin{equation}
\int\mathcal{D}\phi\,e^{-S_{\text{LD}}[\phi]}=\Braket{\prod_{\ell=1}^{m}e^{-\phi/Q}(\zeta_\ell)}_{(\Sigma,x^*g)}\,,
\end{equation}
where the correlation function is taken in the linear dilaton theory of slope $\mathcal{Q}$ with respect to the background metric $h=x^*g$.

Putting everything together, we find that the long-string free energy is given by the integral
\begin{equation}
\begin{split}
\mathcal{F}_g^{(0)}=&e^{k(1-g)}\sum_{m=0}^{\infty}\frac{e^{km/2}}{m!}\int\prod_{\ell=1}^{m}\mathrm{d}^2\xi_\ell\sqrt{g(\xi_\ell)}\\
&\times\sum_{\substack{\text{covering maps}\\x:\Sigma\to X}}\Braket{\prod_{\ell=1}^{m}e^{-\phi/Q}(\zeta_\ell)}_{(\Sigma,x^*g)}Z_C[\Sigma,x^*g]\,.
\end{split}
\end{equation}
The integrand is in principle computable via Wick contractions, assuming that one knows how to calculate the partition function $Z_C$ of the compact CFT on an arbitrary Riemann surface $(\Sigma,x^*g)$. The full second-quantized long-string partition function is then given by
\begin{equation}
\mathfrak{Z}^{(0)}_{\text{string}}=\exp\left(\sum_{g=0}^{\infty}g_s^{2g-2}\mathcal{F}_g^{(0)}\right)\,.
\end{equation}
Since $\mathcal{F}_g^{(0)}$ itself breaks up into a sum over the number of branch points $m$, it is convenient to break the topological expansion of $\mathfrak{Z}^{(0)}$ even further as
\begin{equation}\label{eq:long-string-z-final}
\mathfrak{Z}^{(0)}_{\text{string}}=\exp\left(\sum_{g=0}^{\infty}\sum_{m=0}^{\infty}(g_se^{-k/2})^{2g-2}\frac{e^{km/2}}{m!}\mathcal{F}_{g,m}^{(0)}\right)\,,
\end{equation}
with
\begin{equation}\label{eq:fgm0}
\mathcal{F}_{g,m}^{(0)}=\int\left(\prod_{\ell=1}^{m}\mathrm{d}^2\xi_\ell\sqrt{g}\right)\sum_{\substack{\text{covering maps}\\x:\Sigma\to X}}\Braket{\prod_{\ell=1}^{m}e^{-\phi/Q}(\zeta_\ell)}_{(\Sigma,x^*g)}Z_C[\Sigma,x^*g]\,.
\end{equation}
Again, the sum is over the finite set of distinct branched covering maps $x:\Sigma\to X$ for a fixed set of branch points $\xi_\ell\in X$.

\section{The dual CFT}\label{sec:dual-cft}

In this section we describe the dual CFT proposed by \cite{Eberhardt:2021vsx}, and demonstrate how the Coulomb gas expansion of its partition function precisely reproduces the partition function of the long-string CFT as computed in the previous section. As the CFT of \cite{Eberhardt:2021vsx} is based on a deformation of a symmetric orbifold, we first describe the basics of symmetric orbifolds before discussing the dual CFT in detail. For a concrete introduction to symmetric orbifold CFTs, see Section 2 of \cite{Dei:2019iym}, as well as Section 2.2 of \cite{Kames-King:2023fpa}. Readers familiar with symmetric orbifolds should feel free to skip to Section \ref{subsec:seed-theory}.

\subsection{Symmetric orbifolds}

Given a CFT $\mathcal{S}$ (the `seed' CFT), the $N^{\text{th}}$ symmetric orbifold of $\mathcal{S}$, denoted $\text{Sym}^N(\mathcal{S})$, is the orbifold CFT
\begin{equation}
\text{Sym}^N(\mathcal{S})=(\underbrace{\mathcal{S}\otimes\cdots\otimes \mathcal{S}}_{N\text{ times}})/S_N\,,
\end{equation}
where $S_N$ is the symmetric group on $N$ elements. Intuitively, the symmetric orbifold describes a `second quantized' version of the CFT $\mathcal{S}$, whose untwisted state space consists of states in the $N$ copies of $\mathcal{S}$ which are invariant under permutation.

Since the group of permutations is a gauge symmetry, states/field configurations which are related by a perumtation in $S_N$ are physically equivalent. Thus, as with all orbifolds, one must in principle allow the fields in the theory to obey twisted boundary conditions. That is, given a (non-contractible) loop $\gamma$, we must allow for the possibility that a fundamental field $\boldsymbol{\Phi}$ in the CFT $\mathcal{S}^{\otimes N}$ satisfies the boundary condition
\begin{equation}
\boldsymbol{\Phi}(\gamma\cdot x)=(\pi\cdot\boldsymbol{\Phi})(x)\,.
\end{equation}
Here, we understand $\boldsymbol{\Phi}$ as an $N$-tuple of fields in the seed theory $\mathcal{S}$, where a permutation $\pi$ acts in the obvious way. We also use the shorthand $\gamma\cdot x$ to mean the action of transporting $\boldsymbol{\Phi}$ around a closed loop which starts and ends at the point $x$.

In order to calculate the partition function of the symmetric orbifold, one must sum over all compatible choices of twisted boundary conditions. Concretely, let $X$ be a Riemann surface of genus $G$ with some fixed metric tensor $g_{ij}$. Each independent closed loop on $X$ is labeled by an element of the fundamental group $\pi_1(X)$, and a twisted boundary condition corresponds to a homomorphism $f:\pi_1(X)\to S_N$ which assigns a permutation (i.e. a twist) to each closed loop in $X$. The partition function of the symmetric orbifold is then
\begin{equation}\label{eq:sn-partition-function}
Z_N[X,g]=\frac{1}{N!}\sum_{f:\pi_1(X)\to S_N}Z^{(f)}[X,g]\,,
\end{equation}
where $Z^{(f)}$ is the path integral of $\mathcal{S}^{\otimes N}$ computed with the twisted boundary conditions determined by the homomorphism $f$.

The partition function $Z^{(f)}$ is in turn computed by passing to a covering space $\Sigma$. The surface $\Sigma$ is in turn equipped with an $N$-to-$1$ map $\Gamma:\Sigma\to X$, such that the field configuration of $\mathcal{S}^{\otimes N}$ on $X$ is a single-valued configuration on one copy of $\mathcal{S}$ when lifted to $\Sigma$ \cite{Dixon:1986qv,Hamidi:1986vh,Lunin:2000yv}. Each choice $f$ of twisted boundary conditions determines a covering space along with a covering map $\Gamma:\Sigma\to X$, and the number of times $\Sigma$ covers $X$ is precisely $N$. However, for a given covering space $\Sigma$, there are in principle multiple homomorphisms $f$ corresponding to it. Indeed, the homomorphisms $f$ and $\pi^{-1}\circ f\circ\pi$ for some permutation $\pi$ will always yield the same covering space, since the conjugation by $\pi$ simply `shuffles' the copies of $\mathcal{S}$. In other words, covering spaces $\Gamma:\Sigma\to X$ are in one to one correspondence with homomorphisms $f$ \textit{up to conjugation} by elements of $S_N$.

In the language of covering spaces, the partition function \eqref{eq:sn-partition-function} can be written as
\begin{equation}\label{eq:covering-map-partition}
Z_N[X,g]=\sum_{\Gamma:\Sigma\to X}\frac{1}{|\text{Aut}(\Gamma)|}Z[\Sigma,\Gamma^*g]\,.
\end{equation}
Here, we sum over all covering spaces $\Sigma$ of degree $N$, equipped with a covering map $\Gamma$. The partition function $Z[\Sigma,\Gamma^*g]$ is the partition function of the seed CFT $\mathcal{S}$ on the covering space $\Sigma$, with respect to the pullback metric $\Gamma^*g$. Finally, the automorphism group $\text{Aut}(\Gamma)$ (also known as the group of \textit{Deck transformations}) is the group of continuous maps $\varphi:\Sigma\to\Sigma$ such that $\Gamma\circ\varphi=\Gamma$. See Section 2.2 of \cite{Kames-King:2023fpa} for a detailed derivation of \eqref{eq:covering-map-partition}.

\subsection{The grand canonical ensemble}

The partition function of the symmetric orbifold for a fixed value of $N$ is computed by summing over the partition function of the seed theory $\mathcal{S}$ on all covering spaces $\Sigma$ of degree $N$. It is in practice cumbersome to restrict the degree of the covering map, and so it is convenient to introduce the `grand canonical' partition function
\begin{equation}
\mathfrak{Z}[X,g]=\sum_{N=0}^{\infty}p^NZ_N[X,g]\,.
\end{equation}
Here, $p=e^{2\pi i\sigma}$, where the imaginary part of $\sigma$ plays a role analogous to that of a chemical potential in statistical physics. The convenience of the grand canonical partition function is that it exponentiates into a sum over only connected covering maps:
\begin{equation}\label{eq:grand-canonical-general}
\mathfrak{Z}[X,g]=\exp\left(\sum_{\substack{\Gamma:\Sigma\to X\\\text{connected}}}\frac{p^{N}}{|\text{Aut}(\Gamma)|}Z[\Sigma,\Gamma^*g]\right)\,,
\end{equation}
where $N=\text{deg}(\Gamma)$ is again the number of times $\Sigma$ covers $X$. In the special case that $X$ is a torus with flat metric and modular parameter $t$, one recovers the famous DMVV formula \cite{Dijkgraaf:1996xw}. 

\subsection{The seed theory}\label{subsec:seed-theory}

The CFT of \cite{Eberhardt:2021vsx}, proposed to be dual to pure NS-NS string theory on $\text{AdS}_3\times C$, is based on the symmetric orbifold of the seed theory
\begin{equation}
\mathcal{S}=\mathbb{R}_{\mathcal{Q}}\times C\,.
\end{equation}
Here, by $\mathbb{R}_{\mathcal{Q}}$ we mean a linear dilaton CFT governed by the action
\begin{equation}
S_{\text{LD}}=\frac{1}{4\pi}\int_X\mathrm{d}^2x\sqrt{g}\left(\frac{1}{2}g^{ij}\partial_{i}\phi\partial_{j}\phi-\frac{\mathcal{Q}}{2}R\phi\right)\,,
\end{equation}
where $R$ is the scalar curvature on the Riemann surface $X$. The background charge $\mathcal{Q}$ is determined by the string tension $k$, and is given by \eqref{eq:background-charge-correction}. 

The grand canonical partition function of the symmetric orbifold of $\mathcal{S}$ is determined directly by using the general formula \eqref{eq:grand-canonical-general}:
\begin{equation}
\mathfrak{Z}[X,g]=\exp\left(\sum_{\substack{\Gamma:\Sigma\to X\\\text{connected}}}\frac{p^{N}}{|\text{Aut}(\Gamma)|}\int\mathcal{D}\phi\,e^{-S_{\text{LD}}[\Sigma,\Gamma^*g]}Z_{C}[\Sigma,\Gamma^*g]\right)\,,
\end{equation}
where
\begin{equation}
S_{\text{LD}}[\Sigma,\Gamma^*g]=\frac{1}{4\pi}\int_{\Sigma}\mathrm{d}^2\sigma\sqrt{h}\left(\frac{1}{2}h^{ab}\partial_a\phi\partial_b\phi-\frac{\mathcal{Q}}{2}R_h\phi\right)\,,
\end{equation}
with $h=\Gamma^*g$ the induced metric on the covering surface. Since we only sum over unbranched covering maps which are connected, the genus $g$ of the covering surface $\Sigma$ is determined by the genus $G$ of $X$ through the Riemann-Hurwitz formula
\begin{equation}
(2-2G)N=2-2g\,,
\end{equation}
and so we can organize $\mathfrak{Z}$ into a sum over genera of $g$ of the covering space, namely
\begin{equation}
\mathfrak{Z}[X,g]=\exp\left(\sum_{g=0}^{\infty}\sum_{\Gamma:\Sigma\to X}\frac{p^{\frac{2g-2}{2G-2}}}{|\text{Aut}(\Gamma)|}\int\mathcal{D}\phi\,e^{-S_{\text{LD}}[\Sigma,\Gamma^*g]}Z_{C}[\Sigma,\Gamma^*g]\right)\,.
\end{equation}
We recognize the summand in the above exponential as corresponding to the free energies $\mathcal{F}_{g,0}^{(0)}$ which we computed in Section \ref{sec:sigma-model}, see equation \eqref{eq:fgm0}, and so we have the identification
\begin{equation}
\mathfrak{Z}[X,g]=\exp\left(\sum_{g=0}^{\infty}p^{\frac{2g-2}{2G-2}}\mathcal{F}_{g,0}^{(0)}\right)\,.
\end{equation}
Thus, the grand canonical partition function of the symmetric orbifold of $\mathcal{S}=\mathbb{R}_{\mathcal{Q}}\times C$ matches the $m=0$ truncation of the long-string partition function on $M\times C$ if we make the identification (see also \cite{Eberhardt:2021jvj,Aharony:2024fid})
\begin{equation}\label{eq:chemical-potential-dictionary}
p=(g_se^{-k/2})^{2G-2}\,.
\end{equation}
In the identification, the covering map $\Gamma$ and linear dilaton $\phi$ are directly identified with the embedding coordinates $x$ and the radial direction in $\text{AdS}_3$.

\subsection{The deformation}

In order to obtain a matching of the long-string partition function with a dual CFT quantity, we need to be able to reproduce the free energies $\mathcal{F}_{g,m}^{(0)}$ for $m>0$ from a CFT partition function. This can be done by adding the interaction term \cite{Eberhardt:2021vsx}
\begin{equation}\label{eq:interaction-term-cft}
S_{\text{int}}=\mu\int\mathrm{d}^2x\sqrt{g}\,\sigma_{2}\,e^{-\phi/Q}\,,
\end{equation}
where $\sigma_2$ is the ground state in the $2$-cycle twisted sector of the symmetric orbifold. In the symmetric orbifold with a fixed value of $N$, the partition function of the deformed theory is in principle calculated by the expectation value
\begin{equation}\label{eq:deformed-partition-fixed-n}
Z_N[X,g]=\Braket{\exp\left(-\mu\int\mathrm{d}^2x\sqrt{g}\,\sigma_2\,e^{-\phi/Q}\right)}\,,
\end{equation}
where the expectation value is taken in the fixed symmetric orbifold $\text{Sym}^N(\mathcal{S})$. In a coulomb gas approximation, one can then compute $Z_N$ by dropping down appropriate powers of the interaction Lagrangian in the path integral. For fixed $N$, this is complicated by the fact that one includes both disconnected and connected covering maps in the expansion. However, it was shown in \cite{Aharony:2024fid} that the perturbative series defining \eqref{eq:deformed-partition-fixed-n} exponentiates in the grand canonical ensemble and becomes\footnote{From now on we ignore the factor of $|\text{Aut}(\Gamma)|$ and leave it implicit that we count covering maps up to automorphism.}
\begin{equation}\label{eq:grand-canonical-deformed}
\begin{split}
\mathfrak{Z}_{\text{CFT}}^{\text{pert}}[X,g]=\exp\Bigg(&\sum_{m=0}^{\infty}\frac{(-\mu)^m}{m!}\int\left(\prod_{\ell=1}^{m}\mathrm{d}^2\xi_\ell\sqrt{g}\right)\\
&\times\sum_{\text{covering maps}}p^N\Braket{\prod_{\ell=1}^{m}e^{-\phi/Q}(\zeta_\ell)}_{(\Sigma,\Gamma^*g)}Z_C[\Sigma,\Gamma^*g]\Bigg)\,.
\end{split}
\end{equation}
This follows from equation (3.12) of \cite{Aharony:2024fid} by taking $J_2=-\mu\sqrt{g}\,e^{-\phi/Q}$ and $J_w=0$ for $w\neq 2$. Here, the sum is over all covering maps and covering spaces $\Sigma\to X$ which are branched over the points $\xi_\ell$ on $X$, and $\zeta_\ell$ are the preimages of the branch points on the covering surface $\Sigma$. That the screening operators $e^{-\phi/Q}$ are inserted at the branch points on $\Sigma$ is a consequence of the interaction operator \eqref{eq:interaction-term-cft} living in the 2-cycle twisted sector. We use the notation $\mathfrak{Z}_{\text{CFT}}^{\text{pert}}$ to 1) distinguish the partition function from the string partition function defined in Section \ref{sec:sigma-model}, and 2) to emphasize that it is computed in conformal perturbation theory.

Again, by the Riemann-Hurwitz formula, we can relate the degree $N$ of the covering map to the genera of the base space $X$ and the covering space $\Sigma$, as well as the number $m$ of branch points in the covering map:
\begin{equation}
(2-2G)N=2-2g+m\,.
\end{equation}
Thus, we can write $\mathfrak{Z}$ in a topological expansion as
\begin{equation}
\begin{split}
\mathfrak{Z}_{\text{CFT}}^{\text{pert}}[X,g]=\exp\Bigg(&\sum_{g=0}^{\infty}\sum_{m=0}^{\infty}\frac{(-\mu)^m}{m!}p^{\frac{2-2g+m}{2-2G}}\int\left(\prod_{\ell=1}^{m}\mathrm{d}^2\xi_\ell\sqrt{g}\right)\\
&\times\sum_{\text{covering maps}}\Braket{\prod_{\ell=1}^{m}e^{-\phi/Q}(\zeta_\ell)}_{(\Sigma,\Gamma^*g)}Z_C[\Sigma,\Gamma^*g]\Bigg)\,,
\end{split}
\end{equation}
where now the sum is over covering maps branched over $\xi_\ell$ with a fixed genus. The deformed CFT partition function thus precisely reproduces the long-string partition function \eqref{eq:long-string-z-final} upon making the identification\footnote{Since we have ignored overall constants in the path integral, it is possible that this dictionary is only valid up to multiplicative factors not depending on $g_s$.}
\begin{equation}
\mu=-g_s\,,\quad p=(g_se^{-k/2})^{2G-2}\,.
\end{equation}

We should note that, strictly speaking, all of these manipulations are purely formal. Specifically, since the interaction operator $\sigma_2e^{-\phi/Q}$ is not normalizable, conformal perturbation theory alone cannot be used to compute \eqref{eq:deformed-partition-fixed-n}, and so the arguments of \cite{Aharony:2024fid} to write the grand canonical partition function \eqref{eq:grand-canonical-deformed} are invalid. However, this does not imply that \eqref{eq:grand-canonical-deformed} is useless or does not tell us anything about the partition function of the deformed CFT. What \eqref{eq:grand-canonical-deformed} tells us is information about various poles in the `true' partition function, which occur when the momentum conservation condition
\begin{equation}
\frac{m}{Q}=\mathcal{Q}(1-g)
\end{equation}
is satisfied for some nonnegative integer $m$.\footnote{For generic values of $k$, it is possible that this is never satisfied. Indeed, the conservation equation is equivalent to $2(g-1)(k-3)/(k-2)\in\mathbb{Z}_{\geq 0}$. By taking $k\in\mathbb{Q}$ and $k>3$ or $k<2$, we can guarantee that such a genus $g$ exists. Alternatively, instead of computing the empty partition function, one could compute correlators of local operators which have nonzero $\phi$-momentum, as was done, for example, in \cite{Eberhardt:2021vsx,Hikida:2023jyc,Knighton:2023mhq,Knighton:2024qxd}.} This is analogous to the case of Liouville theory, where the Coulomb gas expansion of the exponential interaction operator gives only part of the answer, namely a subset of the poles appearing in a given correlation function \cite{Dorn:1994xn,Zamolodchikov:1995aa}.

Thus, since \eqref{eq:grand-canonical-deformed} does not compute the full CFT partition function (and, indeed, the full partition function is not known), it cannot capture the full physics of string theory in $\text{AdS}_3$, but rather only a subset of it. The fact that \eqref{eq:grand-canonical-deformed} matches the long-string sector of string theory on $M\times C$, then, is indicative of the fact that the long-string sector of string theory is dual to the Coulomb gas expansion of the CFT of \cite{Eberhardt:2021vsx}. This was predicted in \cite{Knighton:2024qxd}, and was demonstrated there for genus zero correlation functions on global Euclidean $\text{AdS}_3$.

\section{The holographic Weyl anomaly}\label{sec:Weyl-anomaly}

As discussed in the previous section, the CFT which captures the long-string sector of string theory on $\text{AdS}_3\times C$ is a deformation of a symmetric orbifold CFT whose seed theory is a linear dilaton which captures the radial direction of $\text{AdS}_3$. However, unlike in traditional holographic dualities, the dual field theory is not a single CFT, but rather a grand canonical ensemble of symmetric orbifolds with different values of the central charge. This is not a new idea, and can be seen most easily in the operator formalism, in which the DDF operators corresponding to the boundary Virasoro algebra has a central charge which depends on the winding sector of the worldsheet \cite{Giveon:1998ns,Kutasov:1999xu}. In this section, we will show how this observation can be made precise in the path integral formalism developed in this paper.

From the point of view of AdS/CFT, the idea of a worldsheet theory being dual to an ensemble of CFTs with differing central charges is rather counter-intuitive. Indeed, the usual holographic dictionary relates a given theory of quantum gravity in $\text{AdS}_{d+1}$ to a CFT on the boundary with a fixed number of degrees of freedom, usually related to $g_s$ by 't Hooft diagramatics. For example, type IIB string theory on $\text{AdS}_5\times\text{S}^5$ is dual to $\mathcal{N}=4$ super Yang-Mills theory with gauge group $\text{SU}(N)$, where $N\sim 1/g_s$. In this example, there is no sense in which one `averages over $N$'. In light of this observation, string theory on $\text{AdS}_3\times C$ with pure NS-NS flux seems to exhibit an unusual behavior which is absent in all other known UV-complete AdS/CFT dualities, namely that the dual field theory is a grand canonical ensemble of CFTs with varying central charges.

In the path integral language, the role of the central charge of a 2D CFT is in the determination of the Weyl anomaly, i.e. the non-invariance of the path integral under a Weyl transformation $g\to e^{2\omega}g$. Thus, to gain insight into the role of the central charge in the long-string worldsheet theory, it is fruitful to examine the dependence of the worldsheet path integral on the Weyl class of metric chosen on the boundary.

Specifically, we consider the simultaneous transformation
\begin{equation}\label{eq:weyl-transformation}
g_{ij}\to e^{2\omega}g_{ij}\,,\quad \phi\to\phi-\frac{2}{Q}\omega\,,
\end{equation}
for some function $\omega$ on the conformal boundary of $M$. This is the set of transformations which, to leading order in $1/r$, keeps the metric \eqref{eq:fefferman-graham} and $B$-field \eqref{eq:fefferman-graham-b-field} invariant. While this transformation is a symmetry of the classical background on which the string propagates, it is not a symmetry of the full quantum worldsheet theory, as we will now show.

Recall that the worldsheet sigma model near the conformal boundary of $M$ is based on the first-order action\footnote{Here, we have used the equation of motion $Dx=0$ of $\beta$ to make the replacement $f=x^*(\sqrt{g}R)$ in equation \eqref{eq:sigma-model-action-long-string}, see Appendix \ref{app:measure} and the discussion around equation \eqref{eq:f-substitution}.}
\begin{equation}
\begin{split}
S&=\frac{1}{4\pi}\int_{\Sigma}\mathrm{d}^2\sigma\bigg(\frac{1}{2}\sqrt{h}h^{ab}\partial_a\phi\partial_b\phi-\frac{Q}{2}\sqrt{h}R_h\phi+\frac{1}{Q}x^*(\sqrt{g}R)\phi\\
&\hspace{6cm}+\sqrt{h}\braket{\beta,Dx}-k\sqrt{h}\braket{\mathrm{d}x,\mathrm{d}x}^{(2)}\bigg)\,.
\end{split}
\end{equation}
All terms in the second line are invariant under the Weyl transformation \eqref{eq:weyl-transformation}, and so all variations come from the first line. Using the transformation property
\begin{equation}
\sqrt{g}R\to \sqrt{g}R-2\sqrt{g}\nabla_g\omega
\end{equation}
of the Ricci scalar under a conformal change, we find that the `classical' variation of the sigma model action is
\begin{equation}
\begin{split}
\delta S_{\text{cl}}=\frac{1}{4\pi}\int_{\Sigma}\mathrm{d}^2\sigma\bigg(&\frac{2}{Q}\left(\sqrt{h}\nabla_h\omega-x^*(\sqrt{g}\nabla_g\omega)\right)\phi+\frac{2}{Q^2}\sqrt{h}h^{ab}\partial_a\omega\partial_b\omega\\
&\hspace{0.25cm}+\sqrt{h}R_h\omega-\frac{2}{Q^2}x^*(\sqrt{g}R)\omega-\frac{4}{Q^2}x^*(\sqrt{g}g^{ij}\partial_i\omega\partial_j\omega)\bigg)\,.
\end{split}
\end{equation}
Furthermore, there is a `quantum' anomaly coming from the path integral measure of the field $x$, which we calculate in Appendix \ref{app:measure}, and which takes the form
\begin{equation}
\delta S_{\text{q}}=-\frac{1}{2\pi}\int_{\Sigma}\left(\sqrt{h}h^{ab}\partial_a\omega\partial_b\omega+\frac{1}{2}\sqrt{h}R_h\omega+x^*(\sqrt{g}R)\omega\right)\,.
\end{equation}
Thus, the full anomaly is captured by the variation
\begin{equation}
\begin{split}
\delta S_{\text{cl}}+\delta S_{\text{q}}=\frac{1}{4\pi}\int_{\Sigma}\mathrm{d}^2\sigma\bigg(&\frac{2}{Q}\left(\sqrt{h}\nabla_h\omega-x^*(\sqrt{g}\nabla_g\omega)\right)\phi+\frac{2}{Q^2}\sqrt{h}h^{ab}\partial_a\omega\partial_b\omega\\
&\hspace{1cm}-kx^*(\sqrt{g}R)\omega-\left(2+\frac{4}{Q^2}\right)x^*(\sqrt{g}g^{ij}\partial_i\omega\partial_j\omega)\bigg)
\end{split}
\end{equation}
in the quantum effective action of the worldsheet theory. Here, we have used explicitly that $Q=\sqrt{2/(k-2)}$. Finally, we can use the fact that the worldsheet path integral localizes onto the set of worldsheet metrics which live in the same Weyl class as the pullback metric $x^*g$. Specifically, we can use the following identities, which hold when $h$ and $x^*g$ are in the same conformal class:
\begin{equation}
\begin{split}
\sqrt{h}\nabla_h\omega&=x^*(\sqrt{g}\nabla_g\omega)\,,\\
\sqrt{h}h^{ab}\partial_a\omega\partial_b\omega&=x^*(\sqrt{g}g^{ij}\partial_i\omega\partial_j\omega)\,.
\end{split}
\end{equation}
Substituting into the anomaly, the integrand simplifies and we find
\begin{equation}
\begin{split}
\delta S_{\text{cl}}+\delta S_{\text{q}}&=-\frac{k}{4\pi}\int_{\Sigma}\mathrm{d}^2\sigma \,x^*(\sqrt{g}g^{ij}\partial_i\omega\partial_j\omega+\sqrt{g}R\omega)\\
&=-\frac{Nk}{4\pi}\int_{X}\mathrm{d}^2x\left(\sqrt{g}g^{ij}\partial_i\omega\partial_j\omega+\sqrt{g}R\omega\right)\,,
\end{split}
\end{equation}
where $N$ is the degree of the map $x$. Using the Riemann-Hurwitz formula, we can package this into a transformation law for the free energies $\mathcal{F}_{g,m}^{(0)}$ under holographic Weyl transformations:
\begin{tcolorbox}
\begin{equation}
\mathcal{F}_{g,m}^{(0)}\to\exp\left(\frac{Nk}{4\pi}\int_{X}\mathrm{d}^2x\left(\sqrt{g}g^{ij}\partial_i\omega\partial_j\omega+\sqrt{g}R\omega\right)\right)\mathcal{F}_{g,m}^{(0)}\,,
\end{equation}
\end{tcolorbox}
\noindent where $N$ is fixed by $g,G,m$ using \eqref{eq:Riemann-Hurwitz}.

The upshot of the above analysis is that the holographic Weyl anomaly in the worldsheet path integral matches the Weyl anomaly of a 2D CFT on the boundary $X$, whose central charge is given by
\begin{equation}
c=6Nk\,,
\end{equation}
Since the worldsheet path integral is given by a sum over topological sectors, each labeled by a different value of the winding number $N$, we find that the worldsheet theory of long strings cannot be dual to a single CFT on the conformal boundary with a fixed central charge, but rather to an ensemble of CFTs whose central charge is determined by the winding number of the worldsheet around the boundary.

This is in stark contrast to, say, the case of a worldsheet theory dual to a CFT of fixed central charge $c$. In this case, based on intuition from supergravity arguments \cite{Henningson:1998gx}, it is suspected that the holographic Weyl anomaly is captured entirely in the transformation law of the sphere partition function
\begin{equation}
\mathcal{F}_0\to\mathcal{F}_0+\frac{\alpha}{24\pi}\int_X\mathrm{d}^2x\left(\sqrt{g}g^{ij}\partial_i\omega\partial_j\omega+\sqrt{g}R\omega\right)\,,
\end{equation}
for some parameter $\alpha$, so that the full partition function
\begin{equation}
Z_{\text{string}}=\exp\left(g_s^{-2}\mathcal{F}_0+\mathcal{F}_1+g_s^2\mathcal{F}_2+\cdots\right)
\end{equation}
picks up a holographic Weyl anomaly with fixed central charge $\alpha/g_s^2$. In contrast, the long-string sector of bosonic string theory on $\text{AdS}_3\times C$ has a holographic Weyl anomaly for every worldsheet topology, and is not captured by the sphere partition function alone.

In \cite{Kim:2015gak}, it was suggested that one can pass from the grand canonical ensemble defined by perturbative string theory to the usual microcanonical (fixed $N$) ensemble by introducing a chemical potential for the winding number $N$ and performing a Legendre transform. In \cite{Eberhardt:2021jvj,Eberhardt:2021vsx,Aharony:2024fid}, this proposal was refined by suggesting that the two ensembles are instead related by a Laplace transform. Let us briefly discuss how this works in our language. We first assume that the boundary has genus $G\neq 1$.\footnote{A minor modification of this argument also works for $G=1$, see \cite{Eberhardt:2021jvj}.} Then one can add a topological term
\begin{equation}
\delta S=\frac{2\pi i\lambda}{2G-2}\int_{\Sigma}x^*(\sqrt{g}R)
\end{equation}
to the worldsheet action. Since $\delta S=-2\pi i\lambda N$ when evaluated in the $N$-winding sector, the addition of the topological term has the effect of modifying the chemical potential \eqref{eq:chemical-potential-dictionary} for the long strings to
\begin{equation}
p=(g_se^{-k/2})^{2G-2}e^{2\pi i\lambda}\,.
\end{equation}
As such, the integral transform
\begin{equation}
\int_{0}^{1}\mathrm{d}\lambda\,e^{-2\pi iN\lambda}\mathfrak{Z}_{\text{string}}^{(0)}(\lambda)
\end{equation}
transforms with the Weyl anomaly of a single CFT of central charge $6kN$, and indeed reproduces the partition function of the deformed symmetric orbifold at fixed $N$. For large $N$, this reproduces the Legendre transform of \cite{Kim:2015gak} to leading order in $1/N$, see Section 3 of \cite{Aharony:2024fid}.

\section{Partition functions on various backgrounds}\label{sec:backgrounds}

In this section, we explore the partition function of the long-string sigma model on various backgrounds of the form $M\times C$, where $M$ is some hyperbolic 3-manifold. In contrast to the more formal manipulations of Section \ref{sec:sigma-model}, in this section we will largely study the worldsheet theory using the more traditional approach of computing the worldsheet CFT partition function and integrating over the moduli space $\mathcal{M}_g$ of curves. Specifically, in the cases of thermal $\text{AdS}_3$ and global $\text{AdS}_3$, we find agreement with the general principles of Section \ref{sec:sigma-model}, namely that the path integral localizes to the moduli space of holomorphic maps from the worldsheet to the boundary. We also comment on the role of Euclidean wormholes in the long-string worldsheet theory, and show that the dual CFT is still a grand canonical ensemble of CFTs, but with the ensemble chosen in such a way that the partition function still factorizes.

\subsection[Thermal \texorpdfstring{$\text{AdS}_3$}{AdS3}/Euclidean BTZ]{\boldmath Thermal \texorpdfstring{$\text{AdS}_3$}{AdS3}/Euclidean BTZ}

The simplest example background is thermal $\text{AdS}_3$, whose partition function is dual to the torus partition function in the dual CFT. While the one-loop partition function of bosonic string theory on thermal $\text{AdS}_3$ is well-known \cite{Maldacena:2000hw,Maldacena:2000kv}, we will repeat the calculation of the long-string partition function here. The result agrees with that of \cite{Maldacena:2000kv}, assuming that one restricts to the continuous series representations of $\mathfrak{sl}(2,\mathbb{R})_k$.

In Euclidean signature, thermal $\text{AdS}_3$ is a solid torus. The boundary $X$ is the two-torus $\mathbb{T}^2$, and in a particular Weyl frame, the boundary metric is simply the flat metric $g_{ij}=\delta_{ij}$. In this case, we can pick global complex coordinates $\gamma,\bar\gamma$ on the boundary. If the boundary carries conformal structure labeled by the modular parameter $t$, the complex coordinates can be chosen to satisfy the periodicity conditions
\begin{equation}
\gamma\sim\gamma+1\sim\gamma+t\,.
\end{equation}

Choosing local complex coordinates on the worldsheet as well, we can write the sigma model in the form
\begin{equation}
S=\frac{1}{2\pi}\int_{\Sigma}\mathrm{d}^2z\left(\frac{1}{2}\partial\phi\overline{\partial}\phi-\frac{Q}{4}\sqrt{h}R_h\phi+\beta\overline{\partial}\gamma+\bar\beta\partial\bar\gamma\right)\,.
\end{equation}
Now, the equations of motion for $\beta,\bar\beta$ imply that $\gamma:\Sigma\to\mathbb{T}^2$ is a holomorphic map. Since the target is compact, $\gamma$ can live in various winding sectors. Let $\rho\in\pi_1(\Sigma,p)$ be some closed loop based at a point $z$ on the worldsheet. Single-valuedness of $\gamma$ on the torus implies that
\begin{equation}
\gamma(\rho\cdot z)=\gamma(z)+mt+n\,.
\end{equation}
for integers $m,n$. For a spherical worldsheet, there is an obstruction, as there are no nonconstant covering maps $\text{S}^2\to\mathbb{T}^2$. As a consequence, we have
\begin{equation}
\mathcal{F}_0^{(0)}=0\,.
\end{equation}

Turning to the one-loop ($g=1$) partition function, we can use the usual prescription of 
perturbative string theory to compute the free energy. We do this by gauge-fixing the worldsheet metric $h$ to be that of a flat torus with modulus $\tau$. The cost of gauge-fixing is to introduce the usual $b,c$ ghost system of bosonic string theory. The final free energy is then computed as an integral over the fundamental domain $\mathscr{F}$ in the upper-half plane
\begin{equation}
\mathcal{F}_1^{(0)}=\int_{\mathscr{F}}\frac{\mathrm{d}^2\tau}{\tau_2}Z_{bc}Z_{\gamma,\bar\gamma}Z_{\phi}Z_C\,.
\end{equation}

Let us first compute the path integral of the $\beta,\gamma$ system. Since $\gamma,\bar\gamma$ are the global complex coordinates of a map $\mathbb{T}^2_\tau\to\mathbb{T}^2_t$ from the worldsheet to the boundary torus, it must obey the periodicity conditions
\begin{equation}
\gamma(z+\tau)=at+b\,,\quad\gamma(z+1)=ct+d\,.
\end{equation}
for integers $a,b,c,d$. The non-degeneracy of this map implies that $ad-bc\neq 0$. Each non-degenerate choice of $a,b,c,d$ will label a particular winding sector in the path integral, and in the end we will need to sum over all such winding sectors.

We can compute the path integral of the $\beta,\gamma$ system by using the trick of \cite{Hashimoto:2019wct}. Specifically, we can add to the action the deformation
\begin{equation}\label{eq:lambda-deformation}
\delta S=-\frac{\lambda}{2\pi}\int\mathrm{d}^2z\beta\bar\beta\,.
\end{equation}
Integrating out $\beta,\bar\beta$ leads to the effective action
\begin{equation}
S_{\lambda}=\frac{1}{2\pi}\int\mathrm{d}^2z\left(\frac{1}{2}\partial\phi\overline{\partial}\phi+\frac{1}{\lambda}\bar\partial\gamma\partial\bar\gamma\right)\,.
\end{equation}
Since the action is quadratic, the path integral for the map $x=(\gamma,\bar\gamma)$ can then be performed using the saddle point approximation. In the winding sector labeled by $(a,b,c,d)$, the classical solutions to the $\gamma,\bar\gamma$ equations of motion are
\begin{equation}
\gamma=\frac{[(at+b)-(ct+d)\bar\tau]z-[(at+b)-(ct+d)\bar\tau]\bar{z}}{2i\tau_2}+\gamma_0\,,
\end{equation}
where $\gamma_0$ is some constant living in the boundary torus. Thus, the $\gamma,\bar\gamma$ path integral can be read off (using for example equations  (7.3.6), (8.2.11) of \cite{Polchinski:1998rq}) as
\begin{equation}
\frac{t_2}{4\lambda\tau_2|\eta(\tau)|^4}\sum_{\substack{a,b,c,d\in\mathbb{Z}\\ ad-bc\neq 0}}\exp\left(-\frac{\pi|(at+b)-(ct+d)\tau|^2}{2\lambda\tau_2}\right)\,.
\end{equation}
The factors of $\eta(\tau)$ come as usual from the one-loop determinant of the operator $\partial\bar\partial$, while the factor of $t_2$ comes from the integral over the zero mode $\gamma_0$, which takes values in the boundary torus.

For small $\lambda$, this is a sum of Gaussian-shaped distributions around the points $\tau=(at+b)/(ct+d)$ in the upper-half plane $\mathbb{H}^2$. Taking the limit $\lambda\to 0^+$ (effectively undoing the deformation \eqref{eq:lambda-deformation}) yields the sum of delta functions (see \cite{Dei:2024sct})
\begin{equation}
\frac{1}{|\eta(\tau)|^4}\frac{t_2}{2|ct+d|^2}\delta^{(2)}\left(\tau-\frac{at+b}{ct+d}\right)\,.
\end{equation}
The delta function condition is precisely the condition that the worldsheet covers the boundary holomorphically, as explained in more general grounds in Section \ref{sec:sigma-model}.

As usual, the $b,c$ system removes two bosonic oscillators from the set of degrees of freedom, and the partition function is
\begin{equation}
Z_{bc}=|\eta(\tau)|^4\,.
\end{equation}
Thus, we have
\begin{equation}
\mathcal{F}_1^{(0)}=\frac{1}{2}\sum_{\substack{a,b,c,d\in\mathbb{Z}\\ ad-bc\neq 0}}\frac{t_2}{|ct+d|^2}\int_{\mathscr{F}}\frac{\mathrm{d}^2\tau}{\tau_2}Z_{\phi}(\tau)Z_{C}(\tau)\delta^{(2)}\left(\tau-\frac{at+b}{ct+d}\right)\,.
\end{equation}
Now, we can perform the integral over the fundamental domain $\mathscr{F}$ with use of the delta function. However, for a given $t$, the combination $(at+b)/(ct+d)$ only lies in $\mathscr{F}$ for specific integers $a,b,c,d$. If $ad-bc<0$, then $(at+b)/(ct+d)$ lives in the lower-half plane, and the delta function has no support, and so we can restrict the sum to only integer matrices with positive determinant.  Since there is a symmetry $(a,b,c,d)\to(-a,-b,-c,-d)$, we can restrict to $a>0$ and simply multiply by 2. Thus,
\begin{equation}
\mathcal{F}_1^{(0)}=\sum_{\substack{a,b,c,d\in\mathbb{Z}\\ ad-bc>0\\a>0}}\frac{t_2}{|ct+d|^2}\int_{\mathscr{F}}\frac{\mathrm{d}^2\tau}{\tau_2}Z_{\phi}(\tau)Z_{C}(\tau)\delta^{(2)}\left(\tau-\frac{at+b}{ct+d}\right)\,.
\end{equation}
Finally, using the `unfolding trick', we can trade the integral over $\mathscr{F}$ for an integral over the full upper-half plane $\mathbb{H}$, the the expense of summing over the equivalence classes
\begin{equation}
\begin{pmatrix}
a & b\\ c & d
\end{pmatrix}\sim\gamma
\begin{pmatrix}
a & b\\ c & d
\end{pmatrix}\,,
\end{equation}
where $\gamma\in\text{PSL}(2,\mathbb{Z})$ is a modular transformation (see, for example, \cite{Eberhardt:2020bgq}). In fact, one can always pick the equivalence classes to take the form
\begin{equation}
\begin{pmatrix}
a & b\\ 0 & d
\end{pmatrix}\,,\quad b\in\{0,\ldots,d-1\}\,,
\end{equation}
with $a,d>0$. Therefore, since $Z_{\phi}$ and $Z_{C}$ are modular functions, we have
\begin{equation}
\begin{split}
\mathcal{F}_1^{(0)}&=\sum_{a,d=1}^{\infty}\sum_{b=0}^{d-1}\frac{t_2}{d^2}\int_{\mathbb{H}}\frac{\mathrm{d}^2\tau}{\tau_2}Z_{\phi}(\tau)Z_C(\tau)\delta^{(2)}\left(\tau-\frac{a\tau+b}{d}\right)\\
&=\sum_{a,d=1}^{\infty}\sum_{b=0}^{d-1}\frac{1}{ad}Z_{\phi}\left(\frac{at+b}{d}\right)Z_C\left(\frac{at+b}{d}\right)\,.
\end{split}
\end{equation}

Recall that $\mathcal{S}=\mathbb{R}_{\mathcal{Q}}\times C$ is the seed CFT of the symmetric orbifold introduced in Section \ref{sec:dual-cft}. Then the partition function of of the seed CFT is
\begin{equation}
Z_{\mathcal{S}}(t)=Z_\phi(t)Z_C(t)\,,
\end{equation}
and so we can write
\begin{equation}
\begin{split}
\mathcal{F}_1^{(0)}&=\sum_{a,d=1}^{\infty}\sum_{b=0}^{d-1}\frac{1}{ad}Z_{\mathcal{S}}\left(\frac{at+b}{d}\right)\\
&=\sum_{N=1}^{\infty}T_NZ_{\mathcal{S}}(t)\,,
\end{split}
\end{equation}
where $T_N$ are the Hecke operators acting on modular functions. Recalling that the sphere free energy vanishes, the full long string partition function on Euclidean thermal $\text{AdS}_3$ is given by
\begin{equation}
\begin{split}
\mathfrak{Z}_{\text{string}}^{(0)}&=\exp\left(g_s^{-2}\cdot 0+\sum_{N=1}^{\infty}T_NZ_{\mathcal{S}}(t)+\mathcal{O}(g_s^2)\right)\\
&=\exp\left(\sum_{N=1}^{\infty}T_NZ_{\mathcal{S}}(t)\right)+\mathcal{O}(g_s^2)\,.
\end{split}
\end{equation}
The first term is precisely the grand canonical partition function of the symmetric orbifold of $\mathcal{S}$ \cite{Dijkgraaf:1996xw}. The correction terms, of order $g_s^2$, break the symmetric orbifold structure, and holographically arise from the addition of the deformation operator \eqref{eq:interaction-term-cft} in the dual CFT.

\subsubsection*{Higher genus}

Let us also say some words about the higher-loop partition functions. While we will not directly compute the partition function at higher-genus, we will comment on the origin of the localization property of the moduli space integral. Specifically, we will show how the $(3g-3)$-dimensional integral over the moduli space $\mathcal{M}_g$ becomes a $(2g-2)$-dimensional integral over the moduli space of branched covering maps $\gamma:\Sigma\to\mathbb{T}^2_t$.

First, let us focus on the computation of the $\beta,\gamma$ system. Even on a higher-genus worldsheet, we can use the same trick as in the previous section and consider the deformation \eqref{eq:lambda-deformation}. Upon integrating out $\beta,\bar\beta$, we are left with the effective action
\begin{equation}
\frac{1}{2\pi\lambda}\int_{\Sigma}\mathrm{d}^2z\,|\overline{\partial}\gamma|^2\,.
\end{equation}
Just as before, since the action is quadratic in $\gamma$, we can exactly evaluate the path integral over $\gamma,\bar\gamma$ via the saddle-point approximation. The classical saddles are harmonic functions on $\Sigma$ which are single-valued up to the identification
\begin{equation}
\gamma\sim\gamma+1\sim\gamma+t\,.
\end{equation}
Just as in the case of a genus-one worldsheet, it is possible to compute all saddle-point solutions to this action, a computation which we perform in Appendix \ref{app:harmonic-functions}. The upshot is that the saddle-point approximation to the $\gamma$ path integral is proportional to the sum
\begin{equation}
\sum_{\varphi\in\mathbb{Z}^{2\times 2g}}\frac{t_2}{\big|\det\overline{\partial}'_{\Sigma}\big|^2}\exp\left(-\frac{v_{\varphi}^{\dagger}\text{Im}(\Omega)^{-1}v_{\varphi}}{8\pi\lambda}\right)\,,
\end{equation}
where $\Omega$ is the period matrix of the worldsheet $\Sigma$, $\det\overline{\partial}'_{\Sigma}$ is the determinant of the antiholomorphic derivative on functions minus the zero mode, and the sum is over all integer $2\times 2g$ matrices
\begin{equation}
\varphi=
\begin{pmatrix}
 d_1 & \cdots d_g & b_1 & \cdots & b_g\\
 c_1 & \cdots c_g & a_1 & \cdots & a_g
\end{pmatrix}\,.
\end{equation}
The vectors $v_{\varphi}$ are given by
\begin{equation}
(v_{\varphi})_{i}=\Omega_{ij}(c_jt+d_j)-(a_it+b_i)\,.
\end{equation}
As in the case of the torus partition function, the factor of $t_2$ comes from the integration over the zero mode $\gamma=\text{const}$.

Taking the limit $\lambda\to 0^+$ gives the sum of delta functions setting $v_{\varphi}=0$, namely
\begin{equation}
\sum_{\varphi\in\mathbb{Z}^{2\times 2g}}\frac{t_2\text{det}(\text{Im}(\Omega))}{\big|\det\overline{\partial}'_{\Sigma}\big|^2}\prod_{i=1}^{g}\delta^{(2)}\left(\Omega_{ij}(c_jt+d_j)-(a_it+b_i)\right)\,.
\end{equation}
The delta functions in the above sum impose a set of $g$ constraints on the period matrix $\Omega$ of the worldsheet. Alternatively, one can think of these constraints as $g$ conditions on the worldsheet complex structure moduli. Thus, the integral over the moduli space $\mathcal{M}_g$ localizes to a subspace of dimension
\begin{equation}
\text{dim}(\mathcal{M}_g)-g=2g-3\,.
\end{equation}
This counting can be understood in the language of Section \ref{sec:sigma-model} as follows. A covering map $\gamma:\Sigma\to\mathbb{T}^2_t$ has precisely $2g-2$ branch points, as can be read off from the Riemann-Hurwitz formula \eqref{eq:Riemann-Hurwitz}.\footnote{Alternatively, we simply note that $\partial\gamma$ is a globally-defined holomorphic $(1,0)$-form on $\Sigma$ and therefore has $2g-2$ zeroes.} These covering spaces are found by choosing $2g-2$ points $\xi_{\ell}\in\mathbb{T}^2_t$ for which $\gamma^{-1}$ looks like
\begin{equation}
\gamma^{-1}(x)\sim (x-\xi_{\ell})^{1/2}
\end{equation}
in a neighborhood of $\xi_{\ell}$. The space of all such covering maps is locally labeled by the coordinates $\xi_{\ell}$, and so the space of all holomorphic branched coverings $\gamma:\Sigma\to\mathbb{T}^2_t$ has dimension $2g-2$. Now, by adding a constant to $\gamma$, we can shift the branch points by the same constant, i.e. there is a symmetry
\begin{equation}
(\xi_1,\ldots,\xi_{2g-2})\to(\xi_1+\gamma_0,\ldots,\xi_{2g-2}+\gamma_0)\,.
\end{equation}
Sets of branch points related by this symmetry yield the same Riemann surface $\Sigma$ as their branched cover. Thus, the subspace of $\mathcal{M}_g$ of surfaces which admit branched covers to $\mathbb{T}^2_t$ has dimension $2g-3$, as predicted by the worldsheet calculation.

\subsection[Global \texorpdfstring{$\text{AdS}_3$}{AdS3}]{\boldmath Global \texorpdfstring{$\text{AdS}_3$}{AdS3}}\label{sec:global-ads3}

We now turn our attention to global $\text{AdS}_3$. In Euclidean signature, global $\text{AdS}_3$ is nothing more than the upper half-space $\mathbb{H}^3$, whose boundary is a two-sphere $\text{S}^2$.

To begin, we first note that the long-string partition function formally vanishes on global $\text{AdS}_3$ for $k> 3$. This can be seen from the action \eqref{eq:linear-dilaton-action-string} governing the radial field $\phi$ on a generic background. Integrating over the zero mode of $\phi$ and using the formal identity
\begin{equation}
\int_{-\infty}^{\infty}\mathrm{d}x\,e^{-\alpha x}\propto\delta(\alpha)\,,
\end{equation}
we see that the long-string free energies $\mathcal{F}_{g,m}^{(0)}$ pick up the momentum-conserving delta function
\begin{equation}
\delta\left(Q(1-g)-\frac{2N}{Q}(1-G)\right)\,.
\end{equation}
For global $\text{AdS}_3$, whose boundary has genus $G=0$, the delta function demands
\begin{equation}
N=\frac{Q^2}{2}(1-g)=\frac{1-g}{k-2}\,.
\end{equation}
Thus, the long-string partition function on global $\text{AdS}_3$ formally vanishes unless $(1-g)/(k-2)$ is an integer. While this imposes a constraint on the allowed values of $k$, it is still possible to obtain a nonvanishing result for any $k$ by, say, inserting a screening charge, or by studying correlation functions instead of the partition function \cite{Eberhardt:2021vsx,Knighton:2023mhq,Knighton:2024qxd}.

\begin{figure}
\centering
\begin{tikzpicture}
\fill[RoyalBlue, opacity = 0.2] (0,0) circle (3);
\fill[white] (0,0.5) circle (2.5);
\draw (0,0.5) circle (2.5);
\draw (0,1) circle (2);
\draw (0,1.5) circle (1.5);
\draw (0,2) circle (1);
\draw (0,2.5) circle (0.5);
\draw (0,2.75) circle (0.25);
\draw (0,3) -- (0,-3);
\draw (0.75,3) [partial ellipse = 180:334:0.75 and 0.75];
\draw (-0.75,3) [partial ellipse = 0:-154:0.75 and 0.75];
\draw (1.5,3) [partial ellipse = 180:307:1.5 and 1.5];
\draw (-1.5,3) [partial ellipse = 0:-127:1.5 and 1.5];
\draw (3,3) [partial ellipse = 180:270:3 and 3];
\draw (-3,3) [partial ellipse = 0:-90:3 and 3];
\draw (7,3) [partial ellipse = 180:226.5:7 and 7];
\draw (-7,3) [partial ellipse = 0:-46.5:7 and 7];
\draw[ultra thick] (0,0) circle (3);
\draw[thick, -latex] (4,2) to[out = -120, in = 0] (2,1.25);
\node[above] at (4,2) {Constant $\Phi$};
\draw[thick, -latex] (4,-2) to[out = 120, in = 30] (1.5,-1);
\node[below] at (4,-2) {Constant $(\gamma,\bar{\gamma})$};
\end{tikzpicture}
\caption{Euclidean $\text{AdS}_3$ in Poincar\'e coordinates. The shaded region denotes the set $\Phi>\Phi_0$ for some large $\Phi_0$. Near the north pole $\gamma=\infty$, the coordinates are not defined. Figure adapted with permission from \cite{Knighton:2024qxd}.}
\label{fig:global-ads3}
\end{figure}
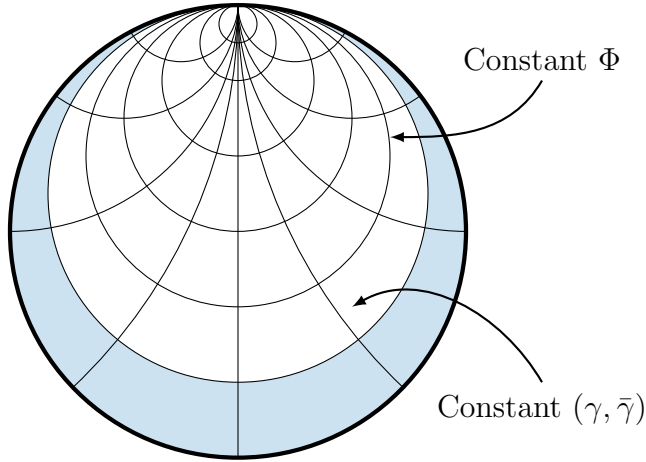

With the preliminary discussion out of the way, we can proceed to calculating the partition function on global $\text{AdS}_3$. First, we pick a coordinate system. It is convenient to choose Poincar\'e coordinates, for which the metric is
\begin{equation}
\mathrm{d}s^2=\frac{L^2}{r^2}\left(\mathrm{d}r^2+\mathrm{d}\gamma\mathrm{d}\bar\gamma\right)\,,
\end{equation}
where, as usual, $(\gamma,\bar\gamma)$ are complex coordinates on the boundary. These coordinates are globally defined in the bulk, but at the boundary there is a coordinate singularity as one approaches the north pole on the boundary sphere (see Figure \ref{fig:global-ads3}). This singularity is reflected in the fact that the boundary scalar curvature is a delta-function localized at the north pole, specifically
\begin{equation}
\sqrt{g}R=8\pi\delta^{(2)}(\gamma,\infty)\,.
\end{equation}
Now, given a worldsheet $\Sigma$ of genus $g$, we know from the general discussion in Section \ref{sec:sigma-model} that the worldsheet path integral is an integral over the moduli space of holomorphic maps $\gamma:\Sigma\to\text{S}^2$. Such a holomorphic map can equivalently be thought of as a meromorphic function on $\Sigma$, with the poles of $\gamma$ being identified with the preimages of $\infty$. The degree $N$ of the covering map is then identified nicely with the number of poles of the corresponding meromorphic function. 

Let us label by $\lambda_a$ the locations of the poles of $\gamma$ on the worldsheet. The holomorphic derivative $\partial\gamma$ defines a meromorphic $(1,0)$-form on $\Sigma$ with double poles at $\lambda_a$. In general, the number $Z(\omega)$ of zeroes and the number $P(\omega)$ poles of a $(1,0)$-form $\omega$ are constrained by the relation
\begin{equation}
Z(\omega)=P(\omega)+2g-2
\end{equation}
Thus, the one-form $\partial\gamma$ must have $2N+2g-2$ zeroes. A (simple) zero of $\partial\gamma$ in turn defines a branch point of $\gamma$, i.e. a point $\zeta_\ell$ on the worldsheet such that
\begin{equation}
\gamma(z)\sim\xi_\ell+b_{\ell}(z-\zeta_\ell)^2+\cdots
\end{equation}
in a neighborhood of $\zeta_\ell$, where $\xi_\ell=\gamma(\zeta_\ell)$ is the image of the branch point on the boundary sphere. The number $m$ of such branch points and the degree $N$ of the covering map $\gamma$ are thus related by the Riemann-Hurwitz relation \eqref{eq:Riemann-Hurwitz}
\begin{equation}
2N=2-2g+m\,.
\end{equation}

Once the choice of poles $\lambda_a$ has been made, we can write the pullback of the boundary curvature as
\begin{equation}
x^*(\sqrt{g}R)=8\pi\sum_{a=1}^{N}\delta^{(2)}(z,\lambda_a)\,.
\end{equation}
Thus, the worldsheet sigma model action takes the form
\begin{equation}
S=\frac{1}{2\pi}\int_{\Sigma}\mathrm{d}^2z\left(\frac{1}{2}\partial\phi\bar\partial\phi-\frac{Q}{4}\sqrt{h}R_h\phi+\beta\overline{\partial}\gamma+\bar\beta\partial\bar\gamma\right)+\sum_{a=1}^{N}\frac{1}{Q}\phi(\lambda_a)\,.
\end{equation}
Now, the path integral over $\gamma$ is an integral over the space of maps from the worldsheet to the sphere. Alternatively, upon integrating out $\beta$, this is an integral over the space of holomorphic functions with with fixed poles $\lambda_a$, followed by an integral over the locations of the poles $\lambda_a$. This decomposition of $\gamma$ into a finite piece (a function which is holomorphic on $\Sigma\setminus\{\lambda_1,\ldots,\lambda_N\}$) and its singularities (the poles at $\lambda_a$) is somewhat subtle. One way to implement it, and the method proposed by \cite{Eberhardt:2019ywk,Dei:2023ivl,Hikida:2023jyc,Knighton:2023mhq,Knighton:2024qxd} (see also \cite{Frenkel:2005ku}), is to treat the points $\lambda_a$ as arising from the insertion of an operator which has a simple pole with $\gamma$. While we will not derive the form of this operator rigorously (although it in principle should be possible), we simply quote the proposal of \cite{Hikida:2023jyc,Knighton:2023mhq}, namely that the poles in $\gamma$ arise from the insertion of the operator
\begin{equation}
\left|\oint\gamma\right|^{-2(k-1)}\delta(\beta)\delta(\bar\beta)\,.
\end{equation}
As explained in \cite{Knighton:2023mhq,Knighton:2024qxd}, the delta function $\delta(\beta)$ imposes that $\gamma$ has a simple pole where it is inserted, and the contour integral reads off the residue of that pole.

With this proposal in place, the long string partition function on global $\text{AdS}_3$ is given by the integral
\begin{equation}
\mathcal{F}_g^{(0)}=\sum_{N=1}^{\infty}\frac{e^{kN}}{N!}\int_{\mathcal{M}_{g}}\mathrm{d}\mu\,\Bigg\langle\prod_{a=1}^{N}\int\mathrm{d}\lambda_a\left|\oint\gamma\right|^{-2(k-1)}\delta(\beta)\delta(\bar\beta)e^{-\phi/Q}(\lambda_a)\Bigg\rangle\,.
\end{equation}
The integral over the moduli space $\mathcal{M}_g$ is weighted by the usual measure $\mathrm{d}\mu$ found from the $c=-26$ conformal ghost system in bosonic string theory. The screening operators located at the poles $\lambda_a$ are formally similar to the screening operators introduced by \cite{Giribet:2001ft,Iguri:2007af} in the study of the $\text{SL}(2,\mathbb{R})$ WZW model. Here, we see them arise naturally as a result of the coupling of the radial scalar $\phi$ to the curvature of the boundary sphere.

While we will not compute the above integral explicitly, we note that it in principle should be possible given the technology developed in \cite{Hikida:2023jyc,Knighton:2024qxd}. It would constitute a good sanity check of the analysis of Section \ref{sec:sigma-model} if the resulting integral reproduced the Coulomb gas expansion of the dual CFT on a covering surface of genus $g$.

\subsubsection*{The sphere partition function}

Formally, the expression derived above for the long-string partition function is valid for genera $g\geq 1$. For almost all locally-$\text{AdS}_3$ backgrounds, the boundary has genus $G\geq 1$, and so the sphere partition function automatically vanishes (as there are no covering maps from a sphere to a surface of genus $G\geq 1$). However, global $\text{AdS}_3$ is the only hyperbolic 3-manifold whose boundary is a sphere, and as such is the only Euclidean $\text{AdS}_3$ background whose long-string partition function potentially contains a sphere contribution. It is therefore worth taking a closer look at the sphere partition function $\mathcal{F}^{(0)}_0$ in this background.

The sphere partition function in string theory is subtle due to the existence of conformal killing vectors on the worldsheet. Specifically, there is one conformal class of metrics on the sphere, and the group $\text{PSL}(2,\mathbb{C})$ of M\"obius transformations fix this conformal class. As such, the group $\text{PSL}(2,\mathbb{C})$ represents a degeneracy of physically-equivalent worldsheets, which must be identified in the path integral. Formally, this can be done by picking a specific metric (say, the almost-flat metric $\mathrm{d}s^2_{\Sigma}=\mathrm{d}z\mathrm{d}\bar{z}$) and dividing by the volume $|\text{PSL}(2,\mathbb{C})|$ of the group of conformal killing vectors. This leads to the partition function
\begin{equation}\label{eq:sphere-partition-psl}
\mathcal{F}_0^{(0)}=\frac{1}{|\text{PSL}(2,\mathbb{C})|}\int\mathcal{D}(\beta,\gamma,\phi)e^{k\text{deg}(\gamma)}e^{-S}Z_C\,,
\end{equation}
where $Z_C$ is the sphere partition function of the compact CFT with respect to the (almost) flat metric $\mathrm{d}z\mathrm{d}\bar{z}$.

Since $\text{PSL}(2,\mathbb{C})$ is non-compact, its volume is infinite. In addition, unlike the slightly simpler case of the group $\text{SL}(2,\mathbb{R})$ of conformal killing vectors of the disk, there is no known way of consistently regularizing its volume (see \cite{Eberhardt:2021ynh} and references therein). However, in the specific case of long strings in $\text{AdS}_3$, there is a saving grace, namely that the factor of $|\text{PSL}(2,\mathbb{C})|$ is exactly canceled by the volume of the space of M\"obius transformations on the boundary.\footnote{The author is grateful to Vit Sriprachyakul and Jakub Vo\v{s}mera for initial discussions on this point.} We will work this result out in detail now.

As usual, the integral over the Lagrange multiplier $\beta$ yields the delta function $\delta(\overline{\partial}\gamma)$ demanding that $\gamma$ is a holomorphic map $\gamma:\text{S}^2\to\text{S}^2$. Equivalently, $\gamma$ can be thought of as a meromorphic function on the complex plane with at most a pole of finite order at infinity. Assuming that $\gamma$ approaches a constant at infinity (which is generically true), we have
\begin{equation}
\gamma(z)=\frac{P_N(z)}{(z-\lambda_1)\cdots(z-\lambda_N)}\,,
\end{equation}
where $\lambda_a$ are the locations of the poles of $\gamma$, and $P_N$ is some polynomial of degree $N$. Similarly, the derivative of $\gamma$ satisfies
\begin{equation}
\partial_z\gamma(z)=\frac{Q_{2N-2}(z)}{(z-\lambda_1)^2\cdots(z-\lambda_N)^2}\,,
\end{equation}
where $Q_{2N-2}$ is a polynomial of degree $2N-2$. That $Q_{2N-2}$ has that specific degree can be seen as follows. If $\gamma\sim a+b/z+\cdots$ at $z\to\infty$, then $\partial\gamma\sim -b/z^2+\cdots$, which tells us that the leading term in $Q_{2N-2}$ goes like $z^{2N-2}$. Factoring $Q_{2N-2}$ gives
\begin{equation}
\partial\gamma(z)\sim C\frac{(z-\zeta_1)\cdots(z-\zeta_m)}{(z-\lambda_1)^2\cdots(z-\lambda_N)^2}\,.
\end{equation}
For a fixed set of branch points $\zeta_\ell$, the locations of the poles $\lambda_a$ are almost uniquely determined. Indeed, in order for $\partial\gamma$ to be the derivative of a meromorphic function, we need
\begin{equation}
\mathop{\text{Res}}_{z=\lambda_a}\partial\gamma=0
\end{equation}
for each $a=1,\ldots,N$. This leads to $N$ equations constraining the poles $\lambda_a$, known as the `scattering equations' \cite{Roumpedakis:2018tdb}:
\begin{equation}
\sum_{b\neq a}\frac{2}{\lambda_b-\lambda_a}=\sum_{\ell=1}^{m}\frac{1}{\zeta_\ell-\lambda_a}\,,\quad a=1,\ldots,N\,.
\end{equation}
These equations are all independent, as the sum of the residues of the poles of any one-form vanish on a compact surface. Thus, we find that only $N-1$ of the poles are determined by the locations of the branch points $\zeta_\ell$. In addition to the normalization factor $C$, as well as the constant of integration, this yields a three-dimensional space of covering maps for each fixed choice of branch points $\zeta_\ell$.

The three-dimensional space of covering maps for a fixed set of branch points can be seen as a result of the $\text{PSL}(2,\mathbb{C})$ symmetry of the boundary. Indeed, if $\gamma$ is a covering map $\gamma:\text{S}^2\to\text{S}^2$ which is branched at the points $\zeta_\ell$ on the worldsheet, then
\begin{equation}
\gamma'=\frac{a\gamma+b}{c\gamma+d}\,,\quad\begin{pmatrix}
a & b\\
c & d
\end{pmatrix}\in\text{PSL}(2,\mathbb{C})
\end{equation}
is as well. In fact, the worldsheet theory is invariant under the simultaneous set of transformations
\begin{equation}
\gamma\to\frac{a\gamma+b}{c\gamma+d}\,,\quad\beta\to(c\gamma+d)^2\beta\,,\quad\phi\to\phi+\frac{2}{Q}\log|c\gamma+d|^2\,.
\end{equation}
Thus, we can trade the integral over all holomorphic maps $\gamma:\text{S}^2\to\text{S}^2$ for an integral over the $\text{PSL}(2,\mathbb{C})$ orbits of such maps by compensating with a factor of $|\text{PSL}(2,\mathbb{C})|$. Put another way, we can write \eqref{eq:sphere-partition-psl} as
\begin{equation}
\int'\mathcal{D}(\beta,\gamma,\phi)e^{k\text{deg}(\gamma)}e^{-S}Z_C\,,
\end{equation}
where the primed integral represents integration only the quotient of the space of fields $(\beta,\gamma,\phi)$ by $\text{PSL}(2,\mathbb{C})$. After integrating out $\beta$ and fixing the degree $N=\text{deg}(\gamma)$, the $\gamma$ integral is over the space
\begin{equation}
\{\gamma:\text{S}^2\to\text{S}^2|\overline{\partial}\gamma=0\}/\text{PSL}(2,\mathbb{C})\,,
\end{equation}
which is a finite dimensional moduli space of dimension $2N-2$, which by the Riemann Hurwitz formula is the number of branch points of $\gamma$. Thus, the sphere partition function of long strings in global $\text{AdS}_3$ is well-defined, and does not suffer from the usual pathology of the sphere partition function on a compact manifold.

Mathematically, the above discussion can be summarized as follows. While the moduli space $\mathcal{M}_0$ of genus-zero curves has negative virtual dimension (arising from the group $\text{PSL}(2,\mathbb{C})$ of conformal killing vectors on $\text{S}^2$), the Kontsevich moduli space $\mathcal{M}_0(\text{S}^2,N)$ of maps of genus-zero curves into $\text{S}^2$ is well-defined and has non-negative complex dimension $2N-2$. Since the latter is the moduli space of worldsheet instantons on the boundary of global $\text{AdS}_3$, the sphere partition function is well-defined.

\subsection{Euclidean wormholes}

\begin{figure}
\centering
\begin{tikzpicture}
\draw[thick, fill = RoyalBlue, fill opacity = 0.2] (0,5) to[out = -10, in = 180] (3,4) to[out = 0, in = 190] (6,5) -- (6,0) to[out = 170, in = 0] (3,1) to[out = 180, in = 10] (0,0) -- (0,5);
\begin{scope}
\draw[thick, fill = white] (0,0) to[out = 0, in = -90] (1,1) to[out = 90, in = -90] (0.5,2.5) to[out = 90, in = -90] (1,4) to[out = 90, in = 0] (0,5) to[out = 180, in = 90] (-1,4) to[out = -90, in = 90] (-0.5,2.5) to[out = -90, in = 90] (-1,1) to[out = -90, in = 180] (0,0);
\draw[thick] (-0.1,1) [partial ellipse = -80:80:0.2 and 0.4];
\draw[thick] (0.1,1) [partial ellipse = 115:245:0.175 and 0.35];
\draw[thick] (-0.1,3.75) [partial ellipse = -80:80:0.2 and 0.4];
\draw[thick] (0.1,3.75) [partial ellipse = 115:245:0.175 and 0.35];
\end{scope}
\begin{scope}[xshift = 6cm, rotate = 180, yshift = -5cm]
\draw[thick, fill = white] (0,0) to[out = 0, in = -90] (1,1) to[out = 90, in = -90] (0.5,2.5) to[out = 90, in = -90] (1,4) to[out = 90, in = 0] (0,5) to[out = 180, in = 90] (-1,4) to[out = -90, in = 90] (-0.5,2.5) to[out = -90, in = 90] (-1,1) to[out = -90, in = 180] (0,0);
\draw[thick] (-0.1,1) [partial ellipse = -80:80:0.2 and 0.4];
\draw[thick] (0.1,1) [partial ellipse = 115:245:0.175 and 0.35];
\draw[thick] (-0.1,3.75) [partial ellipse = -80:80:0.2 and 0.4];
\draw[thick] (0.1,3.75) [partial ellipse = 115:245:0.175 and 0.35];
\end{scope}
\node at (0,2.5) {$X_1$};
\node at (6,2.5) {$X_2$};
\node at (3,2.5) {$M$};
\end{tikzpicture}
\caption{A Euclidean wormhole $M$ with two asymptotic boundaries $X_1$ and $X_2$. A long string can be supported at either boundary.}
\label{fig:wormhole}
\end{figure}
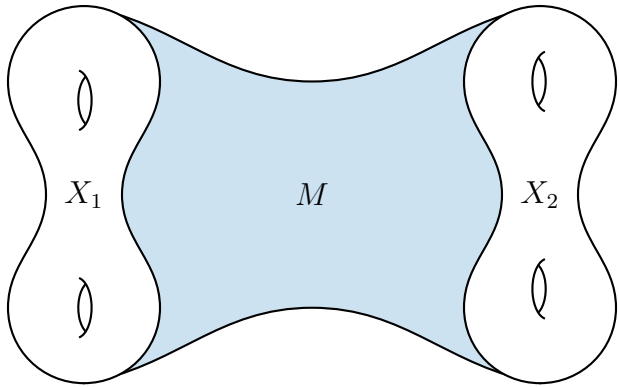

Throughout this work, we have implicitly assumed that the boundary $X$ of the bulk spacetime $M$ is connected. This is the case for many backgrounds of interest, such as Euclidean global and thermal $\text{AdS}_3$, as well as the Euclidean BTZ black hole. However, we can also consider Euclidean wormholes -- connected hyperbolic 3-manifolds $M$ with disconnected boundaries, see Figure \ref{fig:wormhole}. One such example is the famous Maldacena-Maoz wormhole \cite{Maldacena:2004rf}, whose metric tensor can be put in the form
\begin{equation}
\mathrm{d}s^2=L^2(\mathrm{d}\rho^2+\cosh^2{\rho}\,\mathrm{d}s_{X}^2)\,.
\end{equation}
Here, $X$ is some Riemann surface of genus $G>1$ and $\mathrm{d}s^2_{X}$ is a metric of constant negative curvature on $X$. This geometry has two boundaries located at $\rho=\pm\infty$. While the two boundaries of the Maldacena-Maoz wormhole are identical (up to orientation), there exist other hyperbolic 3-manifolds with arbitrarily many boundary components which are non-identical.

Let $M$ be a hyperbolic 3-manifold whose boundary is a union $X_1\sqcup\cdots\sqcup X_n$ of Riemann surfaces. A long string on $M$ can wind any one of these boundary components, but cannot wind more than one at the same time. Near a given boundary component $X_r$, we can consider a set of coordinates such that the metric and $B$-field on $M$ take the form \eqref{eq:fefferman-graham} and \eqref{eq:fefferman-graham-b-field} for a boundary metric $g_r$. Thus, a long string winding the boundary $X_r$ is described by the sigma model action \eqref{eq:sigma-model-action-long-string} with $X=X_r$ and $g=g_r$. Since the long-string can live on any one of the boundary components, in the path integral we must sum over boundary components. That is, the long-string free energy is a sum over the free energies of a long string on the single boundaries $X_r$:
\begin{equation}
\mathcal{F}_{g}^{(0)}=\sum_{r=1}^{n}\mathcal{F}_{g}^{(0)}[X_r,g_r]\,.
\end{equation}
As a consequence, the long-string partition function factorizes among the boundary components:
\begin{equation}
\mathfrak{Z}^{(0)}_{\text{string}}=\prod_{r=1}^{n}\mathfrak{Z}^{(0)}_{\text{string}}[X_r,g_r]\,.
\end{equation}
In this case, the boundary CFT describing the long strings is not the grand canonical ensemble of the symmetric orbifold on $X_1\sqcup\cdots\sqcup X_n$. Instead, it is a product of grand canonical ensembles on each boundary component $X_r$, with chemical potential
\begin{equation}
p_r=(g_se^{-k/s})^{2G_r-2}\,,
\end{equation}
where $G_r$ is the genus of the boundary component $X_r$. That there is a separate grand canonical ensemble for each boundary component is necessary for the final CFT partition function to factorize.\footnote{This idea was first proposed in \cite{Eberhardt:2021jvj} in the context of the $k=1$ string on $\text{AdS}_3\times\text{S}^3\times\mathbb{T}^4$.}

\section{Discussion}\label{sec:discussion}

In this paper, we computed the partition function of a gas of long strings in $M\times C$, where $M$ is a Euclidean locally-$\text{AdS}_3$ manifold and $C$ is some compact CFT needed to compensate for the Weyl anomaly. We found that, when one includes a careful sum/integral over worldsheet instanton sectors, the worldsheet path integral of a single long string produces the connected part of the Coulomb gas integrals of the proposed dual CFT of \cite{Eberhardt:2021vsx}. We note that this computation is perturbatively exact and valid for every worldsheet and spacetime topology, as well as at every order in conformal perturbation theory in the dual CFT.

The holographic dictionary between the long-string sector and the perturbative sector of the dual CFT is worth emphasizing. We found that the linear dilaton in the dual CFT is precisely realized by the radial direction in $\text{AdS}_3$ (as already noted in \cite{Seiberg:1999xz}). Furthermore, we found that the integrals arising in conformal perturbation theory in the dual CFT naturally arise in terms of the integral over the worldsheet instanton moduli space. Finally, since the Weyl anomaly of the worldsheet theory depends on the number of times the worldsheet wraps the asymptotic boundary, we strengthened the claim that the long-string theory is not dual to a single field theory, but rather a grand canonical ensemble of symmetric orbifolds of central charge $6kN$ \cite{Kim:2015gak}.

\vspace{0.5cm}

An important point which we have not discussed in the bulk of this work is the qualitative difference between the behaviors of the long-string partition functions at $k<3$ and $k>3$. Consider specficially the zero-mode integral of $\phi$ on the worldsheet. Since we work with strings close to the boundary $\phi=\infty$, this integral schematically takes the form
\begin{equation}\label{eq:zero-mode-integral-phi}
\int^{\infty}\mathrm{d}\phi_0\,\exp\left(\left(Q(1-g)-\frac{2N}{Q}(1-G)\right)\phi_0\right)
\end{equation}
in the $N$-winding sector.\footnote{This can be seen by inserting a constant value of $\phi_0$ in the action \eqref{eq:linear-dilaton-action-string}.} A necessary condition for the covnergence of this integral is that
\begin{equation}
Q(g-1)>\frac{2N}{Q}(G-1)\,.
\end{equation}
Using the Riemann-Hurwitz formula \eqref{eq:Riemann-Hurwitz}, and $Q=\sqrt{2/(k-2)}$, this bound can be rewritten as
\begin{equation}
-(k-3)(g-1)>m\,.
\end{equation}
Thus, we see that if $k>3$, the zero mode integral of $\phi_0$ formally diverges for any worldsheet of genus $g\geq 1$. On the other hand, for $k<3$, the zero mode integral converges for positive worldsheet genus (but diverges for $g=0$). In the language of \cite{Seiberg:1999xz}, the $k>3$ theory is `singular'.

One could argue that the divergence of the zero mode integral signals a sickness of the worldsheet theory as well as of the dual CFT for $k>3$. Indeed, as was noted in \cite{Balthazar:2021xeh,Martinec:2021vpk}, for $k<3$ the background charge $\mathcal{Q}$ of the linear dilaton $\phi$ in the dual CFT is positive, yielding a strong-coupling sector which is effectively shielded by the exponential wall $e^{-\phi/Q}\sigma_2$. On the other hand, for $k>3$, the strong-coupling region lives near the boundary, and so perturbative string theory should not be trusted.

One could take this statement as evidence that pure NS-NS string theory for $k>3$ is simply a sick theory, or at the very least does not define a background around which perturbative string theory is valid. While this view is understandable, it is also true that the divergence in the integral \eqref{eq:zero-mode-integral-phi} is mild, and can be formally regularized via the analytic continuation
\begin{equation}
\int_{0}^{\infty}\mathrm{d}x\,e^{-\alpha x}:=\frac{1}{\alpha}\,,\quad\alpha\neq 0\,,
\end{equation}
which extends the range of validity from $\text{Re}(\alpha)>0$ to the whole complex plane minus the origin. Thus, it might be reasonable to analytically continue the zero mode integral \eqref{eq:zero-mode-integral-phi} to all values of $k\neq 3$.

\vspace{0.5cm}

We close our discussion with a list of potential avenues for future research.

\paragraph{Local correlators:} In this work we formally computed the long-string free energy and matched it to the perturbative expansion of the boundary CFT partition function. Of course, AdS/CFT doesn't just predict equality of partition functions, but also predicts a matching at the level of correlation functions. Recent work has been extremely fruitful in computing tree-level \cite{Eberhardt:2019ywk,Dei:2020zui,Dei:2021xgh,Dei:2021yom,Dei:2022pkr,Dei:2023ivl,Knighton:2023mhq,Knighton:2024qxd,Sriprachyakul:2024gyl,Yu:2024kxr} and higher-loop level \cite{Eberhardt:2020akk,Hikida:2020kil,Knighton:2020kuh} correlation functions of spectrally-flowed operators in global Euclidean $\text{AdS}_3$, and all computations are consistent with a long-string CFT described by a deformed symmetric orbifold CFT. However, little is known about correlation functions of local operators on locally-$\text{AdS}_3$ spacetimes with nontrivial topologies (see, however, \cite{Eberhardt:2021jvj} in the case of the `minimal tension' string in $\text{AdS}_3\times\text{S}^3\times\mathbb{T}^4$). We believe that the covariant long-string sigma model described in the present paper will be useful in writing down and computing the string path integral associated to spectrally-flowed vertex operators in nontrivial topologies. This is likely to involve a covariant generalization of the delta-function operators introduced in \cite{Dei:2023ivl,Knighton:2023mhq}.

\paragraph{Superstrings:} Another obvious avenue for generalizing the work of this paper is the treatment of superstrings on backgrounds like $\text{AdS}_3\times C$ with pure NS-NS flux. The dual CFT for generic NS-NS flux was guessed already in \cite{Eberhardt:2021vsx} for the most famous cases of $C=\text{S}^3\times\mathbb{T}^4$ and $C=\text{S}^3\times\text{K}3$, and recently three-point functions of long-string states on both sides were shown to match \cite{Yu:2024kxr}. In addition, recent work \cite{Sriprachyakul:2024gyl} has suggested a simple generalization to all consistent compactifications $C$ which schematically looks like
\begin{equation}\label{eq:long-superstring-cft}
\text{Sym}(\mathbb{R}_{\mathcal{Q}}^{(1)}\times C)+\mu\int\sigma_{2,\alpha}\,,
\end{equation}
where $\mathbb{R}^{(1)}_{\mathcal{Q}}$ is an $\mathcal{N}=1$ supersymmetric linear dilaton theory with slope\footnote{The computation of this slope appeared already in \cite{Seiberg:1999xz}.}
\begin{equation}
\mathcal{Q}=-\sqrt{\frac{2(k-1)^2}{k}}
\end{equation}
and $\sigma_{2,\alpha}$ is some BPS operator in the twist-2 sector of the symmetric orbifold which carries momentum $\alpha=\sqrt{k/2}$. In \cite{Sriprachyakul:2024gyl,Yu:2024kxr}, the matching of string correlators with correlators in the CFT \eqref{eq:long-superstring-cft} was based on explicit calculations of long-string correlation functions in the RNS formalism.

Just as the present work provides a concrete derivation of the bosonic long-string CFT from the worldsheet path integral, it seems likely that a similar approach is viable for deriving the long-superstring CFT \eqref{eq:long-superstring-cft}. In the bosonic case, a fundamental component of the analysis carried out in this paper was showing that the conformal structure of the worldsheet can be identified with that of the boundary in the long-string limit. In the case of superstrings, one would hope that a similar structure would emerge, namely that the superconformal structure on the worldsheet could be obtained from the superconformal structure on the boundary. Given that it is difficult to write superstring theories which treat target-space and worldsheet supersymmetry simultaneously in a manifest fashion, this may prove difficult.

% \begin{figure}
% \centering
% \begin{tikzpicture}
% \draw[thick] (0,0) circle (2);
% \draw[Maroon, very thick] (-0.05,1.025) -- (-0.05,-1.025);
% \draw[Maroon, very thick] (0.05,1.025) -- (0.05,-1.025);
% \draw[thick] (45:2) -- (90:1) -- (135:2);
% \draw[thick] (-45:2) -- (-90:1) -- (-135:2);
% \node[above right] at (45:2) {$\sigma_{w_1,q_1}$};
% \node[above left] at (135:2) {$\sigma_{w_2,q_2}$};
% \node[below left] at (-135:2) {$\sigma_{w_3,q_3}$};
% \node[below right] at (-45:2) {$\sigma_{w_4,q_4}$};
% \end{tikzpicture}
% \caption{A long-string correlator which contains a short-string `resonance' in the bulk.}
% \label{fig:bulk-resonance}
% \end{figure}

\paragraph{Short strings and bound states:} In this work we have carefully derived the effective CFT of long-strings in locally-$\text{AdS}_3$ spacetimes. This effective theory has many interesting properties. For example, it is universal (in the sense that all long-string CFTs are structurally similar, and depend very little on the details of the compactification), and background-independent (in the sense that it only depends on the conformal boundary, and not on the details of the bulk). One major drawback is that, by its very nature, it does not seem to describe the propagation of short strings in the bulk. This is partly puzzling since, by AdS/CFT, there should be a `full' dual CFT which describes both long and short strings. However, the long-string CFT is already a full-fledged 2D CFT which reproduces at least a subset of the physics of strings in $\text{AdS}_3$ and which, in a given winding sector, reproduces the central charge $c=6kN$ predicted by the Brown-Henneaux formula. With all of this in mind, it seems difficult to consider a minimal modification of the long-string CFT that can include short strings (for example, by tensoring with a `short-string' CFT).

One possible resolution to this puzzle suggested in \cite{Eberhardt:2021vsx} is that short strings in the bulk may arise as nonlocal `bound states' in the long-string CFT. Evidence for this claim was provided by noting that local correlators in the long-string CFT contain `resonances', similar to the kinematic poles of QFT scattering amplitudes, hinting at the existence of on-shell intermediate states in the bulk of $\text{AdS}_3$. This proposal would imply, then, that the complicated dynamics of short strings propagating in the bulk can be described simply by bound states in the potential of the long-string CFT. Even more surprising is the case in which there are multiple bulk manifolds for a given boundary. In this case, the complicated details of summing over bulk geometries with a fixed boundary topology would somehow result in something as simple as computing a resonance in an interacting CFT. Thus, while the details of how the short-string sector actually enters the dual CFT are still poorly understood, this direction seems very promising for future investigation.

\paragraph{Acknowledgements} I would like to thank Alejandra Castro, Andrea Dei, Lorenz Eberhardt, Alex Frenkel, Emil Martinec, Kiarash Naderi, Volker Schomerus, Sean Seet, Vit Sriprachyakul, David Skinner, and Jakub Vo\v{s}mera for useful discussions. I also thank the organizers of the Institut-Pascal workshop `Speakable and Unspeakable in Quantum Gravity', where many of these ideas first took shape. I especially thank Andrea Dei, Vit Sriprachyakul, and Jakub Vo\v{s}mera for helpful comments on an early version of the manuscript. This work was supported by STFC consolidated grants ST/T000694/1 and ST/X000664/1.

\appendix

\section{The path integral anomaly}\label{app:measure}

In this appendix we derive the transformation law of the path integral measure $\mathcal{D}x$ in the long-string theory on a hyperbolic 3-manifold $M$ under Weyl transformations of the boundary metric. Specifically, we consider the path integral measure $\mathrm{d}\mu[g]$ defined implicitly by the inner product
\begin{equation}
\braket{\delta x_1,\delta x_2}=\int_{\Sigma}\mathrm{d}^2\sigma\sqrt{h}g_{ij}(x)\delta x_1^i\delta x_2^j
\end{equation}
on the tangent space to the space of maps $\text{Map}(\Sigma\to X)$, where $X$ is the conformal boundary of $M$. Our goal is to show that, on the support of the delta function $\delta(Dx)$ in the path integral \eqref{eq:long-string-free-energy}, we have
\begin{equation}\label{eq:anomalous-transformation-appendix}
\frac{\mathrm{d}\mu[e^{2\omega}g]}{\mathrm{d}\mu[g]}=\exp\left(-\frac{1}{2\pi}\int_{\Sigma}\mathrm{d}^2\sigma\left(\sqrt{h}h^{ab}\partial_a\omega\partial_b\omega+\frac{1}{2}\sqrt{h}R_h\omega+x^*(\sqrt{g}R)\omega\right)\right)\,,
\end{equation}
where $R$ is the scalar curvature of the reference metric $g$, and $x^*$ denotes the pullback from $X$ to $\Sigma$.

As a starting point, we will consider the flat metric $g_{ij}=\delta_{ij}$. In this case, the transformation of the measure is actually known, and the result is (see, for example, \cite{Gerasimov:1990fi,Ishibashi:2000fn})
\begin{equation}\label{eq:flat-metric-weyl-anomaly}
\frac{\mathrm{d}\mu[e^{2\omega}\delta]}{\mathrm{d}\mu[\delta]}=\exp\left(-\frac{1}{2\pi}\int_{\Sigma}\mathrm{d}^2\sigma\left(\sqrt{h}h^{ab}\partial_a\omega\partial_b\omega+\frac{1}{2}\sqrt{h}R_h\omega\right)\right)\,.
\end{equation}
Now, we use the fact that, locally, for a given metric $g$, there exists a set of coordinates for which $g_{ij}=e^{\rho}\delta_{ij}$. Thus, locally, we have
\begin{equation}
\frac{\mathrm{d}\mu[e^{2\omega}g]}{\mathrm{d}\mu[g]}=\frac{\mathrm{d}\mu[e^{2\omega+\rho}\delta]}{\mathrm{d}\mu[e^{\rho}\delta]}=\left(\frac{\mathrm{d}\mu[e^{2\omega+\rho}\delta]}{\mathrm{d}\mu[\delta]}\right)\left(\frac{\mathrm{d}\mu[e^{\rho}\delta]}{\mathrm{d}\mu[\delta]}\right)^{-1}\,.
\end{equation}
It follows from \eqref{eq:flat-metric-weyl-anomaly} that
\begin{equation}
\frac{\mathrm{d}\mu[e^{2\omega}g]}{\mathrm{d}\mu[g]}=\exp\left(-\frac{1}{2\pi}\int_{\Sigma}\mathrm{d}^2\sigma\left(\sqrt{h}h^{ab}\partial_a\omega\partial_b\omega+\frac{1}{2}\sqrt{h}R_h\omega-\sqrt{h}\nabla_h\rho\,\omega\right)\right)\,,
\end{equation}
where $\nabla_h\rho=h^{-1/2}h^{ab}\partial_a(\sqrt{h}\partial_b\rho)$ is the Laplacian on the worldsheet.

Now, the path integral of the worldsheet theory near the conformal boundary of $M$ localizes to maps for which $x^*g$ and $h$ are in the same conformal class. Alternatively, the integral over worldsheet metrics localizes to those metrics which are in the same conformal class as $x^*g$. In either case, we can assume that $h$ and $x^*g$ are related by a Weyl transformation, which in turn implies that
\begin{equation}
\sqrt{h}\nabla_h\rho=\text{det}(\mathrm{d}x)\,\delta^{ij}\partial_i\partial_j\rho\,,
\end{equation}
where the derivatives on the right-hand-side are with respect to the coordinates on $X$. Now, recall that for a conformal-gauge metric $g_{ij}=e^{\rho}\delta_{ij}$, the scalar curvature is given by
\begin{equation}
\sqrt{g}R=-\delta^{ij}\partial_i\partial_j\rho\,,
\end{equation}
so that $\text{det}(\mathrm{d}x)\,\delta^{ij}\partial_i\partial_j\rho$ is the pullback of $\sqrt{g}R$ onto the worldsheet, where the determinant term comes from the fact that $\sqrt{g}$ is a scalar density, rather than a scalar. Thus, we obtain the transformation rule \eqref{eq:anomalous-transformation-appendix}.

\section{Harmonic maps to the torus}\label{app:harmonic-functions}

In this section we collect various properties of harmonic maps $\gamma:\Sigma\to\mathbb{T}^2_t$ which are used in the computation of the long-string partition function on thermal $\text{AdS}_3$.

Since the torus $\mathbb{T}^2_t$ with the flat metric can be identified by the quotient
\begin{equation}
\mathbb{T}^2_t=\mathbb{C}/(\mathbb{Z}\oplus\tau\mathbb{Z})\,,
\end{equation}
a harmonic map $\gamma:\Sigma\to\mathbb{T}^2_t$ can be thought of as a multivalued function on $\Sigma$ satisfying the following two properties:
\begin{itemize}

    \item $\gamma$ is harmonic: $\partial\overline{\partial}\gamma=0$.

    \item For every loop $\rho\in\pi_1(\Sigma,p)$ based at a point $p\in\Sigma$, we have
    \begin{equation}
    \gamma(\rho\cdot p)=\gamma(p)+a_{\rho}t+b_{\rho}
    \end{equation}
    for some integers $a_{\rho},b_{\rho}$.
    
\end{itemize}
The second condition specifies the data of a homomorphism $\varphi:\pi_1(\Sigma,p)\to\mathbb{Z}^2$. Since the image of $\varphi$ is abelian, we can equally replace $\pi_1(\Sigma,p)$ with its abelianization $\mathbb{Z}^{2g}$. That is, the topological classes of harmonic maps $\gamma$ are labeled by $2\times 2g$ integer matrices
\begin{equation}
\begin{pmatrix}
 d_1 & \cdots d_g & b_1 & \cdots & b_g\\
 c_1 & \cdots c_g & a_1 & \cdots & a_g
\end{pmatrix}\,.
\end{equation}
The periodicity conditions for $\gamma$ are then captured in the integral equations
\begin{equation}\label{eq:app-gamma-periodicity}
\oint_{A_i}\mathrm{d}\gamma=c_it+d_i\,,\quad\oint_{B_i}\mathrm{d}\gamma=a_it+b_i\,,
\end{equation}
where $A_i,B_i$ are a choice of homology cycles on $\Sigma$ satisfying $A_i\cap B_j=\delta_{ij}$.

Now, since $\gamma$ is harmonic, we know that $\partial\gamma$ is a holomorphic $(1,0)$-form on $\Sigma$. Similarly, $\overline{\partial}\gamma$ is an anti-holomorphic $(0,1)$-form. Given a normalized set of $g$ holomorphic $(1,0)$-forms $\omega_i$ satisfying
\begin{equation}
\oint_{A_i}\omega_j=\delta_{ij}\,,\quad \oint_{B_i}\omega_j=\Omega_{ij}\,,
\end{equation}
where $\Omega_{ij}$ is the period matrix of $\Sigma$, we can expand $\partial\gamma$ and $\overline{\partial}\gamma$ in the basis $\omega_i$ and $\bar{\omega}_i$, i.e. we can write (using Einstein summation)
\begin{equation}
\mathrm{d}\gamma=\alpha_i\omega_i+\beta_i\bar{\omega}_i\,.
\end{equation}
The periodicity conditions \eqref{eq:app-gamma-periodicity} determine the coefficients $\alpha_i,\beta_i$ uniquely, since they must satisfy the equations
\begin{equation}
\begin{split}
\oint_{A_i}\mathrm{d}\gamma&=\alpha_i+\beta_i=c_it+d_i\,,\\
\oint_{B_i}\mathrm{d}\gamma&=\Omega_{ij}\alpha_j+\overline{\Omega}_{ij}\beta_j=a_it+b_i\,,
\end{split}
\end{equation}
and thus
\begin{equation}
\begin{split}
\alpha_j&=-\frac{1}{2i}\text{Im}(\Omega)_{ij}^{-1}\left[\overline{\Omega}_{jk}(c_kt+d_k)-(a_jt+b_j)\right]\\
\beta_i&=\frac{1}{2i}\text{Im}(\Omega)^{-1}_{ij}\left[\Omega_{jk}(c_kt+d_k)-(a_jt+b_j)\right]\,.
\end{split}
\end{equation}
From the above expressions, we can readily determine the value of the map $\gamma$ up to an overall constant through the integral
\begin{equation}
\begin{split}
\gamma(p)=-\frac{1}{2i}\int_{p_0}^{p}\text{Im}(\Omega)^{-1}_{ij}\big(&\left[\overline{\Omega}_{jk}(c_kt+d_k)-(a_jt+b_j)\right]\omega_i\\
&-\left[\Omega_{jk}(c_kt+d_k)-(a_jt+b_j)\right]\bar{\omega}_i\big)\,,
\end{split}
\end{equation}
where $p_0$ is some arbitrarily chosen base point which determines the constant of integration\footnote{This integral is independent of the choice of path from $p_0$ to $p$, up to the identification $\gamma\sim\gamma+1\sim\gamma+t$.}. We note that in the case that $\gamma$ is holomorphic, we demand $\beta_i=0$, that is
\begin{equation}\label{eq:period-matrices-holomorphic}
\Omega_{ij}(c_jt+d_j)=(a_jt+d_j)\,.
\end{equation}
This is a series of $g$ equations which restricts the allowed values of the period matrix $\Omega_{ij}$. When $\Omega_{ij}$ is chosen to satisfy the above equations, we can write
\begin{equation}
\gamma=\sum_{i=1}^{g}(c_it+d_i)\int_{p_0}^{p}\omega_i\,.
\end{equation}
We emphasize that such a $\gamma$ can exist only when the period matrix $\Omega$ is fine-tuned by the constraint \eqref{eq:period-matrices-holomorphic}.

In the main text, we are interested in computing the on-shell action
\begin{equation}
\int_{\Sigma}\mathrm{d}^2z|\overline{\partial}\gamma|^2=-\frac{1}{2i}\int_{\Sigma}\overline{\partial}\gamma\wedge\partial\bar\gamma\,.
\end{equation}
This can be computed using the Riemann bilinear relations
\begin{equation}
\int_{\Sigma}\omega\wedge\omega'=\sum_{i=1}^{g}\left(\oint_{A_i}\omega\oint_{B_i}\omega'-\oint_{B_i}\omega\oint_{A_i}\omega'\right)\,,
\end{equation}
which hold for any closed one-forms $\omega,\omega'$ on $\Sigma$. Plugging in $\omega=\overline{\partial}\gamma$ and $\omega'=\partial\bar\gamma$ and recalling that the period matrix of any Riemann surface is symmetric gives
\begin{equation}
\begin{split}
\int_{\Sigma}\mathrm{d}^2z\,|\overline{\partial}\gamma|^2&=\bar{\beta}_i\text{Im}(\Omega)_{ij}\beta_j\\
&=\frac{1}{4}v^{\dagger}\text{Im}(\Omega)^{-1}v\,,
\end{split}
\end{equation}
where
\begin{equation}
v_i=\Omega_{ij}(c_jt+d_j)-(a_it+b_i)\,.
\end{equation}
Since $\text{Im}(\Omega)$ is a positive-definite matrix, this integral is non-negative.

\bibliography{references.bib}

\providecommand{\href}[2]{#2}\begingroup\raggedright\begin{thebibliography}{10}

\bibitem{Maldacena:1997re}
J.~M. Maldacena, ``{The Large $N$ Limit of Superconformal Field Theories and
  Supergravity},'' {\em Int. J. Theor. Phys.} {\bf 38} (1999) 1113--1133,
  \href{http://www.arXiv.org/abs/hep-th/9711200}{{\tt hep-th/9711200}}.
[Adv. Theor. Math. Phys.2,231(1998)].
%%CITATION = HEP-TH/9711200;%%.

\bibitem{Witten:1998qj}
E.~Witten, ``{Anti-de Sitter space and holography},'' {\em Adv. Theor. Math.
  Phys.} {\bf 2} (1998) 253--291,
  \href{http://www.arXiv.org/abs/hep-th/9802150}{{\tt hep-th/9802150}}.

\bibitem{Maldacena:2000hw}
J.~M. Maldacena and H.~Ooguri, ``{Strings in $\mathrm{AdS}_3$ and
  $\mathrm{SL}(2,\mathbb{R})$ WZW model 1.: The Spectrum},'' {\em J. Math.
  Phys.} {\bf 42} (2001) 2929--2960,
\href{http://www.arXiv.org/abs/hep-th/0001053}{{\tt hep-th/0001053}}.
%%CITATION = HEP-TH/0001053;%%.

\bibitem{Knighton:2023mhq}
B.~Knighton, S.~Seet, and V.~Sriprachyakul, ``{Spectral flow and localisation
  in AdS$_{3}$ string theory},'' {\em JHEP} {\bf 05} (2024) 113,
  \href{http://www.arXiv.org/abs/2312.08429}{{\tt 2312.08429}}.

\bibitem{Knighton:2024qxd}
B.~Knighton and V.~Sriprachyakul, ``{Unravelling AdS$_3$/CFT$_2$ near the
  boundary},'' \href{http://www.arXiv.org/abs/2404.07296}{{\tt 2404.07296}}.

\bibitem{Sriprachyakul:2024gyl}
V.~Sriprachyakul, ``{Superstrings near the conformal boundary of $\rm
  AdS_3$},'' \href{http://www.arXiv.org/abs/2405.03678}{{\tt 2405.03678}}.

\bibitem{Seiberg:1999xz}
N.~Seiberg and E.~Witten, ``{The D1 / D5 System and Singular CFT},'' {\em JHEP}
  {\bf 04} (1999) 017, \href{http://www.arXiv.org/abs/hep-th/9903224}{{\tt
  hep-th/9903224}}.

\bibitem{Hosomichi:1999uj}
K.~Hosomichi and Y.~Sugawara, ``{Multistrings on AdS(3) x S**3 from matrix
  string theory},'' {\em JHEP} {\bf 07} (1999) 027,
  \href{http://www.arXiv.org/abs/hep-th/9905004}{{\tt hep-th/9905004}}.

\bibitem{Hikida:2000ry}
Y.~Hikida, K.~Hosomichi, and Y.~Sugawara, ``{String theory on AdS(3) as
  discrete light cone Liouville theory},'' {\em Nucl. Phys. B} {\bf 589} (2000)
  134--166, \href{http://www.arXiv.org/abs/hep-th/0005065}{{\tt
  hep-th/0005065}}.

\bibitem{Argurio:2000tb}
R.~Argurio, A.~Giveon, and A.~Shomer, ``{Superstrings on Ad$S_3$ and Symmetric
  Products},'' {\em JHEP} {\bf 12} (2000) 003,
  \href{http://www.arXiv.org/abs/hep-th/0009242}{{\tt hep-th/0009242}}.

\bibitem{Eberhardt:2021vsx}
L.~Eberhardt, ``{A perturbative CFT dual for pure NS\textendash{}NS AdS$_{3}$
  strings},'' {\em J. Phys. A} {\bf 55} (2022), no.~6, 064001,
  \href{http://www.arXiv.org/abs/2110.07535}{{\tt 2110.07535}}.

\bibitem{Balthazar:2021xeh}
B.~Balthazar, A.~Giveon, D.~Kutasov, and E.~J. Martinec, ``{Asymptotically free
  AdS$_{3}$/CFT$_{2}$},'' {\em JHEP} {\bf 01} (2022) 008,
  \href{http://www.arXiv.org/abs/2109.00065}{{\tt 2109.00065}}.

\bibitem{Eberhardt:2019qcl}
L.~Eberhardt and M.~R. Gaberdiel, ``{String theory on $\mathrm{AdS}_3$ and the
  symmetric orbifold of Liouville theory},'' {\em Nucl. Phys. B} {\bf 948}
  (2019) 114774, \href{http://www.arXiv.org/abs/1903.00421}{{\tt 1903.00421}}.

\bibitem{Dei:2021xgh}
A.~Dei and L.~Eberhardt, ``{String correlators on AdS$_{3}$: three-point
  functions},'' {\em JHEP} {\bf 08} (2021) 025,
  \href{http://www.arXiv.org/abs/2105.12130}{{\tt 2105.12130}}.

\bibitem{Dei:2021yom}
A.~Dei and L.~Eberhardt, ``{String correlators on AdS$_{3}$: four-point
  functions},'' {\em JHEP} {\bf 09} (2021) 209,
  \href{http://www.arXiv.org/abs/2107.01481}{{\tt 2107.01481}}.

\bibitem{Bufalini:2022toj}
D.~Bufalini, S.~Iguri, and N.~Kovensky, ``{A proof for string three-point
  functions in AdS$_{3}$},'' {\em JHEP} {\bf 02} (2023) 246,
  \href{http://www.arXiv.org/abs/2212.05877}{{\tt 2212.05877}}.

\bibitem{Dei:2022pkr}
A.~Dei and L.~Eberhardt, ``{String correlators on $\text{AdS}_3$: Analytic
  structure and dual CFT},'' {\em SciPost Phys.} {\bf 13} (2022), no.~3, 053,
  \href{http://www.arXiv.org/abs/2203.13264}{{\tt 2203.13264}}.

\bibitem{Hikida:2023jyc}
Y.~Hikida and V.~Schomerus, ``{Engineering perturbative string duals for
  symmetric product orbifold CFTs},'' {\em JHEP} {\bf 06} (2024) 071,
  \href{http://www.arXiv.org/abs/2312.05317}{{\tt 2312.05317}}.

\bibitem{Yu:2024kxr}
Z.-f. Yu and C.~Peng, ``{Correlators of long strings on
  AdS$_3\times$S$^3\times$T$^4$},''
  \href{http://www.arXiv.org/abs/2408.16712}{{\tt 2408.16712}}.

\bibitem{Gawedzki:1991yu}
K.~Gawedzki, ``{Noncompact WZW Conformal Field Theories},'' in {\em {Nato
  Advanced Study Institute: New Symmetry Principles in Quantum Field Theory}},
  pp.~0247--274.
\newblock 10, 1991.
\newblock \href{http://www.arXiv.org/abs/hep-th/9110076}{{\tt hep-th/9110076}}.

\bibitem{Maldacena:2000kv}
J.~M. Maldacena, H.~Ooguri, and J.~Son, ``{Strings in $\mathrm{AdS}_3$ and the
  $\mathrm{SL}(2,\mathbb{R})$ WZW model. Part 2. Euclidean black hole},'' {\em
  J. Math. Phys.} {\bf 42} (2001) 2961--2977,
\href{http://www.arXiv.org/abs/hep-th/0005183}{{\tt hep-th/0005183}}.
%%CITATION = HEP-TH/0005183;%%.

\bibitem{Dorn:1994xn}
H.~Dorn and H.~J. Otto, ``{Two and three point functions in Liouville
  theory},'' {\em Nucl. Phys. B} {\bf 429} (1994) 375--388,
  \href{http://www.arXiv.org/abs/hep-th/9403141}{{\tt hep-th/9403141}}.

\bibitem{Zamolodchikov:1995aa}
A.~B. Zamolodchikov and A.~B. Zamolodchikov, ``{Structure Constants and
  Conformal Bootstrap in Liouville Field Theory},'' {\em Nucl. Phys. B} {\bf
  477} (1996) 577--605, \href{http://www.arXiv.org/abs/hep-th/9506136}{{\tt
  hep-th/9506136}}.

\bibitem{Lunin:2000yv}
O.~Lunin and S.~D. Mathur, ``{Correlation Functions for $ M^N/S^N$
  Orbifolds},'' {\em Commun. Math. Phys.} {\bf 219} (2001) 399--442,
  \href{http://www.arXiv.org/abs/hep-th/0006196}{{\tt hep-th/0006196}}.

\bibitem{Lunin:2001pw}
O.~Lunin and S.~D. Mathur, ``{Three point functions for M(N) / S(N) orbifolds
  with N=4 supersymmetry},'' {\em Commun. Math. Phys.} {\bf 227} (2002)
  385--419, \href{http://www.arXiv.org/abs/hep-th/0103169}{{\tt
  hep-th/0103169}}.

\bibitem{Maldacena:2001km}
J.~M. Maldacena and H.~Ooguri, ``{Strings in $\mathrm{AdS}_3$ and the
  $\mathrm{SL}(2,\mathbb{R})$ WZW model. Part 3. Correlation functions},'' {\em
  Phys. Rev.} {\bf D65} (2002) 106006,
\href{http://www.arXiv.org/abs/hep-th/0111180}{{\tt hep-th/0111180}}.
%%CITATION = HEP-TH/0111180;%%.

\bibitem{Eberhardt:2019ywk}
L.~Eberhardt, M.~R. Gaberdiel, and R.~Gopakumar, ``{Deriving the
  AdS$_{3}$/CFT$_{2}$ correspondence},'' {\em JHEP} {\bf 02} (2020) 136,
  \href{http://www.arXiv.org/abs/1911.00378}{{\tt 1911.00378}}.

\bibitem{Eberhardt:2020akk}
L.~Eberhardt, ``{$\mathrm{AdS}_{3}/\mathrm{CFT}_{2}$ at higher genus},'' {\em
  JHEP} {\bf 05} (2020) 150, \href{http://www.arXiv.org/abs/2002.11729}{{\tt
  2002.11729}}.

\bibitem{Gaberdiel:2018rqv}
M.~R. Gaberdiel and R.~Gopakumar, ``{Tensionless string spectra on
  $\mathrm{AdS}_{3}$},'' {\em JHEP} {\bf 05} (2018) 085,
  \href{http://www.arXiv.org/abs/1803.04423}{{\tt 1803.04423}}.

\bibitem{Eberhardt:2018ouy}
L.~Eberhardt, M.~R. Gaberdiel, and R.~Gopakumar, ``{The Worldsheet Dual of the
  Symmetric Product CFT},'' {\em JHEP} {\bf 04} (2019) 103,
  \href{http://www.arXiv.org/abs/1812.01007}{{\tt 1812.01007}}.

\bibitem{Giribet:2018ada}
G.~Giribet, C.~Hull, M.~Kleban, M.~Porrati, and E.~Rabinovici, ``{Superstrings
  on $\mathrm{AdS}_{3}$ at $k = 1$},'' {\em JHEP} {\bf 08} (2018) 204,
  \href{http://www.arXiv.org/abs/1803.04420}{{\tt 1803.04420}}.

\bibitem{Dei:2020zui}
A.~Dei, M.~R. Gaberdiel, R.~Gopakumar, and B.~Knighton, ``{Free field
  world-sheet correlators for ${\rm AdS}_3$},'' {\em JHEP} {\bf 02} (2021) 081,
  \href{http://www.arXiv.org/abs/2009.11306}{{\tt 2009.11306}}.

\bibitem{Gaberdiel:2020ycd}
M.~R. Gaberdiel, R.~Gopakumar, B.~Knighton, and P.~Maity, ``{From Symmetric
  Product CFTs to ${\rm AdS}_3$},''
  \href{http://www.arXiv.org/abs/2011.10038}{{\tt 2011.10038}}.

\bibitem{Knighton:2020kuh}
B.~Knighton, ``{Higher genus correlators for tensionless AdS$_{3}$ strings},''
  {\em JHEP} {\bf 04} (2021) 211,
  \href{http://www.arXiv.org/abs/2012.01445}{{\tt 2012.01445}}.

\bibitem{Eberhardt:2020bgq}
L.~Eberhardt, ``{Partition functions of the tensionless string},'' {\em JHEP}
  {\bf 03} (2021) 176, \href{http://www.arXiv.org/abs/2008.07533}{{\tt
  2008.07533}}.

\bibitem{Eberhardt:2021jvj}
L.~Eberhardt, ``{Summing over Geometries in String Theory},'' {\em JHEP} {\bf
  05} (2021) 233, \href{http://www.arXiv.org/abs/2102.12355}{{\tt 2102.12355}}.

\bibitem{Gaberdiel:2021kkp}
M.~R. Gaberdiel, B.~Knighton, and J.~Vo\v{s}mera, ``{D-branes in AdS$_{3}$
  \texttimes{} S$^{3}$ \texttimes{} \ensuremath{\mathbb{T}}$^{4}$ at k = 1 and
  their holographic duals},'' {\em JHEP} {\bf 12} (2021) 149,
  \href{http://www.arXiv.org/abs/2110.05509}{{\tt 2110.05509}}.

\bibitem{Gaberdiel:2022oeu}
M.~R. Gaberdiel and B.~Nairz, ``{BPS correlators for AdS$_{3}$/CFT$_{2}$},''
  {\em JHEP} {\bf 09} (2022) 244,
  \href{http://www.arXiv.org/abs/2207.03956}{{\tt 2207.03956}}.

\bibitem{McStay:2023thk}
N.~M. McStay and R.~A. Reid-Edwards, ``{Symmetries and Covering Maps for the
  Minimal Tension String on $\text{AdS}_3 \times \text{S}^3 \times
  \text{T}^4$},'' \href{http://www.arXiv.org/abs/2306.16280}{{\tt 2306.16280}}.

\bibitem{Dei:2023ivl}
A.~Dei, B.~Knighton, and K.~Naderi, ``{Solving AdS$_{3}$ string theory at
  minimal tension: tree-level correlators},'' {\em JHEP} {\bf 09} (2024) 135,
  \href{http://www.arXiv.org/abs/2312.04622}{{\tt 2312.04622}}.

\bibitem{Gaberdiel:2023lco}
M.~R. Gaberdiel, R.~Gopakumar, and B.~Nairz, ``{Beyond the Tensionless Limit:
  Integrability in the Symmetric Orbifold},''
  \href{http://www.arXiv.org/abs/2312.13288}{{\tt 2312.13288}}.

\bibitem{Knighton:2024ybs}
B.~Knighton, ``{A note on background independence in $\text{AdS}_3$ string
  theory},'' \href{http://www.arXiv.org/abs/2404.19571}{{\tt 2404.19571}}.

\bibitem{Pakman:2009zz}
A.~Pakman, L.~Rastelli, and S.~S. Razamat, ``{Diagrams for Symmetric Product
  Orbifolds},'' {\em JHEP} {\bf 10} (2009) 034,
  \href{http://www.arXiv.org/abs/0905.3448}{{\tt 0905.3448}}.

\bibitem{Kim:2015gak}
J.~Kim and M.~Porrati, ``{On the central charge of spacetime current algebras
  and correlators in string theory on AdS$_{3}$},'' {\em JHEP} {\bf 05} (2015)
  076, \href{http://www.arXiv.org/abs/1503.07186}{{\tt 1503.07186}}.

\bibitem{Graham:1991jqw}
C.~R. Graham and J.~M. Lee, ``{Einstein metrics with prescribed conformal
  infinity on the ball},'' {\em Adv. Math.} {\bf 87} (1991), no.~2, 186--225.

\bibitem{Brown:1986nw}
J.~D. Brown and M.~Henneaux, ``{Central Charges in the Canonical Realization of
  Asymptotic Symmetries: An Example from Three-Dimensional Gravity},'' {\em
  Commun. Math. Phys.} {\bf 104} (1986) 207--226.

\bibitem{Witten:1988xj}
E.~Witten, ``{Topological Sigma Models},'' {\em Commun. Math. Phys.} {\bf 118}
  (1988) 411.

\bibitem{Giveon:1998ns}
A.~Giveon, D.~Kutasov, and N.~Seiberg, ``{Comments on String Theory on
  $\mathrm{AdS}_3$},'' {\em Adv. Theor. Math. Phys.} {\bf 2} (1998) 733--782,
  \href{http://www.arXiv.org/abs/hep-th/9806194}{{\tt hep-th/9806194}}.

\bibitem{Kutasov:1999xu}
D.~Kutasov and N.~Seiberg, ``{More Comments on String Theory on
  $\mathrm{AdS}_3$},'' {\em JHEP} {\bf 04} (1999) 008,
  \href{http://www.arXiv.org/abs/hep-th/9903219}{{\tt hep-th/9903219}}.

\bibitem{Wakimoto:1986gf}
M.~Wakimoto, ``{Fock representations of the affine lie algebra A1(1)},'' {\em
  Commun. Math. Phys.} {\bf 104} (1986) 605--609.

\bibitem{Dei:2019iym}
A.~Dei and L.~Eberhardt, ``{Correlators of the Symmetric Product Orbifold},''
  {\em JHEP} {\bf 01} (2020) 108,
  \href{http://www.arXiv.org/abs/1911.08485}{{\tt 1911.08485}}.

\bibitem{Kames-King:2023fpa}
J.~Kames-King, A.~Kanargias, B.~Knighton, and M.~Usatyuk, ``{The lion, the
  witch, and the wormhole: ensemble averaging the symmetric product
  orbifold},'' {\em JHEP} {\bf 07} (2024) 236,
  \href{http://www.arXiv.org/abs/2306.07321}{{\tt 2306.07321}}.

\bibitem{Dixon:1986qv}
L.~J. Dixon, D.~Friedan, E.~J. Martinec, and S.~H. Shenker, ``{The Conformal
  Field Theory of Orbifolds},'' {\em Nucl. Phys. B} {\bf 282} (1987) 13--73.

\bibitem{Hamidi:1986vh}
S.~Hamidi and C.~Vafa, ``{Interactions on Orbifolds},'' {\em Nucl. Phys. B}
  {\bf 279} (1987) 465--513.

\bibitem{Dijkgraaf:1996xw}
R.~Dijkgraaf, G.~W. Moore, E.~P. Verlinde, and H.~L. Verlinde, ``{Elliptic
  genera of symmetric products and second quantized strings},'' {\em Commun.
  Math. Phys.} {\bf 185} (1997) 197--209,
  \href{http://www.arXiv.org/abs/hep-th/9608096}{{\tt hep-th/9608096}}.

\bibitem{Aharony:2024fid}
O.~Aharony and E.~Y. Urbach, ``{Type II string theory on
  AdS3\texttimes{}S3\texttimes{}T4 and symmetric orbifolds},'' {\em Phys. Rev.
  D} {\bf 110} (2024), no.~4, 046028,
  \href{http://www.arXiv.org/abs/2406.14605}{{\tt 2406.14605}}.

\bibitem{Henningson:1998gx}
M.~Henningson and K.~Skenderis, ``{The Holographic Weyl anomaly},'' {\em JHEP}
  {\bf 07} (1998) 023, \href{http://www.arXiv.org/abs/hep-th/9806087}{{\tt
  hep-th/9806087}}.

\bibitem{Hashimoto:2019wct}
A.~Hashimoto and D.~Kutasov, ``{$ T\overline{T},J\overline{T},T\overline{J} $
  partition sums from string theory},'' {\em JHEP} {\bf 02} (2020) 080,
  \href{http://www.arXiv.org/abs/1907.07221}{{\tt 1907.07221}}.

\bibitem{Polchinski:1998rq}
J.~Polchinski, {\em {String theory. Vol. 1: An introduction to the bosonic
  string}}.
\newblock Cambridge Monographs on Mathematical Physics. Cambridge University
  Press, 12, 2007.

\bibitem{Dei:2024sct}
A.~Dei, B.~Knighton, K.~Naderi, and S.~Sethi, ``{Tensionless AdS$_3$/CFT$_2$
  and Single Trace $T\overline{T}$},''
  \href{http://www.arXiv.org/abs/2408.00823}{{\tt 2408.00823}}.

\bibitem{Frenkel:2005ku}
E.~Frenkel and A.~Losev, ``{Mirror symmetry in two steps: A-I-B},'' {\em
  Commun. Math. Phys.} {\bf 269} (2006) 39--86,
  \href{http://www.arXiv.org/abs/hep-th/0505131}{{\tt hep-th/0505131}}.

\bibitem{Giribet:2001ft}
G.~Giribet and C.~A. Nunez, ``{Correlators in $\mathrm{AdS}_3$ string
  theory},'' {\em JHEP} {\bf 06} (2001) 010,
  \href{http://www.arXiv.org/abs/hep-th/0105200}{{\tt hep-th/0105200}}.

\bibitem{Iguri:2007af}
S.~Iguri and C.~A. Nunez, ``{Coulomb integrals for the
  $\mathrm{SL}(2,\mathbb{R})$ WZW model},'' {\em Phys. Rev. D} {\bf 77} (2008)
  066015, \href{http://www.arXiv.org/abs/0705.4461}{{\tt 0705.4461}}.

\bibitem{Eberhardt:2021ynh}
L.~Eberhardt and S.~Pal, ``{The disk partition function in string theory},''
  {\em JHEP} {\bf 08} (2021) 026,
  \href{http://www.arXiv.org/abs/2105.08726}{{\tt 2105.08726}}.

\bibitem{Roumpedakis:2018tdb}
K.~Roumpedakis, ``{Comments on the S$_{N}$ orbifold CFT in the large
  $N$-limit},'' {\em JHEP} {\bf 07} (2018) 038,
  \href{http://www.arXiv.org/abs/1804.03207}{{\tt 1804.03207}}.

\bibitem{Maldacena:2004rf}
J.~M. Maldacena and L.~Maoz, ``{Wormholes in AdS},'' {\em JHEP} {\bf 02} (2004)
  053, \href{http://www.arXiv.org/abs/hep-th/0401024}{{\tt hep-th/0401024}}.

\bibitem{Martinec:2021vpk}
E.~J. Martinec, ``{AdS3's with and without BTZ's},''
  \href{http://www.arXiv.org/abs/2109.11716}{{\tt 2109.11716}}.

\bibitem{Hikida:2020kil}
Y.~Hikida and T.~Liu, ``{Correlation functions of symmetric orbifold from
  $\mathrm{AdS}_{3}$ string theory},'' {\em JHEP} {\bf 09} (2020) 157,
  \href{http://www.arXiv.org/abs/2005.12511}{{\tt 2005.12511}}.

\bibitem{Gerasimov:1990fi}
A.~Gerasimov, A.~Morozov, M.~Olshanetsky, A.~Marshakov, and S.~L. Shatashvili,
  ``{Wess-Zumino-Witten model as a theory of free fields},'' {\em Int. J. Mod.
  Phys. A} {\bf 5} (1990) 2495--2589.

\bibitem{Ishibashi:2000fn}
N.~Ishibashi, K.~Okuyama, and Y.~Satoh, ``{Path integral approach to string
  theory on AdS(3)},'' {\em Nucl. Phys. B} {\bf 588} (2000) 149--177,
  \href{http://www.arXiv.org/abs/hep-th/0005152}{{\tt hep-th/0005152}}.

\end{thebibliography}\endgroup
\bibliographystyle{utphys.bst}

\end{document}